\documentclass[aps,prx,epsf,twocolumn,showpacs]{revtex4-2}
\usepackage [latin1]{inputenc}
\usepackage{CJK,amsmath,amssymb,graphics,epsfig,epstopdf,color,verbatim,ulem,braket,tabularx,float, lipsum, multirow,setspace,siunitx}
\usepackage[colorlinks,linkcolor=blue,citecolor=blue,urlcolor=blue]{hyperref}

\begin{document}
\begin{CJK*}{GBK}{song}
\title{\mbox{Trilayer  multi-orbital models of $\mathrm{La_{4}Ni_{3}O_{10}}$} }

\author{Cui-Qun Chen}
\thanks{These authors contributed equally to this work}
\author{Zhihui Luo}
\thanks{These authors contributed equally to this work}
\author{Meng Wang}
\author{W\'ei W\'u}
\author{Dao-Xin Yao}
\email{yaodaox@mail.sysu.edu.cn}
\affiliation{Center for Neutron Science and Technology, Guangdong Provincial Key Laboratory of Magnetoelectric Physics and Devices, State Key Laboratory of Optoelectronic Materials and Technologies, School of Physics, Sun Yat-Sen University, Guangzhou, 510275, China}

\begin{abstract}
Recently, the discovery of superconductivity in   Ruddlesden-Popper (RP) $\mathrm{La_4Ni_3O_{10}}$  under pressure has   further expanded the realm of   nickelate-based superconductor family. In this paper, we performed a first-principle study of $\mathrm{La_4Ni_3O_{10}}$ for both $P2_1/a$ phase at ambient pressure and $I4/mmm$ phase at high pressure, with $U$=0, 3.5\ eV.
Our results confirmed the characteristic upward shift of Ni-$d_{z^2}$ bonding band under pressure.
Moreover, our analysis of electronic spectrum and orbital occupancy unveil the dynamic mechanism of electronic reconstructions under pressure, embedded in a critical dual effect.
Based on our results, we further proposed a trilayer two-orbital model by performing Wannier downfolding on Ni-$e_g$ orbitals. 
Our model  reveals four Fermi surface sheets with $\alpha,\beta,\beta^\prime,\gamma$ pockets, bearing resemblance to  that of bilayer $\mathrm{La_3Ni_2O_7}$.
According to the model, our calculated spin susceptibility under random phase approximation shows that $d_{x^2-y^2}$ orbital is also important for the magnetic fluctuation in RP series.
Finally, a high energy sixteen-orbital model with direct $dp,pp$ hoppings is proposed, which  
implies that $\mathrm{La_4Ni_3O_{10}}$ also lies in charge-transfer picture within  Zaanen-Sawatzky-Allen scheme. Our exposition of electronic reconstructions and multi-orbital models shed light on theoretical electronic correlation study and experimental exploration of lower pressure superconductor in RP series.

\end{abstract}

\date{\today}
\maketitle
\end{CJK*}

\section{Introduction}
The recently consecutive discoveries of  nickelate-based superconductors from infinite-layer nickelates $R\mathrm{NiO_2}$ ($R$=La,Nd,Pr) \citep{NdNiO2,LaNiO2,PrNiO2} to Ruddlesden-Popper (RP) series nickelates ${\rm La}_{n+1}{\rm Ni}_n{\rm O}_{3n+1}$ (n=2,3) \citep{bilayernature,wang2023pressureinduced,JunHou117302,zhu2024superconductivity,zhou2023evidence,zhang2023superconductivity,QingLi17401}, have drawn intensive investigations on the field of high-$T_c$ superconductivity~ \citep{bilayermodel,luo2023hightc,RN20,RN24,RN25,PhysRevResearch.1.032046,PhysRevB.109.045151,ChinPhysLett.40.127401,PhysRevB.75.012414,PhysRevB.101.064513,PhysRevB.108.L140504,wu2023charge,botzel2024theory,PhysRevB.100.205138,PhysRevX.10.011024,PhysRevLett.124.207004,PhysRevB.101.020501,PhysRevResearch.2.023112,PhysRevB.101.041104,PhysRevResearch.2.013219,RN16,RN17,RN18,RN19,PhysRevLett.129.027002,PhysRevB.101.060504,luo2023superconductivity,fan2023superconductivity,PhysRevB.108.125105,PhysRevB.108.L140505,gu2023effective,yang2023orbitaldependent,ZHANG2024147,YUAN2024127511}, particularly in the aspect of how correlations renormalize the low-lying electronic states and potentially drive the unconventional pairing in these compounds \citep{PhysRevB.101.060504,PhysRevLett.124.207004,PhysRevB.101.020501,PhysRevResearch.2.023112,PhysRevB.101.041104,PhysRevResearch.2.013219,RN16,RN17,RN18,RN19,PhysRevLett.129.027002,luo2023superconductivity,yang2023orbitaldependent,fan2023superconductivity,PhysRevB.108.125105,PhysRevB.108.L140505}. Also, for $R\mathrm{NiO_2}$, the sample-dependent factors, such as defeats, impurity and domains deserve careful consideration \citep{RN23,LaNiO2}. 
For RP series nickelates, pressure is undoubtedly in the centre of the roadmap, as the observations of zero-resistance always correlate with a  structure transition, and further promote the electronic reconstructions \citep{bilayernature,zhu2024superconductivity,zhang2023superconductivity,QingLi17401,bilayermodel,jiang2023pressure}. 
The reconstructions exhibit a quite general trend among this series, which is largely manifested in the upward shift of Ni-$d_{z^2}$ states towards Fermi energy \citep{bilayernature,bilayermodel,experimental}. 
It is widely believed that such upward shift should be responsible for the emergence of superconductivity.
However, there still lacks of a comprehension of the dynamic mechanism of these reconstructions, which is indeed fundamental and indispensable.
Furthermore,  inspections of layer dependence is another important aspect. These are crucial to unveil   superconducting mechanism in RP series.

In bilayer $\mathrm{La_3Ni_2O_7}$, the structure at ambient pressure is characterized by octahedral-distorted $Amam$ (space group: 63), which transits to octahedral-regular $Fmmm$ phase (space group: 69)  roughly in the range of 10$\sim$15\ GPa \citep{zhou2023evidence,wang2023pressureinduced,JunHou117302} within orthorhombic lattice. Surprisingly, this range highly coincides with  the full development of superconducting $T_c$ from 0 to 80\ K~\citep{bilayernature,wang2023pressureinduced,JunHou117302}.
For pressure up to $\sim$19\ GPa, a higher symmetric $I4/mmm$ (space group: 139) phase within tetragonal lattice is emergent~\citep{wang2023structure,geisler2023structural},  where $T_c$  exhibits certain amount of decrease~\citep{bilayernature,wang2023pressureinduced,JunHou117302}.
In trilayer $\mathrm{La_4Ni_3O_{10}}$, an analogous monoclinic $P2_1/a$ (space group: 14) phase is characterized at ambient pressure, which transits to $I4/mmm$ phase roughly in the range of 12$\sim$15\ GPa. However, the maximal $T_c$ merely reaches 20$\sim$30\ K under a much  higher pressure of about 43\ GPa \citep{zhu2024superconductivity}.
These further suggest some critical roles of pressure to be clear in the series.

In this paper, we perform a comprehensive first-principle study of trilayer $\mathrm{La_4Ni_3O_{10}}$ for both $P2_1/a$ phase at ambient pressure (AP) and $I4/mmm$ phase at high pressure (HP). Our resulting electronic structure and microscopic multi-orbital models strongly suggest an alike superconducting mechanism  as compared with $\mathrm{La_3Ni_2O_7}$. 
Moreover, based on our DFT results, we clarify the dynamic mechanism of such electronic reconstructures, which is of importance for a comprehensive understanding of superconductivity in RP series.

The paper is organized as follows.
In section.~\ref{sec:structure},  we firstly analyse the structure difference of P2$_1$/a phase at AP and $I4/mmm$ phase at HP, which can provide readers a general understanding of the pressure effect on the  lattice level. 
On this basis, in section.~\ref{sec:method}, we present the detailed settings of our DFT calculations.
In section.~\ref{sec:A}, we present our main DFT results, which include band structures and electronic spectra for both phases, with $U$=0, 3.5\ eV.
The charge transfer processes and valence under pressure are also particularly analysed in this section.
In section.~\ref{sec:B}, we establish an effective trilayer two-orbital model based on our DFT results.
In section.~\ref{sec:C}, we investigate the  spin susceptibility based on the model.
In section.~\ref{sec:D}, we further proposed a high energy sixteen-orbital model.
In section.~\ref{sec:disscussion}, we provide  discussions regarding the dynamic mechanism of electronic reconstructions and layer dependence, followed by the summary in section.~\ref{Sec:Summary}.

\section{Structure transition under pressure}
\label{sec:structure}
\begin{figure}
\noindent \begin{centering}
\includegraphics[width=1.0\columnwidth,height=1.0\columnwidth,keepaspectratio]{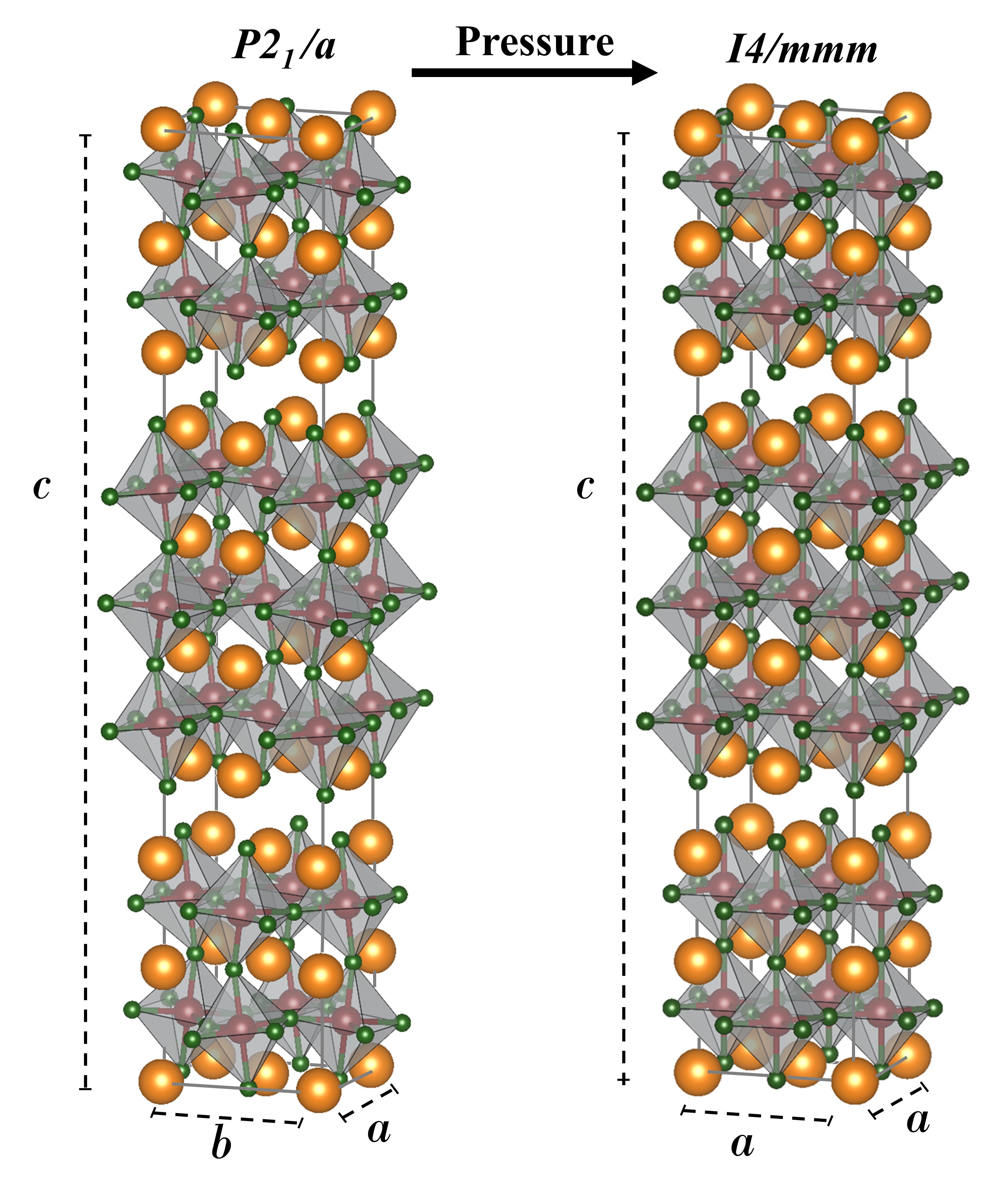}\caption{ 
Structure of the trilayer nickelate $\rm{La_4Ni_3O_{10}}$ for both $P2_1/a$ and $I4/mmm$ phases.
The red, green and orange balls represent the nickel, oxygen and lanthanum atoms, respectively. The grey area shapes the Ni-O octahedra.  $a,b,c$ denote the lattice constants. \label{fig1}}
\par\end{centering}
\end{figure}

In trilayer $\mathrm{La_4Ni_3O_{10}}$, each Ni ion is surrounded by six oxygen sites forming NiO$_6$ octahedron. The corner-shared stacking of these octahedrons gives rise to NiO$_2$ trilayers, as well as the important inter-layer Ni-O-Ni bond along $c$ axis. 
Both structures of AP $P2_1/a$ and HP $I4/mmm$ phase are demonstrated in Fig.~\ref{fig1}. 
The schematic reveals two major differences: (1) For AP phase, there is perceivable octahedral tiltings as compared with HP phase; 
(2) Also, both phases are differentiated  in the lattice constants under a $\sqrt{2}\times\sqrt{2}$ lattice, in which AP phase shows  orthorhombic distortion with unequal $a,b$ lengths, while the HP phase possesses square structure. 
According to Ref.~\cite{experimental}, the lattice constants are $a$=5.4675, $b$=5.4164, $c$=27.9564 \AA 
\ for AP, and $a$=5.1769, $c$=26.2766 \AA\ for HP at 44.3\ GPa, which corresponds to 15\% of the lattice collapse.  
It should be noted that, there are also several  other types of structure distortions in RP series, such as octahedral rotation,  bond disproportionation of NiO$_6$ octahedron.
But their implications on electronic structure are quite insignificant compared with  octahedral tilting~\citep{geisler2023structural}.
In fact, for $\rm La_4Ni_3O_{10}$, there exists   another phase of  $Bmab$ (space group: 64) at AP, which survives in high temperature.
This phase differs from $P2_1/a$ phase 
 by a relative octahedral rotation,  while both band structures are almost indistinguishable around $E_{\rm F}$~\cite{PhysRevB.105.085150}.

\section{METHOD Details}

\label{sec:method}

Density functional theory (DFT) calculations were performed by Vienna ab initio simulation package (VASP)~\citep{VASP1,VASP2}, in which
the projector augmented wave (PAW) \citep{PAW1,PAW2} method within
the framework of the local density approximation (LDA) \citep{LDA}
exchange correlation potential are applied.
The energy cutoff of the plane-wave expansion was set as 600 eV and a $\Gamma$-centered
$20\times20\times19$ Monkhorst Pack k-mesh grid was adopted.
In structural relaxations, we adopted the experimental refined lattice constants for both phases~\citep{experimental}, which is $P2_1/a$ phase at ambient pressure and $I4/mmm$ phase at pressure of 44.3\ GPa. 
The convergence criterion of force was set to be 0.001\ eV/{\rm \AA}  and total energy convergence criterion was set to be $10^{-7}$eV.
In band structure calculation,  we adopted a  $\sqrt{2}$$\times$$\sqrt{2}$ primitive cell for $I4/mmm$ phase in order to have a direct comparison with $P2_1/a$ phase, as both contain 6 Ni atoms.
To obtain the projected tight-binding models, we further performed Wannier downfolding  as implemented by WANNIER90~\citep{w90} package, in which  the good convergences were reached.


\section{RESULTS}

The DFT results are carefully examined, which are in agreement with previous theoretical calculations \citep{PhysRevB.105.085150, experimental} as well as the reported ARPES results \citep{NCARPES}. On this basis, we adopted a very large {\rm k}-mesh size as mentioned before in order to determine the precise Fermi level, which is important given the flat band feature in this material.

\subsection{Electronic structure }

\label{sec:A}

\begin{figure*}
\noindent \begin{centering}
\includegraphics[width=2.0\columnwidth,height=2.0\columnwidth,keepaspectratio]{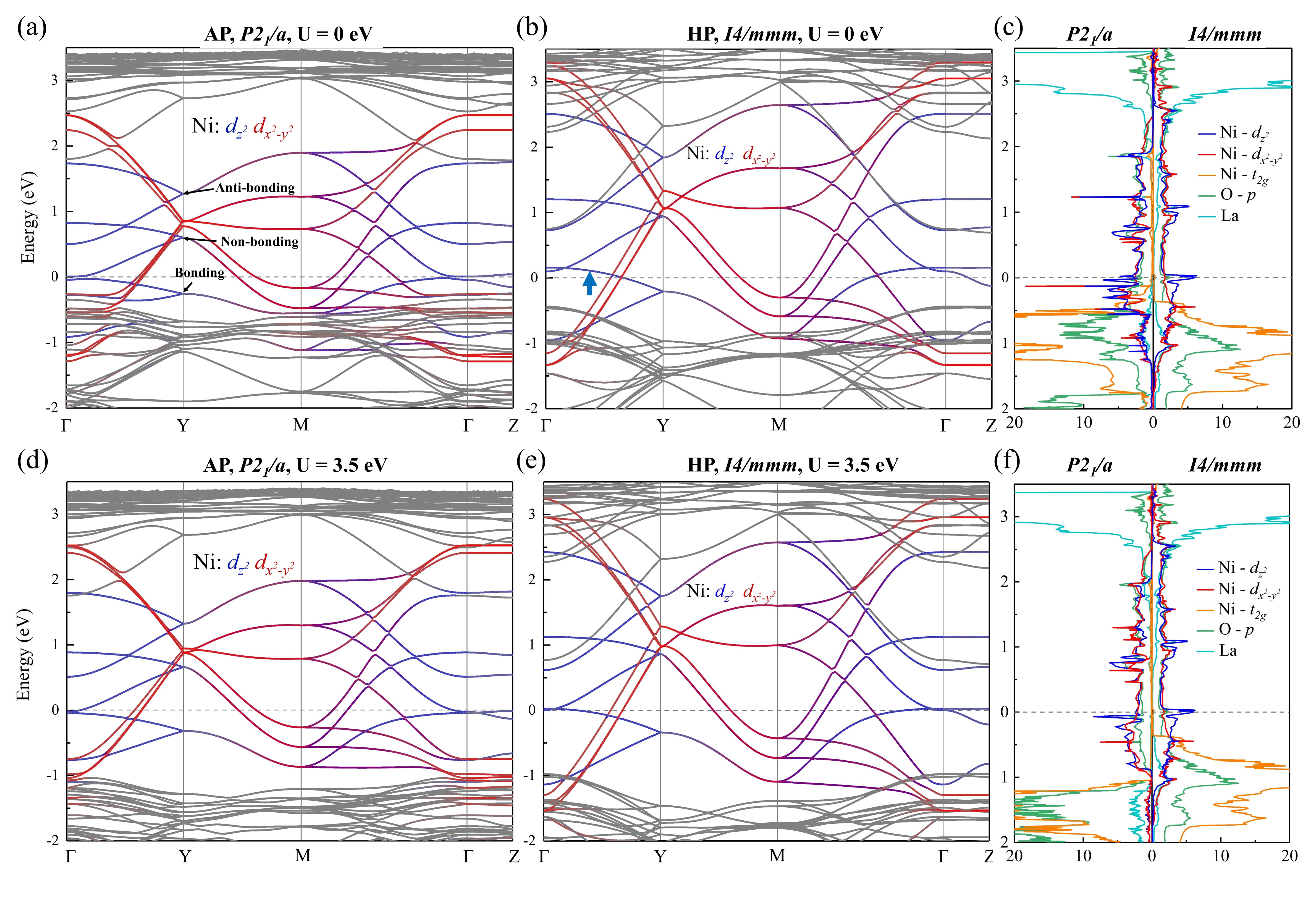}\caption{Electronic structure of $\mathrm{La_4Ni_3O_{10}}$. DFT (LDA) calculated band structures of $\mathrm{La_{4}Ni_{3}O_{10}}$ at $U=0$ eV (a)-(b) and $U=3.5$ eV (d)-(e) for AP $P2_1/a$ phase and HP $I4/mmm$ phase.
The blue and red colors are spectral weight for $d_{3z^2}$
and $d_{x^{2}-y^{2}}$ orbitals, respectively. (c), (f) Projected DOS of $\mathrm{La_{4}Ni_{3}O_{10}}$ at AP (left panel) and HP (right panel). The blue, red, orange, green and cyan colors denote the projection of Ni-$d_{3z^{2}-r^{2}}$, Ni-$d_{x^{2}-y^{2}}$, Ni-$t_{2g}$,  O-$p$ and La orbitals.  \label{fig2}}
\par\end{centering}
\end{figure*}

 In Fig.~\ref{fig2}, we present the electronic structures of $\mathrm{La_4Ni_3O_{10}}$ for both AP $P2_1/a$ and HP $I4/mmm$ phases. Here the upper and lower panels represent the cases of $U=0$ and 3.5\ eV, respectively.
From these band structures we can observe  the dominant Ni-$d_{z^2}$ (blue) and $d_{x^2-y^2}$ (red) orbitals around Fermi energy, which is quite similar to the bilayer system \citep{bilayernature,bilayermodel,PhysRevLett.131.206501}. 
For Ni-$d_{z^2}$ band, it clearly exhibits the characteristic 3-branch structure, which corresponds to bonding, non-bonding and anti-bonding states with increasing energy, as indicated in Fig.~\ref{fig2}(a) \citep{experimental,sakakibara2023theoretical}. 
Apart from Ni-$e_g$ sector, there also appears dense bands (grey) away from $E_{\rm F}$, which can be found in the density of states (DOS) [Fig.~\ref{fig2}(c),(f)] associating with Ni-$t_{2g}$ sector around -1\ eV (orange) and La sector around 3\ eV (cyan).

Next, we discuss the electronic reconstructions under pressure. As expected that pressure is prone to  enhance the itinerancy of electrons, which in our case corresponds to a ratio of $\sim$1.3 to the increase of bandwidth from AP to HP, as illustrated in Fig.~\ref{fig2}(a)-(b)  for $U$=0\ eV and Fig.~\ref{fig2}(d)-(e) for $U$=3.5\ eV.  
As we further zoom in to the Fermi level, for AP phase (left panel), we observe a very narrow gap relating  the bonding and non-bonding bands at $\Gamma$ point \citep{PhysRevB.105.085150, experimental}. For HP phase (middle panel), the bonding band gets upward shifted as a hole pocket for both $U$=0, 3.5\ eV.
It is worth noting that, for $U$=3.5\ eV [Fig.~\ref{fig2}(e)], such a hole pocket is almost flat right at $E_{\rm F}$, giving rise to a sharp peak in DOS, further highlighting a crucial link to the superconductivity \citep{RN26,PhysRevB.75.035110} in RP series.

\begin{table}

\caption{DFT calculated ($U$=0 eV) orbital occupancies of $\mathrm{La_4Ni_3O_{10}}$ for both AP $P2_1/a$ phase and HP $I4/mmm$ phase. 
Here $n^{\rm O}_{p_x/p_y}$ denotes one of the in-plane O-$p_x$, $p_y$ orbitals that is elongated along its adjacent Ni site. For $n^{\rm O}_{p_z}$, the row of ``Inner" and ``Outer" denote the apical O-$p_z$ orbitals that are inside and outside the trilayer, respectively.
Note that for AP phase, each value is averaged appropriately over doubled atoms.
}

\begin{onehalfspace}\label{tab:occupancy}
\noindent \begin{centering}
\begin{tabular}{ccccccc}
\hline \hline
 & Layer & $n_{x^{2}-y^{2}}^{\rm Ni}$ & $n_{z^{2}}^{\rm Ni}$ & $n^{\rm Ni}_{t_{2g}}$ & $n_{p_{x}/p_{y}}^{\rm O}$ &  $n_{p_{z}}^{\rm O}$\tabularnewline
\hline 
\multirow{2}{*}{AP} & Inner & 1.199 & 1.289 &5.874 & 1.695 & 1.656\tabularnewline
\cline{2-7} \cline{3-7} \cline{4-7} \cline{5-7} \cline{6-7} \cline{7-7} 
 & Outer & 1.173 & 1.241 & 5.942 & 1.698 & 1.861\tabularnewline
\hline 
\multirow{2}{*}{HP} & Inner & 1.066 & 1.022 & 5.996 & 1.675 & 1.641\tabularnewline
\cline{2-7} \cline{3-7} \cline{4-7} \cline{5-7} \cline{6-7} \cline{7-7} 
 & Outer & 1.077 & 1.100 &5.993 & 1.696 & 1.843\tabularnewline
\hline \hline
\end{tabular}
\par\end{centering}
\end{onehalfspace}
\end{table}

Given such a critical band upward shift  under pressure, we further address the corresponding charge transfer process. 
In Table.~\ref{tab:occupancy}, we present the orbital occupanices of Ni-$d$, O-$p$ orbitals for both phases at $U$=0 eV. It is clear that, both two $e_{g}$ orbitals are over half-filling at AP, while pressure tends to depress their occupancies, and become almost half-filling at HP.
We noted that the decrease of $n^{\rm Ni}_{x^2-y^2}$ is hardly identified in band structure, although it has a relatively smaller amount than that of $n^{\rm Ni}_{z^2}$.
But for $t_{2g}$ sector, pressure has an opposite effect, in which $n^{\rm Ni}_{t_{2g}}$ is increased from about 5.9 to 
 6, i.e., fully filled at HP.
Such trend can also be observed in Fig.~\ref{fig2}(c),(f), in which the  $t_{2g}$ spectrum is closer to  $E_{\rm F}$ for AP phase.
Also, for oxygen occupanies,  we find their variations are quite insignificant under pressure. 
Finally, the table  reveals a slight charge imbalance between inner and outer layers, and we find that these trends and magnitudes are quite close to that of DFT+DMFT results at HP~\cite{leonov2024electronic}. This hightlights that our results at $U=0$ eV are also valid for the consideration of correlation effect.

\begin{figure}
\noindent \begin{centering}
\includegraphics[width=1.0\columnwidth,height=1.0\columnwidth,keepaspectratio]{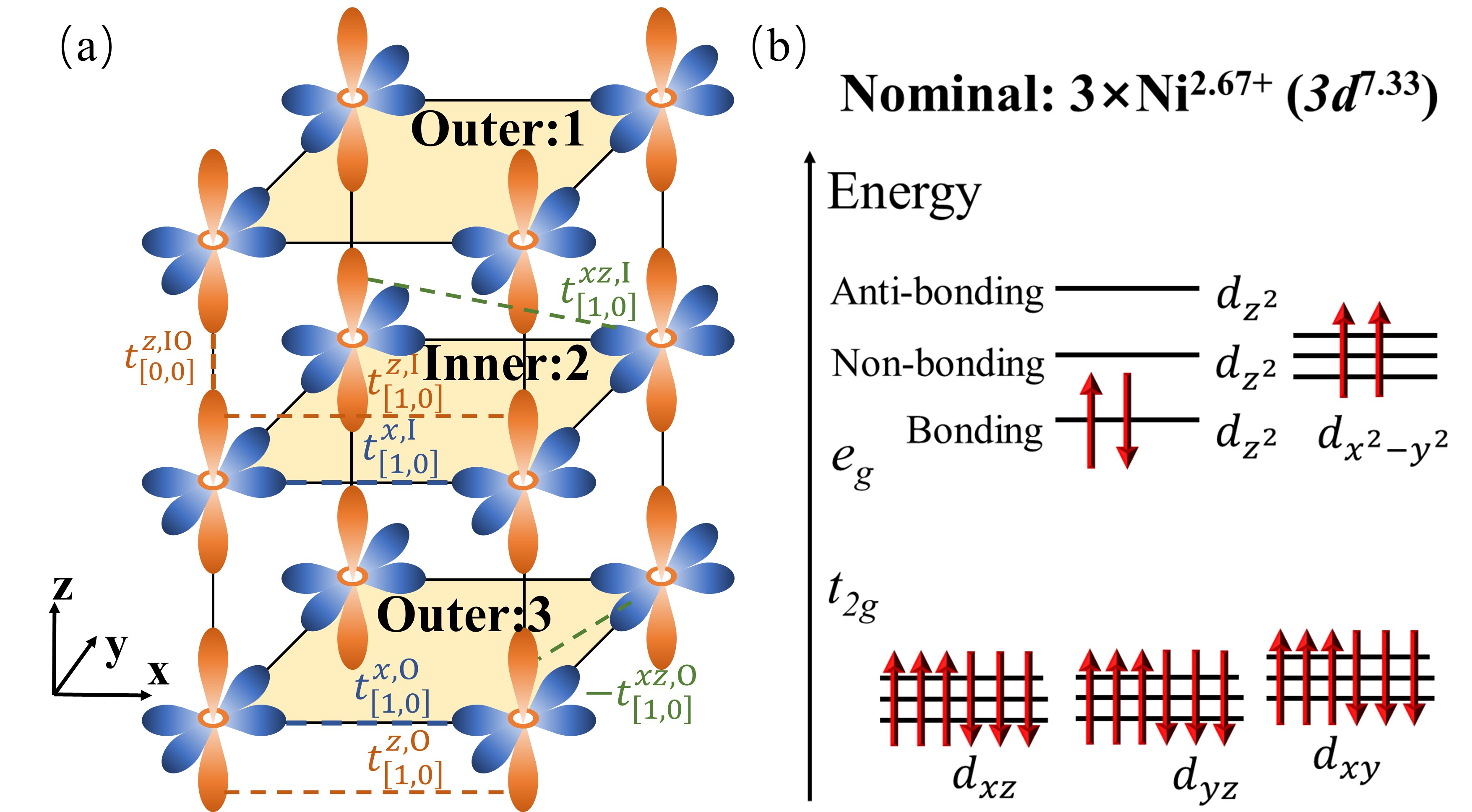}\caption{ (a) Schematic of trilayer $\mathrm{La_{4}Ni_{3}O_{10}}$ lattice.
The blue and orange shapes denote  Ni-$d_{z^{2}}$ and $d_{x^{2}-y^{2}}$  orbitals, respectively. 
Some of the hoppings of trilayer two-orbtal model are drawn, whose values can be found in Table.~\ref{parameters}, also see text for details.  (b) Electronic configuration of $\mathrm{La_4Ni_3O_{10}}$. The arrows indicate nominal configuration $3d^{7.33}$.  \label{fig3}}
\par\end{centering}
\end{figure}

In view of the above analysis, we can quantitatively 
obtain some crucial charge transfer magnitudes from AP to HP based on  Table.~\ref{tab:occupancy}.
Overall, there is an average $3d^{8.36}\rightarrow 3d^{8.14}$ for Ni ions. 
Specifically, 2.44$\rightarrow$2.14 for Ni-$e_g$ and 5.92$\rightarrow$6.0 for Ni-$t_{2g}$ sectors.
However, for oxygen ions, an average $\sim$1.7 valence is persistent.
Here if we assume full ionization of oxygens, we would alternatively obtain an average $d^{7.36}\rightarrow d^{7.14}$, in which the former value is quite close to the nominal $d^{7.33}$ configuration.
These charge transfers imply a critical dual effect of pressure on microscopic electronic structure. On one hand, pressure promotes a charge redistribution within Ni-$3d$ orbitals, which can alter the relative renormalization effects of these orbitals.
In particular, the involvement of $t_{2g}$ sector near $E_{\rm F}$ could enhance the charge fluctuation at AP, which is relevant to the observed charge density wave \citep{2020intertwined,zhu2024superconductivity}.
On the other hand, the electric neutrality indicates a pressure driven charge transfer from Ni to La ions (the transfer to Ni-4$s$ orbitals is negligible).
In fact, from Fig.~\ref{fig2}(a),(b), we can also see a notable drop of the lowest La band from about 2 to 0.7\ eV at $\Gamma$ point, which implies  a general downward shift of La spectrum.
This feature further indicates an enhanced $d$-$f$ hybridization at HP above ${E_{\rm F}}$, 
especially with Ni-$d_{z^2}$ orbital, which might be observed in resonant inelastic x-ray scattering signal \citep{PhysRevB.107.165124}.



\subsection{Trilayer two-orbital model}

\label{sec:B}

\begin{table}
\caption{Tight-binding parameters of trilayer two-orbital model for $\mathrm{La_4Ni_3O_{10}}$ under pressure. 
$t_{[i,j]}^{x/z}$ denotes the hopping term that is connected by $[0,0]$-$[i,j]$ bond within $d_{x^2-y^2}/d_{z^2}$ orbital, while  $t^{xz}_{[i,j]}$ denotes the hybridization between them, all in unit of eV. The symmetrically equivalent terms are not shown for clarity. 
}
\begin{onehalfspace}\label{parameters}
\noindent \centering{}%
\begin{tabular*}{1\columnwidth}{@{\extracolsep{\fill}}cccccc}
\hline \hline 
Layer & $i$ & $j$ & $t_{[i,j]}^{x}$ & $t_{[i,j]}^{z}$& $t_{[i,j]}^{xz}$\tabularnewline
\hline 
\multirow{4}{*}{Inner} & 0 & 0 & 1.094 & 1.081 & 0\tabularnewline
\cline{2-6} \cline{3-6} \cline{4-6} \cline{5-6} \cline{6-6} 
 & 1 & 0 & -0.521 & -0.168 & 0.298\tabularnewline
\cline{2-6} \cline{3-6} \cline{4-6} \cline{5-6} \cline{6-6} 
 & 1 & 1 & 0.069 & -0.018 & 0\tabularnewline
\cline{2-6} \cline{3-6} \cline{4-6} \cline{5-6} \cline{6-6} 
 & 2 & 0 & -0.076 & -0.018 & 0.040\tabularnewline
\hline 
\multirow{4}{*}{Outer} & 0 & 0 & 0.867 & 0.683 & 0\tabularnewline
\cline{2-6} \cline{3-6} \cline{4-6} \cline{5-6} \cline{6-6} 
 & 1 & 0 & -0.511 & -0.143 & 0.274\tabularnewline
\cline{2-6} \cline{3-6} \cline{4-6} \cline{5-6} \cline{6-6} 
 & 1 & 1 & 0.065 & -0.015 & 0\tabularnewline
\cline{2-6} \cline{3-6} \cline{4-6} \cline{5-6} \cline{6-6} 
 & 2 & 0 & -0.074 & -0.017 & 0.039\tabularnewline
\hline 
\multirow{2}{*}{Inner-Outer} & 0 & 0 & 0.035 & -0.738 & 0\tabularnewline
\cline{2-6} \cline{3-6} \cline{4-6} \cline{5-6} \cline{6-6} 
 & 1 & 0 & 0 & 0.033 & -0.057\tabularnewline
\hline 
Outer-outer & 0 & 0 & 0 & -0.078 & 0\tabularnewline
\hline \hline
\end{tabular*}

\end{onehalfspace}
\end{table}

To gain more insights to the electronic property that is directly relevant to the superconductivity, in the following sections, we focus on the HP $I4/mmm$ phase.
Based on the DFT electronic structure, we  further perform Wannier downfolding on the trilayer Ni-$e_g$ orbitals at $U$=0\ eV, which allows us to build an effective trilayer two-orbital model:
\begin{align}
\mathcal{H}&=\mathcal{H}_t+\mathcal{H}_U,\qquad
\mathcal{H}_t=\sum_{{\rm k}\sigma}\Psi_{{\rm k}\sigma}^{\dagger}H({\rm k})\Psi_{{\rm k}\sigma},\\
\mathcal{H}_U&=U\sum_{i}^{\alpha=1,2,3}\left( n_{i\uparrow}^{x_\alpha} n_{i\downarrow}^{x_\alpha}+n_{i\uparrow}^{z_\alpha} n_{i\downarrow}^{z_\alpha} \right)\nonumber\\
&+U^\prime\sum_{i\sigma}^{\alpha=1,2,3}n^{x_\alpha}_{i\sigma}n^{z_\alpha}_{i\bar{\sigma}} +(U^\prime-J_H) \sum_{i\sigma}^{\alpha=1,2,3}n^{x_\alpha}_{i\sigma}n^{z_\alpha}_{i\sigma}\nonumber\\
+ J_H&\sum_{i}^{\alpha=1,2,3}\left( d^\dagger_{ix_\alpha\uparrow} d^\dagger_{ix_\alpha\downarrow} d_{iz_\alpha\downarrow} d_{iz_\alpha \uparrow} - d^\dagger_{ix_\alpha\uparrow} d_{ix_\alpha\downarrow} d^\dagger_{iz_\alpha\downarrow} d_{iz_\alpha\uparrow} +h.c.\right).\nonumber
\end{align}
The model $\mathcal{H}$ is composed of the tight-binding $\mathcal{H}_t$ and Coulomb interaction $\mathcal{H}_U$. The basis is  $\Psi_{\sigma}=(d_{x_1\sigma},d_{x_2\sigma},d_{x_3\sigma},d_{z_1\sigma},d_{z_2\sigma},d_{z_3\sigma})^{T}$,
in which $d_{x_\alpha\sigma}$ denotes annihilation of a $d_{x^2-y^2}$  electron in the $\alpha^{\rm th}$ layer with spin $\sigma$, etc. 
The notation of layer is demonstrated in Fig.~\ref{fig3}(a).
For $\mathcal{H}_U$, there have four terms in full Kanamori form \citep{annurev-conmatphys-020911-125045} with relation $U^\prime=U-2J_H$, and $U,U^\prime,J_H$ are on-site Coulomb repulsion of the same and different orbitals, and Hund's coupling, respectively.

We further express the tight-binding kernel as
\begin{align}\label{eq:htb}
H({\rm k})=\left[
\begin{array}{cc}
     H^x({\rm k})& H^{xz}({\rm k}) \\
     H^{xz}({\rm k})& H^z({\rm k}) 
\end{array}
\right],
\end{align}in which $H^{x/z}({\rm k})$ is the trilayer tight-binding matrix of $d_{x^2-y^2}/d_{z^2}$ orbital, and $H^{xz}({\rm k})$ is the hybridization between them. They are expressed as
\begin{widetext}
\begin{align}
H_{11}^{x/z}&=H_{33}^{x/z}=\epsilon^{x/z,{\rm O}}+2t^{x/z,{\rm O}}_{[1,0]}(\cos k_x+\cos k_y)+4t^{x/z,{\rm O}}_{[1,1]}\cos k_x \cos k_y+2t^{x/z,{\rm O}}_{[2,0]}(\cos 2k_x+\cos 2k_y),\nonumber\\
H_{22}^{x/z}&=\epsilon^{x/z,{\rm I}}+2t^{x/z,{\rm I}}_{[1,0]}(\cos k_x+\cos k_y)+4t^{x/z,{\rm I}}_{[1,1]}\cos k_x \cos k_y+2t^{x/z,{\rm I}}_{[2,0]}(\cos 2k_x+\cos 2k_y),\nonumber\\
H^{x/z}_{12}&=H^{x/z}_{23}=t^{x/z,{\rm IO}}_{[0,0]}+2t^{x/z,{\rm IO}}_{[1,0]}(\cos k_x+\cos k_y),\quad H^{z}_{13}=t^{z,{\rm OO}}_{[0,0]},\\
H^{xz}_{11}&=H^{xz}_{33}=2t^{xz,{\rm O}}_{[1,0]}(\cos k_x-\cos k_y)+2t^{xz,{\rm O}}_{[2,0]}(\cos 2k_x-\cos 2k_y),\nonumber\\
H^{xz}_{22}&=2t^{xz,{\rm I}}_{[1,0]}(\cos k_x-\cos k_y)+2t^{xz,{\rm I}}_{[2,0]}(\cos 2k_x-\cos 2k_y),\nonumber\\
H^{xz}_{12}&=H^{xz}_{23}=2t^{xz,{\rm IO}}_{[1,0]}(\cos k_x-\cos k_y).\nonumber
\end{align}
\end{widetext}

\begin{figure}
\noindent \begin{centering}
\includegraphics[width=1.0\columnwidth,height=1.0\columnwidth,keepaspectratio]{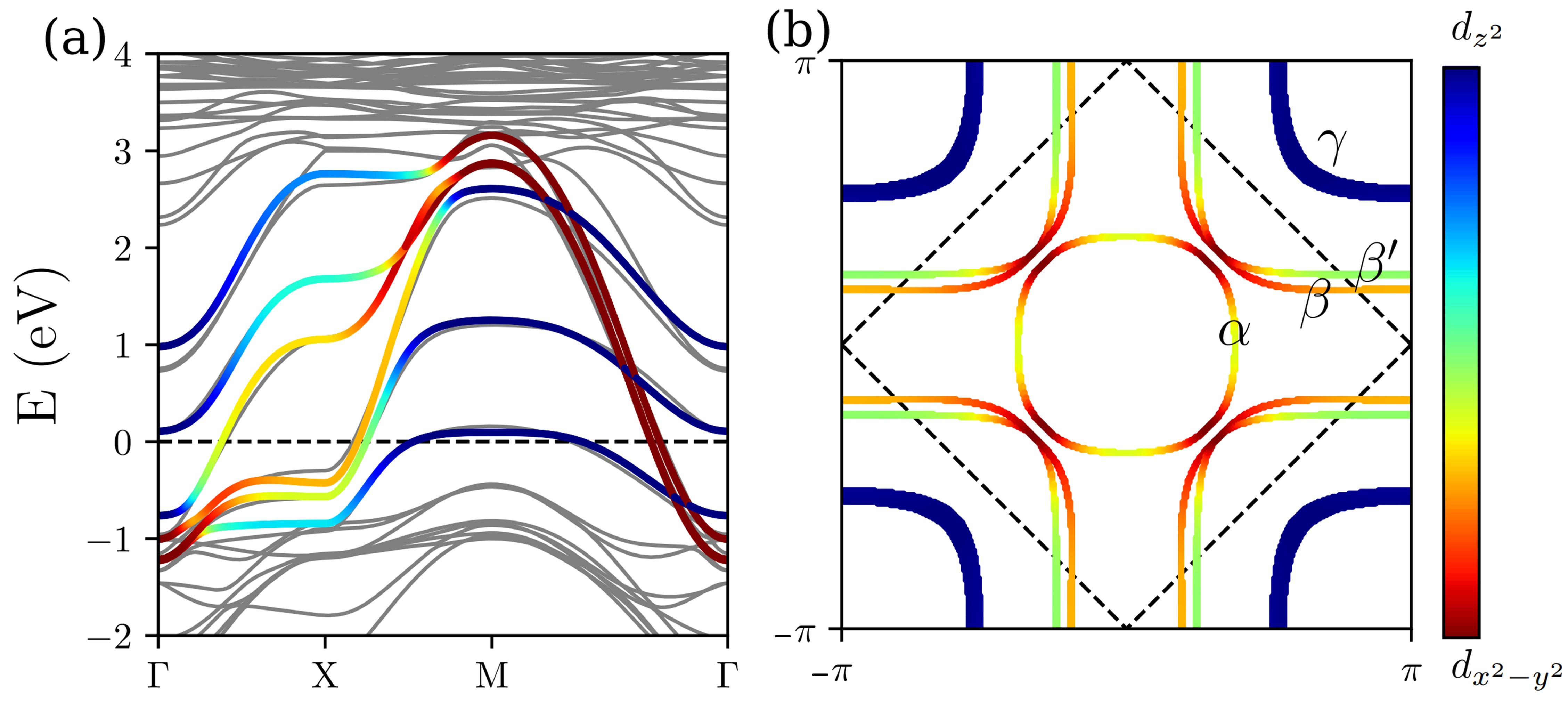}\caption{(a) Band structure and (b) Fermi surface of trilayer two-orbital model. The color bar indicates the orbital weight of Ni-$d_{3z^{2}-r^{2}}$ and $d_{x^{2}-y^{2}}$. The grey lines in (a) are band structure from DFT (LDA, $U=0$ eV). The diamond shape in (b) indicates the folded Brillouin zone. \label{fig4}}
\par\end{centering}
\end{figure}

We consider the hoppings up to the third-nearest neighbor in order to accurately describe the low-lying state of our DFT, which yields a total of 25 parameters listed in Table.~\ref{parameters}. 
Some of the major hoppings are demonstrated in the schematic in Fig.~\ref{fig3}(a). 
Here $t_{[1,0]}^{x,{\rm I}}$ denotes in-plane nearest-neighbor hopping of $d_{x^2-y^2}$ orbitals inside inner layer, and $t^{z,{\rm IO}}_{[0,0]}$ denotes  perpendicular hopping of $d_{z^2}$ orbital between inner and outer layers.
According to the values in Table.~\ref{parameters}, we find that most of the hoppings are enhanced comparably with respect to  bilayer $\mathrm{La_3Ni_2O_7}$. Practically,   $t^x_{[1,0]}$ is enhanced from  -0.483 to -0.521 (inner) or -0.511 (outer), and $t^{z,{\rm IO}}_{[0,0]}$ from  -0.635 to -0.738 \citep{bilayermodel}.
Such enhancement of hopping parameters partially reflect a relative weaker correlation of trilayer than that of  bilayer.

Based on the trilayer two-orbital model, we  calculate the densities of $d_{x^2-y^2}$ and $d_{z^2}$ orbitals, which are 0.62 and 0.54 for inner layer, 0.55 and 0.63 for outer layer, respectively. These correspond to an average $3d^{7.17}$ configuration, in exact agreement with previous analysis.   
According to the configuration, we roughly obtain an schematic depicted in Fig.~\ref{fig3}(b).
Here there actually have 2 electrons on $d_{z^2}$ orbital forming fully-filled bonding state, while the non-bonding state is empty. The remaining 2 electrons indeed  reside on $d_{x^2-y^2}$ orbital with the same spin alignment.

In Fig.~\ref{fig4}, we further show the band structure and Fermi surface of trilayer two-orbital model, which is in good agreement with our DFT results. The model is expressed under the primitive cell that contains only three Ni atoms, hence the flat band will be unfolded from $\Gamma$ to M point, as shown by the  $\gamma$-pocket in Fig.~\ref{fig4}(b). The figure also reveals another three pockets $\alpha,\beta,\beta^\prime$ with $\beta,\beta^\prime$ very close in each other. In general, the Fermi surface profile as well as the orbital weights are quite similar to that of the bilayer $\mathrm{La_3Ni_2O_7}$ \citep{bilayermodel}, as $\gamma$-point is uniquely characterized by $d_{z^2}$ orbital and $\alpha,\beta,\beta^\prime$-pockets by mixing of $d_{z^2}$ and $d_{x^2-y^2}$ orbitals.
Finally ,we note that the band structure shows slight deviation from DFT at $\rm{X}$-point under $E_{\rm F}$, which is a compromise for a more accurate  fitting of the Fermi surface  during our modeling.

To illustrate the bonding feature in $\mathrm{La_4Ni_3O_{10}}$, we perform a rotation of the orbital basis under the trilayer symmetry, in which we define the new basis as
\begin{align}
\Phi&=(c_-,c_+,c)^T,\quad 
c_{\pm}=\frac{1}{\sqrt{2}}(d_1\pm d_3),\quad c=d_2,
\end{align}
with $d$  applied for both $d_x$ and $d_z$ operators.
When along nodal direction ($|k_x|=|k_y|$), the two $e_g$ orbitals are well decoupled as the off-diagonal $H^{xz}({\rm k})=0$ in Eq.~(\ref{eq:htb}).
In this case, we can find that the non-bonding state is decisively characterized by $c_-$ state, namely, a sign-reversed superposition of states from 
 two outer layers, while the bonding and anti-bonding states are associated  with mixing of both $c_{+}$, $c$ states \citep{PhysRevB.104.184518}. 
For the most concerning bonding band of Ni-$d_{z^2}$ orbital, we can find that it is characterized  by the positive superposition of $c_+$ and $c$, which corresponds to a more elongated Wannier wave function  along $c$ axis.

\begin{figure}
\noindent \begin{centering}
\includegraphics[width=1.0\columnwidth,height=1.0\columnwidth,keepaspectratio]{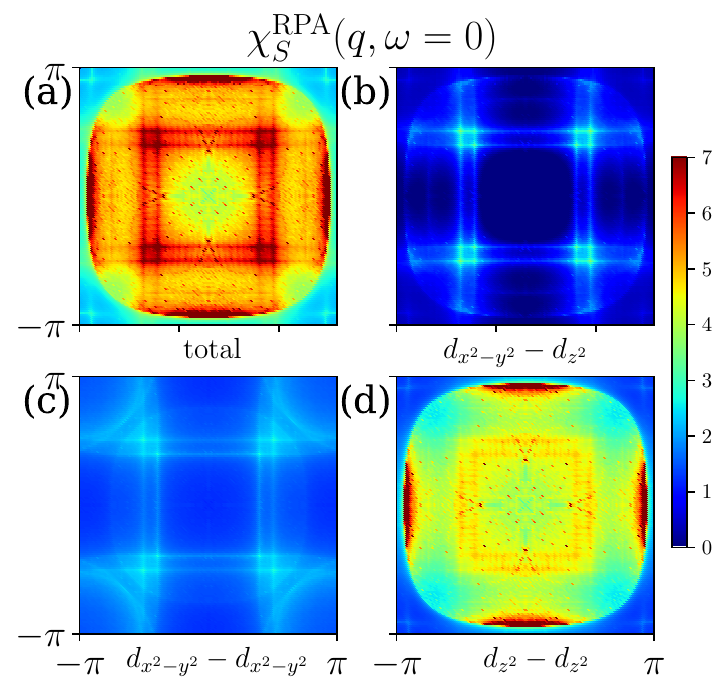}\caption{RPA spin susceptibility $\chi_{_S}^{st,{\rm RPA}}$ of trilayer two-orbital model. $U=3.5,J=U/10$ are adopted in the calculation. (a) Orbital sum $\chi_{_S}^{\rm RPA}=\sum_{st}\chi_{_S}^{st,{\rm RPA}}$; (b-d) Orbital resolved $\chi_{_S}^{st,{\rm RPA}}$. An amplify factor of 2 is used in (b,c). \label{fig5}}
\par\end{centering}
\end{figure}

\subsection{Spin susceptibility}

\label{sec:C}

The multi Fermi surface sheets of $\mathrm{La_4Ni_3O_{10}}$ implies some possible magnetic instabilities that might be observed in experiments. 
In Fig.~\ref{fig5}, we present the spin susceptibility of trilayer two-orbital model at $\omega=0$ under random phase approximation. The total orbital-summed $\chi_{_S}^{\rm RPA}=\sum_{st} \chi_{_S}^{st,{\rm RPA}}$ [Fig.~\ref{fig5}(a)] shows a broad magnetic signal  over the whole Brillouin zone, just like that in Ref.~\cite{leonov2024electronic} as well as in bilayer system \citep{bilayermodel,gu2023effective,botzel2024theory}. Still, we can observe notable enhancement at some particular regimes reflecting fine nesting of Fermi surface, especially at wave vector ${\rm q}=(\pm\frac{\pi}{2},\pm\frac{\pi}{2})$. 
From orbital-resolved $\chi_{_S}^{st,{\rm RPA}}$ [Fig.~\ref{fig5}(b)-(d)], we find that it has a major $d_{z^2}$  character, also, the contribution from $d_{x^2-y^2}$ orbital is  non-negligible.
Remarkably, the Fermi surface in Fig.~\ref{fig4}(b) shows that such wave vectors are more associated with nesting within  $\beta,\beta^\prime$ pockets  which possess quasi-1D feature and strong orbital mixing, instead of $\alpha,\gamma$ pockets. 
This conjecture is also evidenced in $\chi_{_S}^{\rm RPA}$ by the appearance of Moire pattern  along these wave vectors, reflecting the quasi-degeneracy of $\beta,\beta^\prime$ pockets.
This highlights that $d_{x^2-y^2}$ orbital is also important for the magnetic fluctuation of the ground state in $\mathrm{La_4Ni_3O_{10}}$. Despite it has a relatively weak correlation than $d_{z^2}$ orbital \citep{NCARPES}. 
Recently, the magnetic signal at this wave vector is unambiguously revealed in a RIXS experiment for bilayer system, indicating a corresponding double-stripe spin texture~\citep{chen2024electronic}.  
But according to the early experimental probe~\cite{2020intertwined}, the trilayer system in fact possesses an incommensurate magnetic wave vector $(0.62\pi,0.62\pi)$, which is a bit deviated from the bilayer. We conjecture that such deviation is related to the strong interplay with charge density wave, or  possibly charge fluctuation~\cite{tamyuting2015}. 

\begin{figure}
\noindent \begin{centering}
\includegraphics[width=1.0\columnwidth,height=1.1\columnwidth,keepaspectratio]{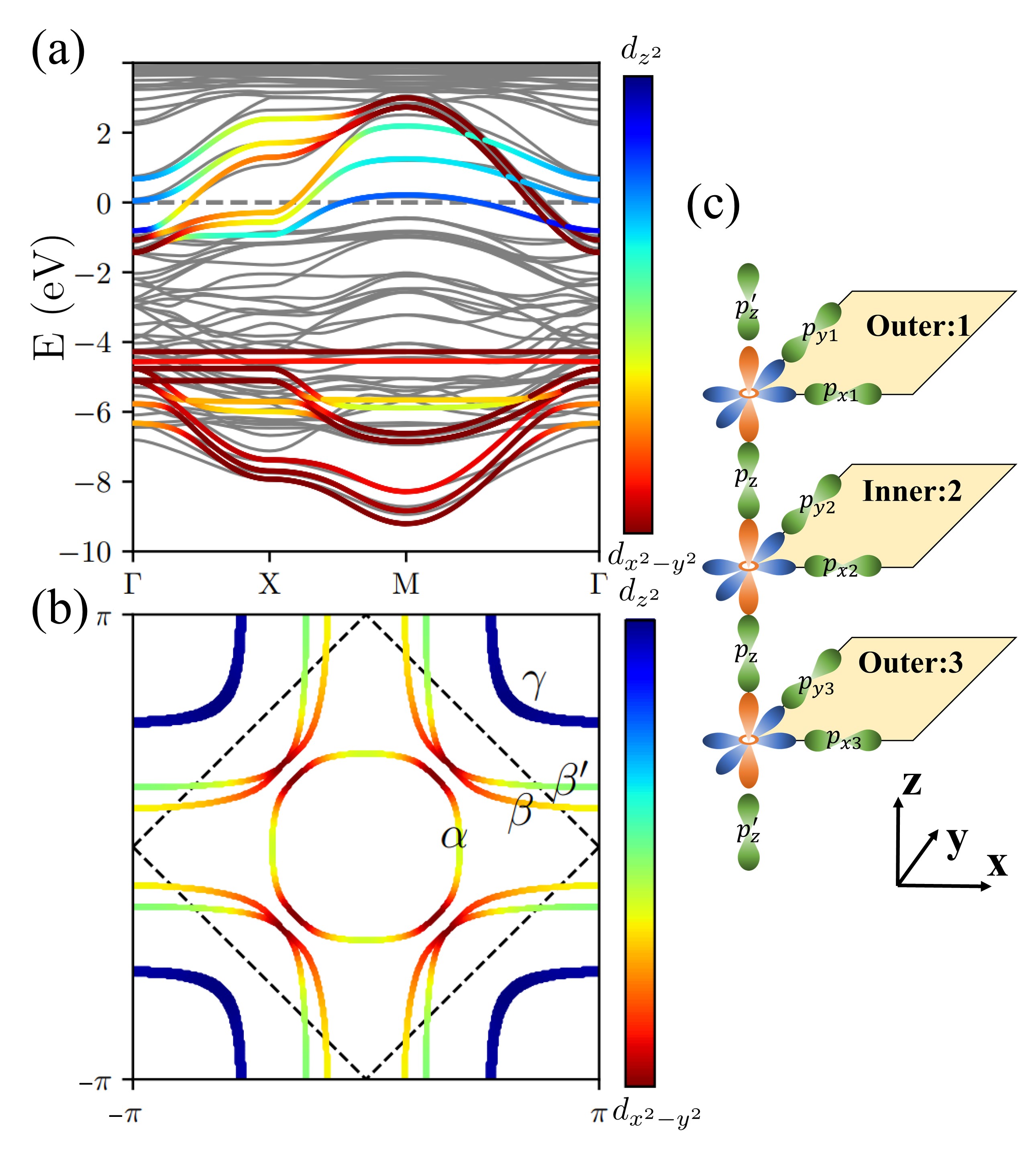}\caption{ (a) Band structure and (b) Fermi surface of sixteen-orbital model. The color bar indicates the orbital weight of Ni-$d_{3z^{2}-r^{2}}$ and $d_{x^{2}-y^{2}}$ orbitals. The grey lines in (a) are band structure from DFT (LDA, $U=0$ eV). The diamond shape in (b) indicates the folded Brillouin zone. (c) Schematic of sixteen orbitals for sixteen-orbital model. \label{fig6}}
\par\end{centering}
\end{figure}

\subsection{Sixteen-orbital model}
In the previous trilayer two-orbital model, the oxygens are implicitly incorporated and Ni-$d$ orbital should be interpreted as ligand-hold state in the sense of Sawatzky's picture~~\citep{ZSA-PhysRevLett.55.418,zhang-rice1988}.
In a dynamic mean-field theory (DMFT) study, the mapping from lattice to impurity problem requires explicit consideration of these conduction degree of freedoms~\citep{pnas_chargetransfer,wu2023charge}. 
Therefore, we further perform Wannier downfolding on both Ni-$e_g$ and O-$p$ orbitals.
\label{sec:D}
For each of the in-plane oxygen, we only pick one of $p_x/p_y$ orbitals that is elongated along its adjacent Ni site, as is usually done in a CuO$_2$ plane \citep{pnas_chargetransfer}. For apical oxygens, only $p_z$ orbitals are selected. These yield a  sixteen-orbital basis as demonstrated in the schematic in Fig.~\ref{fig6}(c).
In this basis, it is sufficient to consider merely 12 hopping parameters, which include the dominated nearest-neighbor $dp$, $dd$ overlaps as listed in Table.~\ref{tab:tb16band}.
In this table, each hopping term is indicated by the orbital pairs which are connected by real-space vector [i,j,k], as can be located in the schematic.
Note that the symmetrically equivalent terms are no shown for clarity.
The site-energies of these orbitals are also presented in the last two panels.
The resulting band structure and Fermi surface  are shown in Fig.~\ref{fig6}(a)-(b), which are all in good agreement with our DFT results.
The band structure covers an energy range of -9$\sim$3\ eV, with strong split into two parts.
We find there appear two flat bands in the lower part, which are originated from two dangling $p_z^\prime$ orbitals with strong coupling to apical $d_{z^2}$ orbitals. 
The huge band split  indicates a significantly large energy scale of $dp$ hybridization, from which the effective direct $dd$ couplings emerge, as illustrated in our trilayer two-orbital model.
This model will be important for further study of strong correlation effects and superexchange couplings.

\begin{table}[t]
\begin{onehalfspace}
\noindent \begin{centering}
\caption{\label{tab:tb16band}Tight-binding parameters of sixteen-orbital model for $\mathrm{La_{4}Ni_{3}O_{10}}$
under pressure, in unit of eV. The corresponding orbitals are displayed in Fig.~\ref{fig6}(b). The symmetrically equivalent terms are no shown
for clarity. }
\renewcommand{\arraystretch}{1.5}
\begin{tabular*}{1\columnwidth}{@{\extracolsep{\fill}}ccccc}
\hline \hline
Hopping & $d_{z_3}$-$p_{z}$ & $d_{z_2}$-$p_{z}$ & $d_{x_2}$-$p_{x_2}$ & $d_{x_1}$-$p_{x_1}$\tabularnewline
 & [0,0,$\frac{1}{2}$] & [0,0,$\frac{1}{2}$] & [$\frac{1}{2}$,0,0] & [$\frac{1}{2}$,0,0]\tabularnewline
\hline 
$t$ & 1.911 & 1.523 & 1.681 & 1.676 \tabularnewline
\hline \hline
Hopping & $d_{z_1}$-$p_{z}^{\prime}$ & $d_{z_2}$-$p_{x_2}$ & $d_{z_1}$-$p_{x_1}$ & $p_{x_1}$-$p_{y_1}$\tabularnewline
 & [0,0,$\frac{1}{2}$] & [$\frac{1}{2}$,0,0] & [$\frac{1}{2}$,0,0] & [$\frac{1}{2}$,$\frac{1}{2}$,0]\tabularnewline
\hline 
$t$ & 1.591 & -0.938 & -0.887 & 0.604\tabularnewline
\hline \hline
Hopping & $p_{x_2}$-$p_{y_2}$ & $p_{z}$-$p_{x_1}$ & $p_{z}$-$p_{x_2}$ & $p_{z}^{\prime}$-$p_{x_1}$\tabularnewline
 & [$\frac{1}{2}$,$\frac{1}{2}$,0] & [$\frac{1}{2}$,0,$\frac{1}{2}$] & [$\frac{1}{2}$,0,$\frac{1}{2}$] & [$\frac{1}{2}$,0,$-\frac{1}{2}$]\tabularnewline
\hline 
$t$ & 0.553 & 0.597 & 0.577 & -0.465 \tabularnewline
\hline \hline
Site & $\text{\ensuremath{\epsilon_{x_{1}}}}$ & $\epsilon_{z_{1}}$ & $\epsilon_{x_{2}}$ & $\epsilon_{z_{2}}$\tabularnewline
\cline{2-5} \cline{3-5} \cline{4-5} \cline{5-5} 
energy & -1.431 & -1.354 & -1.075 & -1.027\tabularnewline
\hline 
Site & $\epsilon_{px_{1}}$ & $\epsilon_{px_{2}}$ & $\epsilon_{p_{z}}$ & $\epsilon_{p_{z}^{\prime}}$\tabularnewline
\cline{2-5} \cline{3-5} \cline{4-5} \cline{5-5} 
energy & -5.109 & -4.757 & -4.494 & -4.149\tabularnewline
\hline\hline
\end{tabular*}
\par\end{centering}
\end{onehalfspace}
\end{table}

According to the table, the site-energy differences within NiO$_2$ plane are calculated as $\epsilon_{x_2}-\epsilon_{px_2}=$3.42 eV and $\epsilon_{x_1}-\epsilon_{px_1}=$3.68 eV, respectively for inner and outer layers.
Along the apical direction, the energy difference are $\epsilon_{z_2}-\epsilon_{p_z}$=3.47 eV or $\epsilon_{z_1}-\epsilon_{p_z}$=3.14 eV.
Compared with bilayer, these magnitudes are quite similar, and both suggest a relative smaller energy difference between $d_{z^2}$ and apical $p_z$ orbitals than that of cuprate counterpart.
This implies that $\mathrm{La_4Ni_3O_{10}}$ should also lie in charge-transfer picture~\citep{wu2023charge}, and would favor a strong inter-layer superexchange coupling~\citep{PhysRevX.11.011050,lu2023interplay}.
However, for trilayer, the microscopic superexchange couplings is supposed to be more complicated as there involve at least three $d_{z^2}$ and $p_z$ orbitals.
For example, the superexchange process of $d_{z^2}-p_z-d_{z^2}$ between two adjacent layers, which is believed to be dominant in bilayer system \citep{lu2023interplay},  would be depressed in trilayer due to interference from other $d_{z^2}$ and $p_z$ along $c$ axis, and we believe that might be the reason for a  much lower transition temperature in $\mathrm{La_4Ni_3O_{10}}$.
Also, such speculation is in line with the fundamental concept that correlation is depressed in higher  dimensions.


\section{Discussion}
\label{sec:disscussion}
The microscopic charge transfer processes as well as the associating dual effect  reflect two impulses from pressure.
For the first one, the charge transfer from Ni-$e_g$ to $t_{2g}$ under pressure can be directly attributed to the octahedral tilting as illustrated in Section.~\ref{sec:structure}. 
Meanwhile, a lattice that possesses higher symmetry is generally preferred by a uniform pressure, which  corresponds to $I4/mmm$ phase in the case of RP series.
This can be simply understood from the definition of enthalpy $H=U+PV$, which shows that pressure would push the lattice to a higher energy state, where the symmetry unbreaking state is preferred.
For the second one, the partial  charge transfer from Ni-$d$ to La ions  essentially reflect  an urge from lattice collapse, which is  strongly reminiscent of lanthanide contraction effect \citep{JMLawrence_1981}.
In lanthanide series, the removal of correlated $4f$ electron under pressure (or lowering temperature) causes a decrease of screening from nucleus charge, so that the outer valence electrons are sucked in closer to the nucleus, further causing a decrease in ion radius \citep{JMLawrence_1981}. More importantly, the partial filling  of these correlated orbitals gives rise to valence fluctuation phenomenon.
In RP series compounds, an analogous dynamic mechanism should be applied, which suggests a likewise enhanced valence fluctuation under pressure. 
In this sense, the superconductivity can be interpreted as an unexpected outcome of valence fluctuation within Ni-$e_g$ orbitals.

Based on above analysis, we further elucidate the dynamic mechanism of electronic reconstructions under pressure.
As gradually apply pressure onto the lattice, 
both two effects would happen simultaneously, jointly leading to the upward shift of Ni-$e_g$ spectrum. Note that their respective contributions should be different, and according to our results, the second effect has the dominated role.
However, the first effect might also be indispensable, as there is a conjecture that the straightening of apical Ni-O-Ni bond angle can significantly enhance the perpendicular superexchange $J_\bot$, which is crucial for the occurrence of superconductivity.
Once the pressure is high  enough for such bond angle to reach formal 180$^\circ$, it corresponds to a  structure transition.
 For even higher pressure, there appears new situation. In this stage, the  oxygens within NiO$_2$ plane start to  move along $c$ axis, leading to a separation of the basal oxygens and Ni planes, but it maintains invariance of $I4/mmm$ structure~\citep{geisler2023structural}.
This structure variation is related to a further lattice collapse along $a/b$ axis, while the collapse along $c$ axis has largely depleted in early stage.
In this stage, it can be expected that there should have a  charge transfer from Ni-$d_{z^2}$ to $d_{x^2-y^2}$ orbital, further pushing Ni-$d_{z^2}$ spectrum towards higher energy.
Overall, we believe our exposition of the dynamic mechanism of electronic reconstructions is particularly helpful for theoretical understanding of electronic properties RP series,
which can inspire various further theoretical investigations 
on various aspects, such as valence fluctuation, superexchanges, correlation renormalization effect~\citep{yang2023orbitaldependent,tian2023correlation}, orbital selective property \citep{PhysRevB.108.L180510,PhysRevB.108.125105,chen2024electronic,PhysRevB.108.L201108,bilayermodel,PhysRevB.108.L180510}, strange metal behavior~\citep{strange327,zhu2024superconductivity},
and the interplay with spin/charge density wave fluctuations~\citep{2020intertwined}.
Moreover, it sheds light on experimental exploration of new superconductors under low pressure via chemical substitute.

Finally, regarding the layer dependence, our investigations on electronic properties of $\mathrm{La_4Ni_3O_{10}}$ have strongly recommended a similar superconducting  mechanism among RP series, in which the increase of layer seems to trivially play a negative role.
We tend to believe the suppression of $T_c$ is largely related to the decrease of correlation, while the latter can be understood in such way: as the bonding state of $d_{z^2}$ orbital always has a Wannier function that is elongated along $c$ axis all over the layers, the increase of layer indicates a more extended ground state wave function, which in turn weaken the correlation.
If so, it would be unsurprising that superconductivity is fully depressed in quadlayer $\mathrm{La_5Ni_4O_{13}}$ even which possesses an alike Fermi surface profile \citep{PhysRevB.105.085150}.
Nevertheless, such an experimental observation is  still of significance. All these features highlight that RP series is a perfect platform for theoretical understanding of the unconventional superconducting mechanism. Also, the role of the even-odd effect, which governs the excitation gap in local spin-$\frac{N}{2}$ chain, remains to be clear in the series~ \citep{sakakibara2023theoretical,rnxx,RN27}.



\section{Summary}
In summary, we performed a comprehensive DFT calculations on RP series $\mathrm{La_4Ni_3O_{10}}$ for both AP $P2_1/a$ and HP $I4/mmm$ phases, with $U$= 0, 3.5\ eV.
\label{Sec:Summary}
Our results clearly reveal the characteristic upward shift of Ni-$d_{z^2}$ bonding band under pressure,  suggesting a crucial link to the superconductivity in RP series. 
Our analysis of  electronic spectra as well as orbital  occupancies indicate a charge transfer of  $3d^{8.36}\rightarrow3d^{8.14}$ under pressure, specifically, 2.44$\rightarrow$2.14 for Ni-$e_g$ and 5.92$\rightarrow$6.0 for Ni-$t_{2g}$ sectors.
These trends indicate a critical dual effect of pressure, in which pressure on one hand promotes a charge redistribution within Ni-$e_g$ and $t_{2g}$ sectors. On the other hand, pressure drives a partial charge transfer from Ni to La ions. On this basis, in the discussion, we fully unveil the dynamic mechanism of electronic reconstructions under pressure, which sheds light on theoretical understanding of electronic correlations, superconductivity, as well as on experimental exploration of new superconductors with lower pressure.

Based on our DFT results, we proposed a trilayer two-orbital model by performing Wannier downfolding on Ni-$e_g$ orbitals. This model reveals four Fermi pockets $\alpha,\beta,\beta^\prime,\gamma$, which is very close to that of $\mathrm{La_4Ni_3O_{10}}$, suggesting a similar superconducting  mechanism in  RP series.
Also, our calculated RPA spin susceptibility suggests that $d_{x^2-y^2}$ orbitals should also be important for the magnetic fluctuation in RP series.


To gain insights to the charge-transfer property within ZAS scheme, we further proposed a high energy sixteen-orbital model based on Wannier downfolding on both Ni-$e_g$ and O-$p$ orbitals. The model also  well reproduces the low-lying band structure, with merely 12 nearest-neighbor $dd,dp$ hoppings.
The site-energies of the  model show that the  energy differences between Ni-$d$ and O-$p$ orbitals are close  to that in bilayer, which are 3.42 and 3.68 eV, respectively for inner and outer layers between $d_{x^2-y^2}-p_x/p_y$, and 3.47, 3.14\ eV for apical $d_{z^2}-p_z$ orbitals.
These implies that $\mathrm{La_4Ni_3O_{10}}$ also lies in the charge-transfer picture within ZAS scheme.
In short, the sixteen-orbital model as well as the trilayer two-orbital model are important for future study of strong correlation effects and unconventional pairing symmetry.

\textit{Note added.} The original version was submitted on Feb. 7th, 2024. Due to the arXiv technical problem, it was released until Feb. 13th, 2024. During the time, we noticed several independent works ~\citep{wang2024nonfermi,zhang2024prediction,yang2024effective,lu2024superconductivity,labollita2024electronic,tian2024effective,zhang2024spmwave} showing consistency with our results.

\begin{acknowledgments}
We are grateful to Qiang-Hua Wang, Guang-Ming Zhang, and Xunwu Hu for fruitful discussions. Work at Sun Yat-Sen University was supported by the National Key Research and Development Program of China (Grants No. 2022YFA1402802, 2018YFA0306001, 2023YFA1406500), the National Natural Science Foundation of China (Grants
No. 92165204, No. 12174454, No. 11974432,
No. 12274472), the Guangdong Basic and Applied Basic Research Foundation (Grants No. 2022A1515011618, No. 2021B1515120015), Guangzhou Basic and Applied Basic Research Funds (grant nos. 2024A04J6417), Guangdong Provincial Key Laboratory of Magnetoelectric Physics and Devices (Grant No. 2022B1212010008), and Shenzhen International Quantum Academy (Grant No. SIQA202102).
\end{acknowledgments}

\bibliography{la4310}

\begin{thebibliography}{87}%
\makeatletter
\providecommand \@ifxundefined [1]{%
 \@ifx{#1\undefined}
}%
\providecommand \@ifnum [1]{%
 \ifnum #1\expandafter \@firstoftwo
 \else \expandafter \@secondoftwo
 \fi
}%
\providecommand \@ifx [1]{%
 \ifx #1\expandafter \@firstoftwo
 \else \expandafter \@secondoftwo
 \fi
}%
\providecommand \natexlab [1]{#1}%
\providecommand \enquote  [1]{``#1''}%
\providecommand \bibnamefont  [1]{#1}%
\providecommand \bibfnamefont [1]{#1}%
\providecommand \citenamefont [1]{#1}%
\providecommand \href@noop [0]{\@secondoftwo}%
\providecommand \href [0]{\begingroup \@sanitize@url \@href}%
\providecommand \@href[1]{\@@startlink{#1}\@@href}%
\providecommand \@@href[1]{\endgroup#1\@@endlink}%
\providecommand \@sanitize@url [0]{\catcode `\\12\catcode `\$12\catcode
  `\&12\catcode `\#12\catcode `\^12\catcode `\_12\catcode `\%12\relax}%
\providecommand \@@startlink[1]{}%
\providecommand \@@endlink[0]{}%
\providecommand \url  [0]{\begingroup\@sanitize@url \@url }%
\providecommand \@url [1]{\endgroup\@href {#1}{\urlprefix }}%
\providecommand \urlprefix  [0]{URL }%
\providecommand \Eprint [0]{\href }%
\providecommand \doibase [0]{https://doi.org/}%
\providecommand \selectlanguage [0]{\@gobble}%
\providecommand \bibinfo  [0]{\@secondoftwo}%
\providecommand \bibfield  [0]{\@secondoftwo}%
\providecommand \translation [1]{[#1]}%
\providecommand \BibitemOpen [0]{}%
\providecommand \bibitemStop [0]{}%
\providecommand \bibitemNoStop [0]{.\EOS\space}%
\providecommand \EOS [0]{\spacefactor3000\relax}%
\providecommand \BibitemShut  [1]{\csname bibitem#1\endcsname}%
\let\auto@bib@innerbib\@empty
\bibitem [{\citenamefont {Li}\ \emph {et~al.}(2019)\citenamefont {Li},
  \citenamefont {Lee}, \citenamefont {Wang}, \citenamefont {Osada},
  \citenamefont {Crossley}, \citenamefont {Lee}, \citenamefont {Cui},
  \citenamefont {Hikita},\ and\ \citenamefont {Hwang}}]{NdNiO2}%
  \BibitemOpen
  \bibfield  {author} {\bibinfo {author} {\bibfnamefont {D.}~\bibnamefont
  {Li}}, \bibinfo {author} {\bibfnamefont {K.}~\bibnamefont {Lee}}, \bibinfo
  {author} {\bibfnamefont {B.~Y.}\ \bibnamefont {Wang}}, \bibinfo {author}
  {\bibfnamefont {M.}~\bibnamefont {Osada}}, \bibinfo {author} {\bibfnamefont
  {S.}~\bibnamefont {Crossley}}, \bibinfo {author} {\bibfnamefont {H.~R.}\
  \bibnamefont {Lee}}, \bibinfo {author} {\bibfnamefont {Y.}~\bibnamefont
  {Cui}}, \bibinfo {author} {\bibfnamefont {Y.}~\bibnamefont {Hikita}},\ and\
  \bibinfo {author} {\bibfnamefont {H.~Y.}\ \bibnamefont {Hwang}},\ }\bibfield
  {title} {\bibinfo {title} {Superconductivity in an infinite-layer
  nickelate},\ }\href {https://doi.org/10.1038/s41586-019-1496-5} {\bibfield
  {journal} {\bibinfo  {journal} {Nature}\ }\textbf {\bibinfo {volume} {572}},\
  \bibinfo {pages} {624} (\bibinfo {year} {2019})}\BibitemShut {NoStop}%
\bibitem [{\citenamefont {Zeng}\ \emph {et~al.}(2022)\citenamefont {Zeng},
  \citenamefont {Li}, \citenamefont {Chow}, \citenamefont {Cao}, \citenamefont
  {Zhang}, \citenamefont {Tang}, \citenamefont {Yin}, \citenamefont {Lim},
  \citenamefont {Hu}, \citenamefont {Yang},\ and\ \citenamefont
  {Ariando}}]{LaNiO2}%
  \BibitemOpen
  \bibfield  {author} {\bibinfo {author} {\bibfnamefont {S.}~\bibnamefont
  {Zeng}}, \bibinfo {author} {\bibfnamefont {C.}~\bibnamefont {Li}}, \bibinfo
  {author} {\bibfnamefont {L.~E.}\ \bibnamefont {Chow}}, \bibinfo {author}
  {\bibfnamefont {Y.}~\bibnamefont {Cao}}, \bibinfo {author} {\bibfnamefont
  {Z.}~\bibnamefont {Zhang}}, \bibinfo {author} {\bibfnamefont {C.~S.}\
  \bibnamefont {Tang}}, \bibinfo {author} {\bibfnamefont {X.}~\bibnamefont
  {Yin}}, \bibinfo {author} {\bibfnamefont {Z.~S.}\ \bibnamefont {Lim}},
  \bibinfo {author} {\bibfnamefont {J.}~\bibnamefont {Hu}}, \bibinfo {author}
  {\bibfnamefont {P.}~\bibnamefont {Yang}},\ and\ \bibinfo {author}
  {\bibfnamefont {A.}~\bibnamefont {Ariando}},\ }\bibfield  {title} {\bibinfo
  {title} {Superconductivity in infinite-layer nickelate
  {La$_{1-x}$Ca$_x$NiO$_2$} thin films},\ }\href
  {https://doi.org/10.1126/sciadv.abl9927} {\bibfield  {journal} {\bibinfo
  {journal} {Science Advances}\ }\textbf {\bibinfo {volume} {8}},\ \bibinfo
  {pages} {eabl9927} (\bibinfo {year} {2022})}\BibitemShut {NoStop}%
\bibitem [{\citenamefont {Osada}\ \emph {et~al.}(2020)\citenamefont {Osada},
  \citenamefont {Wang}, \citenamefont {Goodge}, \citenamefont {Lee},
  \citenamefont {Yoon}, \citenamefont {Sakuma}, \citenamefont {Li},
  \citenamefont {Miura}, \citenamefont {Kourkoutis},\ and\ \citenamefont
  {Hwang}}]{PrNiO2}%
  \BibitemOpen
  \bibfield  {author} {\bibinfo {author} {\bibfnamefont {M.}~\bibnamefont
  {Osada}}, \bibinfo {author} {\bibfnamefont {B.~Y.}\ \bibnamefont {Wang}},
  \bibinfo {author} {\bibfnamefont {B.~H.}\ \bibnamefont {Goodge}}, \bibinfo
  {author} {\bibfnamefont {K.}~\bibnamefont {Lee}}, \bibinfo {author}
  {\bibfnamefont {H.}~\bibnamefont {Yoon}}, \bibinfo {author} {\bibfnamefont
  {K.}~\bibnamefont {Sakuma}}, \bibinfo {author} {\bibfnamefont
  {D.}~\bibnamefont {Li}}, \bibinfo {author} {\bibfnamefont {M.}~\bibnamefont
  {Miura}}, \bibinfo {author} {\bibfnamefont {L.~F.}\ \bibnamefont
  {Kourkoutis}},\ and\ \bibinfo {author} {\bibfnamefont {H.~Y.}\ \bibnamefont
  {Hwang}},\ }\bibfield  {title} {\bibinfo {title} {A superconducting
  praseodymium nickelate with infinite layer structure},\ }\href
  {https://doi.org/10.1021/acs.nanolett.0c01392} {\bibfield  {journal}
  {\bibinfo  {journal} {Nano Letters}\ }\textbf {\bibinfo {volume} {20}},\
  \bibinfo {pages} {5735} (\bibinfo {year} {2020})}\BibitemShut {NoStop}%
\bibitem [{\citenamefont {Sun}\ \emph {et~al.}(2023)\citenamefont {Sun},
  \citenamefont {Huo}, \citenamefont {Hu}, \citenamefont {Li}, \citenamefont
  {Liu}, \citenamefont {Han}, \citenamefont {Tang}, \citenamefont {Mao},
  \citenamefont {Yang}, \citenamefont {Wang}, \citenamefont {Cheng},
  \citenamefont {Yao}, \citenamefont {Zhang},\ and\ \citenamefont
  {Wang}}]{bilayernature}%
  \BibitemOpen
  \bibfield  {author} {\bibinfo {author} {\bibfnamefont {H.}~\bibnamefont
  {Sun}}, \bibinfo {author} {\bibfnamefont {M.}~\bibnamefont {Huo}}, \bibinfo
  {author} {\bibfnamefont {X.}~\bibnamefont {Hu}}, \bibinfo {author}
  {\bibfnamefont {J.}~\bibnamefont {Li}}, \bibinfo {author} {\bibfnamefont
  {Z.}~\bibnamefont {Liu}}, \bibinfo {author} {\bibfnamefont {Y.}~\bibnamefont
  {Han}}, \bibinfo {author} {\bibfnamefont {L.}~\bibnamefont {Tang}}, \bibinfo
  {author} {\bibfnamefont {Z.}~\bibnamefont {Mao}}, \bibinfo {author}
  {\bibfnamefont {P.}~\bibnamefont {Yang}}, \bibinfo {author} {\bibfnamefont
  {B.}~\bibnamefont {Wang}}, \bibinfo {author} {\bibfnamefont {J.}~\bibnamefont
  {Cheng}}, \bibinfo {author} {\bibfnamefont {D.-X.}\ \bibnamefont {Yao}},
  \bibinfo {author} {\bibfnamefont {G.-M.}\ \bibnamefont {Zhang}},\ and\
  \bibinfo {author} {\bibfnamefont {M.}~\bibnamefont {Wang}},\ }\bibfield
  {title} {\bibinfo {title} {Signatures of superconductivity near 80 {K} in a
  nickelate under high pressure},\ }\href
  {https://doi.org/10.1038/s41586-023-06408-7} {\bibfield  {journal} {\bibinfo
  {journal} {Nature}\ }\textbf {\bibinfo {volume} {621}},\ \bibinfo {pages}
  {493} (\bibinfo {year} {2023})}\BibitemShut {NoStop}%
\bibitem [{\citenamefont {Wang}\ \emph
  {et~al.}(2024{\natexlab{a}})\citenamefont {Wang}, \citenamefont {Wang},
  \citenamefont {Shen}, \citenamefont {Hou}, \citenamefont {Ma}, \citenamefont
  {Shi}, \citenamefont {Ren}, \citenamefont {Gu}, \citenamefont {Ma},
  \citenamefont {Yang}, \citenamefont {Liu}, \citenamefont {Guo}, \citenamefont
  {Sun}, \citenamefont {Zhang}, \citenamefont {Calder}, \citenamefont {Yan},
  \citenamefont {Wang}, \citenamefont {Uwatoko},\ and\ \citenamefont
  {Cheng}}]{wang2023pressureinduced}%
  \BibitemOpen
  \bibfield  {author} {\bibinfo {author} {\bibfnamefont {G.}~\bibnamefont
  {Wang}}, \bibinfo {author} {\bibfnamefont {N.~N.}\ \bibnamefont {Wang}},
  \bibinfo {author} {\bibfnamefont {X.~L.}\ \bibnamefont {Shen}}, \bibinfo
  {author} {\bibfnamefont {J.}~\bibnamefont {Hou}}, \bibinfo {author}
  {\bibfnamefont {L.}~\bibnamefont {Ma}}, \bibinfo {author} {\bibfnamefont
  {L.~F.}\ \bibnamefont {Shi}}, \bibinfo {author} {\bibfnamefont {Z.~A.}\
  \bibnamefont {Ren}}, \bibinfo {author} {\bibfnamefont {Y.~D.}\ \bibnamefont
  {Gu}}, \bibinfo {author} {\bibfnamefont {H.~M.}\ \bibnamefont {Ma}}, \bibinfo
  {author} {\bibfnamefont {P.~T.}\ \bibnamefont {Yang}}, \bibinfo {author}
  {\bibfnamefont {Z.~Y.}\ \bibnamefont {Liu}}, \bibinfo {author} {\bibfnamefont
  {H.~Z.}\ \bibnamefont {Guo}}, \bibinfo {author} {\bibfnamefont {J.~P.}\
  \bibnamefont {Sun}}, \bibinfo {author} {\bibfnamefont {G.~M.}\ \bibnamefont
  {Zhang}}, \bibinfo {author} {\bibfnamefont {S.}~\bibnamefont {Calder}},
  \bibinfo {author} {\bibfnamefont {J.-Q.}\ \bibnamefont {Yan}}, \bibinfo
  {author} {\bibfnamefont {B.~S.}\ \bibnamefont {Wang}}, \bibinfo {author}
  {\bibfnamefont {Y.}~\bibnamefont {Uwatoko}},\ and\ \bibinfo {author}
  {\bibfnamefont {J.-G.}\ \bibnamefont {Cheng}},\ }\bibfield  {title} {\bibinfo
  {title} {Pressure-induced superconductivity in polycrystalline
  {${\mathrm{La}}_{3}{\mathrm{Ni}}_{2}{\mathrm{O}}_{7\ensuremath{-}\ensuremath{\delta}}$}},\
  }\href {https://doi.org/10.1103/PhysRevX.14.011040} {\bibfield  {journal}
  {\bibinfo  {journal} {Phys. Rev. X}\ }\textbf {\bibinfo {volume} {14}},\
  \bibinfo {pages} {011040} (\bibinfo {year} {2024}{\natexlab{a}})}\BibitemShut
  {NoStop}%
\bibitem [{\citenamefont {Hou}\ \emph {et~al.}(2023)\citenamefont {Hou},
  \citenamefont {Yang}, \citenamefont {Liu}, \citenamefont {Li}, \citenamefont
  {Shan}, \citenamefont {Ma}, \citenamefont {Wang}, \citenamefont {Wang},
  \citenamefont {Guo}, \citenamefont {Sun}, \citenamefont {Uwatoko},
  \citenamefont {Wang}, \citenamefont {Zhang}, \citenamefont {Wang},\ and\
  \citenamefont {Cheng}}]{JunHou117302}%
  \BibitemOpen
  \bibfield  {author} {\bibinfo {author} {\bibfnamefont {J.}~\bibnamefont
  {Hou}}, \bibinfo {author} {\bibfnamefont {P.-T.}\ \bibnamefont {Yang}},
  \bibinfo {author} {\bibfnamefont {Z.-Y.}\ \bibnamefont {Liu}}, \bibinfo
  {author} {\bibfnamefont {J.-Y.}\ \bibnamefont {Li}}, \bibinfo {author}
  {\bibfnamefont {P.-F.}\ \bibnamefont {Shan}}, \bibinfo {author}
  {\bibfnamefont {L.}~\bibnamefont {Ma}}, \bibinfo {author} {\bibfnamefont
  {G.}~\bibnamefont {Wang}}, \bibinfo {author} {\bibfnamefont {N.-N.}\
  \bibnamefont {Wang}}, \bibinfo {author} {\bibfnamefont {H.-Z.}\ \bibnamefont
  {Guo}}, \bibinfo {author} {\bibfnamefont {J.-P.}\ \bibnamefont {Sun}},
  \bibinfo {author} {\bibfnamefont {Y.}~\bibnamefont {Uwatoko}}, \bibinfo
  {author} {\bibfnamefont {M.}~\bibnamefont {Wang}}, \bibinfo {author}
  {\bibfnamefont {G.-M.}\ \bibnamefont {Zhang}}, \bibinfo {author}
  {\bibfnamefont {B.-S.}\ \bibnamefont {Wang}},\ and\ \bibinfo {author}
  {\bibfnamefont {J.-G.}\ \bibnamefont {Cheng}},\ }\bibfield  {title} {\bibinfo
  {title} {Emergence of high-temperature superconducting phase in pressurized
  {La$_{3}$Ni$_{2}$O$_7$} crystals},\ }\href
  {https://doi.org/10.1088/0256-307X/40/11/117302} {\bibfield  {journal}
  {\bibinfo  {journal} {Chinese Physics Letters}\ }\textbf {\bibinfo {volume}
  {40}},\ \bibinfo {eid} {117302} (\bibinfo {year} {2023})}\BibitemShut
  {NoStop}%
\bibitem [{\citenamefont {Zhu}\ \emph {et~al.}(2024)\citenamefont {Zhu},
  \citenamefont {Zhang}, \citenamefont {Pan}, \citenamefont {Chen},
  \citenamefont {Peng}, \citenamefont {Chen}, \citenamefont {Ren},
  \citenamefont {Liu}, \citenamefont {Li}, \citenamefont {Xing}, \citenamefont
  {Han}, \citenamefont {Wang}, \citenamefont {Jia}, \citenamefont {Wo},
  \citenamefont {Gu}, \citenamefont {Gu}, \citenamefont {Ji}, \citenamefont
  {Wang}, \citenamefont {Gou}, \citenamefont {Shen}, \citenamefont {Ying},
  \citenamefont {Chen}, \citenamefont {Yang}, \citenamefont {Zheng},
  \citenamefont {Zeng}, \citenamefont {gang Guo},\ and\ \citenamefont
  {Zhao}}]{zhu2024superconductivity}%
  \BibitemOpen
  \bibfield  {author} {\bibinfo {author} {\bibfnamefont {Y.}~\bibnamefont
  {Zhu}}, \bibinfo {author} {\bibfnamefont {E.}~\bibnamefont {Zhang}}, \bibinfo
  {author} {\bibfnamefont {B.}~\bibnamefont {Pan}}, \bibinfo {author}
  {\bibfnamefont {X.}~\bibnamefont {Chen}}, \bibinfo {author} {\bibfnamefont
  {D.}~\bibnamefont {Peng}}, \bibinfo {author} {\bibfnamefont {L.}~\bibnamefont
  {Chen}}, \bibinfo {author} {\bibfnamefont {H.}~\bibnamefont {Ren}}, \bibinfo
  {author} {\bibfnamefont {F.}~\bibnamefont {Liu}}, \bibinfo {author}
  {\bibfnamefont {N.}~\bibnamefont {Li}}, \bibinfo {author} {\bibfnamefont
  {Z.}~\bibnamefont {Xing}}, \bibinfo {author} {\bibfnamefont {J.}~\bibnamefont
  {Han}}, \bibinfo {author} {\bibfnamefont {J.}~\bibnamefont {Wang}}, \bibinfo
  {author} {\bibfnamefont {D.}~\bibnamefont {Jia}}, \bibinfo {author}
  {\bibfnamefont {H.}~\bibnamefont {Wo}}, \bibinfo {author} {\bibfnamefont
  {Y.}~\bibnamefont {Gu}}, \bibinfo {author} {\bibfnamefont {Y.}~\bibnamefont
  {Gu}}, \bibinfo {author} {\bibfnamefont {L.}~\bibnamefont {Ji}}, \bibinfo
  {author} {\bibfnamefont {W.}~\bibnamefont {Wang}}, \bibinfo {author}
  {\bibfnamefont {H.}~\bibnamefont {Gou}}, \bibinfo {author} {\bibfnamefont
  {Y.}~\bibnamefont {Shen}}, \bibinfo {author} {\bibfnamefont {T.}~\bibnamefont
  {Ying}}, \bibinfo {author} {\bibfnamefont {X.}~\bibnamefont {Chen}}, \bibinfo
  {author} {\bibfnamefont {W.}~\bibnamefont {Yang}}, \bibinfo {author}
  {\bibfnamefont {C.}~\bibnamefont {Zheng}}, \bibinfo {author} {\bibfnamefont
  {Q.}~\bibnamefont {Zeng}}, \bibinfo {author} {\bibfnamefont {J.}~\bibnamefont
  {gang Guo}},\ and\ \bibinfo {author} {\bibfnamefont {J.}~\bibnamefont
  {Zhao}},\ }\href@noop {} {\bibinfo {title} {Superconductivity in trilayer
  nickelate {La$_4$Ni$_3$O$_{10}$} single crystals}} (\bibinfo {year} {2024}),\
  \Eprint {https://arxiv.org/abs/2311.07353} {arXiv:2311.07353
  [cond-mat.supr-con]} \BibitemShut {NoStop}%
\bibitem [{\citenamefont {Zhou}\ \emph {et~al.}(2023)\citenamefont {Zhou},
  \citenamefont {Guo}, \citenamefont {Cai}, \citenamefont {Sun}, \citenamefont
  {Wang}, \citenamefont {Zhao}, \citenamefont {Han}, \citenamefont {Chen},
  \citenamefont {Wu}, \citenamefont {Ding}, \citenamefont {Wang}, \citenamefont
  {Xiang}, \citenamefont {kwang Mao},\ and\ \citenamefont
  {Sun}}]{zhou2023evidence}%
  \BibitemOpen
  \bibfield  {author} {\bibinfo {author} {\bibfnamefont {Y.}~\bibnamefont
  {Zhou}}, \bibinfo {author} {\bibfnamefont {J.}~\bibnamefont {Guo}}, \bibinfo
  {author} {\bibfnamefont {S.}~\bibnamefont {Cai}}, \bibinfo {author}
  {\bibfnamefont {H.}~\bibnamefont {Sun}}, \bibinfo {author} {\bibfnamefont
  {P.}~\bibnamefont {Wang}}, \bibinfo {author} {\bibfnamefont {J.}~\bibnamefont
  {Zhao}}, \bibinfo {author} {\bibfnamefont {J.}~\bibnamefont {Han}}, \bibinfo
  {author} {\bibfnamefont {X.}~\bibnamefont {Chen}}, \bibinfo {author}
  {\bibfnamefont {Q.}~\bibnamefont {Wu}}, \bibinfo {author} {\bibfnamefont
  {Y.}~\bibnamefont {Ding}}, \bibinfo {author} {\bibfnamefont {M.}~\bibnamefont
  {Wang}}, \bibinfo {author} {\bibfnamefont {T.}~\bibnamefont {Xiang}},
  \bibinfo {author} {\bibfnamefont {H.}~\bibnamefont {kwang Mao}},\ and\
  \bibinfo {author} {\bibfnamefont {L.}~\bibnamefont {Sun}},\ }\href@noop {}
  {\bibinfo {title} {Evidence of filamentary superconductivity in pressurized
  {La$_3$Ni$_2$O$_7$} single crystals}} (\bibinfo {year} {2023}),\ \Eprint
  {https://arxiv.org/abs/2311.12361} {arXiv:2311.12361 [cond-mat.supr-con]}
  \BibitemShut {NoStop}%
\bibitem [{\citenamefont {Zhang}\ \emph
  {et~al.}(2023{\natexlab{a}})\citenamefont {Zhang}, \citenamefont {Pei},
  \citenamefont {Du}, \citenamefont {Cao}, \citenamefont {Wang}, \citenamefont
  {Wu}, \citenamefont {Li}, \citenamefont {Zhao}, \citenamefont {Li},
  \citenamefont {Cao}, \citenamefont {Zhu}, \citenamefont {Zhang},
  \citenamefont {Yu}, \citenamefont {Cheng}, \citenamefont {Zhao},
  \citenamefont {Chen}, \citenamefont {Guo}, \citenamefont {Yang},\ and\
  \citenamefont {Qi}}]{zhang2023superconductivity}%
  \BibitemOpen
  \bibfield  {author} {\bibinfo {author} {\bibfnamefont {M.}~\bibnamefont
  {Zhang}}, \bibinfo {author} {\bibfnamefont {C.}~\bibnamefont {Pei}}, \bibinfo
  {author} {\bibfnamefont {X.}~\bibnamefont {Du}}, \bibinfo {author}
  {\bibfnamefont {Y.}~\bibnamefont {Cao}}, \bibinfo {author} {\bibfnamefont
  {Q.}~\bibnamefont {Wang}}, \bibinfo {author} {\bibfnamefont {J.}~\bibnamefont
  {Wu}}, \bibinfo {author} {\bibfnamefont {Y.}~\bibnamefont {Li}}, \bibinfo
  {author} {\bibfnamefont {Y.}~\bibnamefont {Zhao}}, \bibinfo {author}
  {\bibfnamefont {C.}~\bibnamefont {Li}}, \bibinfo {author} {\bibfnamefont
  {W.}~\bibnamefont {Cao}}, \bibinfo {author} {\bibfnamefont {S.}~\bibnamefont
  {Zhu}}, \bibinfo {author} {\bibfnamefont {Q.}~\bibnamefont {Zhang}}, \bibinfo
  {author} {\bibfnamefont {N.}~\bibnamefont {Yu}}, \bibinfo {author}
  {\bibfnamefont {P.}~\bibnamefont {Cheng}}, \bibinfo {author} {\bibfnamefont
  {J.}~\bibnamefont {Zhao}}, \bibinfo {author} {\bibfnamefont {Y.}~\bibnamefont
  {Chen}}, \bibinfo {author} {\bibfnamefont {H.}~\bibnamefont {Guo}}, \bibinfo
  {author} {\bibfnamefont {L.}~\bibnamefont {Yang}},\ and\ \bibinfo {author}
  {\bibfnamefont {Y.}~\bibnamefont {Qi}},\ }\href@noop {} {\bibinfo {title}
  {Superconductivity in trilayer nickelate {La$_4$Ni$_3$O$_{10}$} under
  pressure}} (\bibinfo {year} {2023}{\natexlab{a}}),\ \Eprint
  {https://arxiv.org/abs/2311.07423} {arXiv:2311.07423 [cond-mat.supr-con]}
  \BibitemShut {NoStop}%
\bibitem [{\citenamefont {Li}\ \emph {et~al.}(2024{\natexlab{a}})\citenamefont
  {Li}, \citenamefont {Zhang}, \citenamefont {Xiang}, \citenamefont {Zhang},
  \citenamefont {Zhu},\ and\ \citenamefont {Wen}}]{QingLi17401}%
  \BibitemOpen
  \bibfield  {author} {\bibinfo {author} {\bibfnamefont {Q.}~\bibnamefont
  {Li}}, \bibinfo {author} {\bibfnamefont {Y.-J.}\ \bibnamefont {Zhang}},
  \bibinfo {author} {\bibfnamefont {Z.-N.}\ \bibnamefont {Xiang}}, \bibinfo
  {author} {\bibfnamefont {Y.}~\bibnamefont {Zhang}}, \bibinfo {author}
  {\bibfnamefont {X.}~\bibnamefont {Zhu}},\ and\ \bibinfo {author}
  {\bibfnamefont {H.-H.}\ \bibnamefont {Wen}},\ }\bibfield  {title} {\bibinfo
  {title} {Signature of superconductivity in pressurized
  {La$_{4}$Ni$_{3}$O$_{10}$}},\ }\href
  {https://doi.org/10.1088/0256-307X/41/1/017401} {\bibfield  {journal}
  {\bibinfo  {journal} {Chinese Physics Letters}\ }\textbf {\bibinfo {volume}
  {41}},\ \bibinfo {eid} {017401} (\bibinfo {year}
  {2024}{\natexlab{a}})}\BibitemShut {NoStop}%
\bibitem [{\citenamefont {Luo}\ \emph {et~al.}(2023{\natexlab{a}})\citenamefont
  {Luo}, \citenamefont {Hu}, \citenamefont {Wang}, \citenamefont {W\'u},\ and\
  \citenamefont {Yao}}]{bilayermodel}%
  \BibitemOpen
  \bibfield  {author} {\bibinfo {author} {\bibfnamefont {Z.}~\bibnamefont
  {Luo}}, \bibinfo {author} {\bibfnamefont {X.}~\bibnamefont {Hu}}, \bibinfo
  {author} {\bibfnamefont {M.}~\bibnamefont {Wang}}, \bibinfo {author}
  {\bibfnamefont {W.}~\bibnamefont {W\'u}},\ and\ \bibinfo {author}
  {\bibfnamefont {D.-X.}\ \bibnamefont {Yao}},\ }\bibfield  {title} {\bibinfo
  {title} {Bilayer two-orbital model of $\mathrm{La_3Ni_2O_7}$ under
  pressure},\ }\href {https://doi.org/10.1103/PhysRevLett.131.126001}
  {\bibfield  {journal} {\bibinfo  {journal} {Physical Review Letters}\
  }\textbf {\bibinfo {volume} {131}},\ \bibinfo {pages} {126001} (\bibinfo
  {year} {2023}{\natexlab{a}})}\BibitemShut {NoStop}%
\bibitem [{\citenamefont {Luo}\ \emph {et~al.}(2023{\natexlab{b}})\citenamefont
  {Luo}, \citenamefont {Lv}, \citenamefont {Wang}, \citenamefont {W\'u},\ and\
  \citenamefont {Yao}}]{luo2023hightc}%
  \BibitemOpen
  \bibfield  {author} {\bibinfo {author} {\bibfnamefont {Z.}~\bibnamefont
  {Luo}}, \bibinfo {author} {\bibfnamefont {B.}~\bibnamefont {Lv}}, \bibinfo
  {author} {\bibfnamefont {M.}~\bibnamefont {Wang}}, \bibinfo {author}
  {\bibfnamefont {W.}~\bibnamefont {W\'u}},\ and\ \bibinfo {author}
  {\bibfnamefont {D.-X.}\ \bibnamefont {Yao}},\ }\href@noop {} {\bibinfo
  {title} {High-{T$_C$} superconductivity in $\mathrm{La_3Ni_2O_7}$ based on
  the bilayer two-orbital {t-J} model}} (\bibinfo {year}
  {2023}{\natexlab{b}}),\ \Eprint {https://arxiv.org/abs/2308.16564}
  {arXiv:2308.16564 [cond-mat.supr-con]} \BibitemShut {NoStop}%
\bibitem [{\citenamefont {Tam}\ \emph {et~al.}(2022)\citenamefont {Tam},
  \citenamefont {Choi}, \citenamefont {Ding}, \citenamefont {Agrestini},
  \citenamefont {Nag}, \citenamefont {Wu}, \citenamefont {Huang}, \citenamefont
  {Luo}, \citenamefont {Gao}, \citenamefont {Garcia-Fernandez}, \citenamefont
  {Qiao},\ and\ \citenamefont {Zhou}}]{RN20}%
  \BibitemOpen
  \bibfield  {author} {\bibinfo {author} {\bibfnamefont {C.~C.}\ \bibnamefont
  {Tam}}, \bibinfo {author} {\bibfnamefont {J.}~\bibnamefont {Choi}}, \bibinfo
  {author} {\bibfnamefont {X.}~\bibnamefont {Ding}}, \bibinfo {author}
  {\bibfnamefont {S.}~\bibnamefont {Agrestini}}, \bibinfo {author}
  {\bibfnamefont {A.}~\bibnamefont {Nag}}, \bibinfo {author} {\bibfnamefont
  {M.}~\bibnamefont {Wu}}, \bibinfo {author} {\bibfnamefont {B.}~\bibnamefont
  {Huang}}, \bibinfo {author} {\bibfnamefont {H.}~\bibnamefont {Luo}}, \bibinfo
  {author} {\bibfnamefont {P.}~\bibnamefont {Gao}}, \bibinfo {author}
  {\bibfnamefont {M.}~\bibnamefont {Garcia-Fernandez}}, \bibinfo {author}
  {\bibfnamefont {L.}~\bibnamefont {Qiao}},\ and\ \bibinfo {author}
  {\bibfnamefont {K.}~\bibnamefont {Zhou}},\ }\bibfield  {title} {\bibinfo
  {title} {Charge density waves in infinite-layer {NdNiO$_2$} nickelates},\
  }\href {https://doi.org/10.1038/s41563-022-01330-1} {\bibfield  {journal}
  {\bibinfo  {journal} {Nature Materials}\ }\textbf {\bibinfo {volume} {21}},\
  \bibinfo {pages} {1116} (\bibinfo {year} {2022})}\BibitemShut {NoStop}%
\bibitem [{\citenamefont {Liu}\ \emph {et~al.}(2020)\citenamefont {Liu},
  \citenamefont {Ren}, \citenamefont {Zhu}, \citenamefont {Wang},\ and\
  \citenamefont {Yang}}]{RN24}%
  \BibitemOpen
  \bibfield  {author} {\bibinfo {author} {\bibfnamefont {Z.}~\bibnamefont
  {Liu}}, \bibinfo {author} {\bibfnamefont {Z.}~\bibnamefont {Ren}}, \bibinfo
  {author} {\bibfnamefont {W.}~\bibnamefont {Zhu}}, \bibinfo {author}
  {\bibfnamefont {Z.}~\bibnamefont {Wang}},\ and\ \bibinfo {author}
  {\bibfnamefont {J.}~\bibnamefont {Yang}},\ }\bibfield  {title} {\bibinfo
  {title} {Electronic and magnetic structure of infinite-layer {NdNiO$_2$}:
  trace of antiferromagnetic metal},\ }\href
  {https://doi.org/10.1038/s41535-020-0229-1} {\bibfield  {journal} {\bibinfo
  {journal} {npj Quantum Materials}\ }\textbf {\bibinfo {volume} {5}},\
  \bibinfo {pages} {31} (\bibinfo {year} {2020})}\BibitemShut {NoStop}%
\bibitem [{\citenamefont {Liu}\ \emph {et~al.}(2022)\citenamefont {Liu},
  \citenamefont {Sun}, \citenamefont {Huo}, \citenamefont {Ma}, \citenamefont
  {Ji}, \citenamefont {Yi}, \citenamefont {Li}, \citenamefont {Liu},
  \citenamefont {Yu}, \citenamefont {Zhang}, \citenamefont {Chen},
  \citenamefont {Liang}, \citenamefont {Dong}, \citenamefont {Guo},
  \citenamefont {Zhong}, \citenamefont {Shen}, \citenamefont {Li},\ and\
  \citenamefont {Wang}}]{RN25}%
  \BibitemOpen
  \bibfield  {author} {\bibinfo {author} {\bibfnamefont {Z.}~\bibnamefont
  {Liu}}, \bibinfo {author} {\bibfnamefont {H.}~\bibnamefont {Sun}}, \bibinfo
  {author} {\bibfnamefont {M.}~\bibnamefont {Huo}}, \bibinfo {author}
  {\bibfnamefont {X.}~\bibnamefont {Ma}}, \bibinfo {author} {\bibfnamefont
  {Y.}~\bibnamefont {Ji}}, \bibinfo {author} {\bibfnamefont {E.}~\bibnamefont
  {Yi}}, \bibinfo {author} {\bibfnamefont {L.}~\bibnamefont {Li}}, \bibinfo
  {author} {\bibfnamefont {H.}~\bibnamefont {Liu}}, \bibinfo {author}
  {\bibfnamefont {J.}~\bibnamefont {Yu}}, \bibinfo {author} {\bibfnamefont
  {Z.}~\bibnamefont {Zhang}}, \bibinfo {author} {\bibfnamefont
  {Z.}~\bibnamefont {Chen}}, \bibinfo {author} {\bibfnamefont {F.}~\bibnamefont
  {Liang}}, \bibinfo {author} {\bibfnamefont {H.}~\bibnamefont {Dong}},
  \bibinfo {author} {\bibfnamefont {H.}~\bibnamefont {Guo}}, \bibinfo {author}
  {\bibfnamefont {D.}~\bibnamefont {Zhong}}, \bibinfo {author} {\bibfnamefont
  {B.}~\bibnamefont {Shen}}, \bibinfo {author} {\bibfnamefont {S.}~\bibnamefont
  {Li}},\ and\ \bibinfo {author} {\bibfnamefont {M.}~\bibnamefont {Wang}},\
  }\bibfield  {title} {\bibinfo {title} {Evidence for charge and spin density
  waves in single crystals of {La$_3$Ni$_2$O$_7$} and {La$_3$Ni$_2$O$_6$}},\
  }\href {https://doi.org/10.1007/s11433-022-1962-4} {\bibfield  {journal}
  {\bibinfo  {journal} {Science China Physics, Mechanics \& Astronomy}\
  }\textbf {\bibinfo {volume} {66}},\ \bibinfo {pages} {217411} (\bibinfo
  {year} {2022})}\BibitemShut {NoStop}%
\bibitem [{\citenamefont {Hu}\ and\ \citenamefont
  {Wu}(2019)}]{PhysRevResearch.1.032046}%
  \BibitemOpen
  \bibfield  {author} {\bibinfo {author} {\bibfnamefont {L.-H.}\ \bibnamefont
  {Hu}}\ and\ \bibinfo {author} {\bibfnamefont {C.}~\bibnamefont {Wu}},\
  }\bibfield  {title} {\bibinfo {title} {Two-band model for magnetism and
  superconductivity in nickelates},\ }\href
  {https://doi.org/10.1103/PhysRevResearch.1.032046} {\bibfield  {journal}
  {\bibinfo  {journal} {Phys. Rev. Res.}\ }\textbf {\bibinfo {volume} {1}},\
  \bibinfo {pages} {032046} (\bibinfo {year} {2019})}\BibitemShut {NoStop}%
\bibitem [{\citenamefont {Zhang}\ \emph
  {et~al.}(2024{\natexlab{a}})\citenamefont {Zhang}, \citenamefont {Lin},
  \citenamefont {Moreo}, \citenamefont {Maier},\ and\ \citenamefont
  {Dagotto}}]{PhysRevB.109.045151}%
  \BibitemOpen
  \bibfield  {author} {\bibinfo {author} {\bibfnamefont {Y.}~\bibnamefont
  {Zhang}}, \bibinfo {author} {\bibfnamefont {L.-F.}\ \bibnamefont {Lin}},
  \bibinfo {author} {\bibfnamefont {A.}~\bibnamefont {Moreo}}, \bibinfo
  {author} {\bibfnamefont {T.~A.}\ \bibnamefont {Maier}},\ and\ \bibinfo
  {author} {\bibfnamefont {E.}~\bibnamefont {Dagotto}},\ }\bibfield  {title}
  {\bibinfo {title} {Electronic structure, magnetic correlations, and
  superconducting pairing in the reduced ruddlesden-popper bilayer
  {${\mathrm{La}}_{3}{\mathrm{Ni}}_{2}{\mathrm{O}}_{6}$} under pressure:
  Different role of ${d}_{3{z}^{2}\ensuremath{-}{r}^{2}}$ orbital compared with
  {${\mathrm{La}}_{3}{\mathrm{Ni}}_{2}{\mathrm{O}}_{7}$}},\ }\href
  {https://doi.org/10.1103/PhysRevB.109.045151} {\bibfield  {journal} {\bibinfo
   {journal} {Phys. Rev. B}\ }\textbf {\bibinfo {volume} {109}},\ \bibinfo
  {pages} {045151} (\bibinfo {year} {2024}{\natexlab{a}})}\BibitemShut
  {NoStop}%
\bibitem [{\citenamefont {Shen}\ \emph {et~al.}(2023)\citenamefont {Shen},
  \citenamefont {Qin},\ and\ \citenamefont {Zhang}}]{ChinPhysLett.40.127401}%
  \BibitemOpen
  \bibfield  {author} {\bibinfo {author} {\bibfnamefont {Y.}~\bibnamefont
  {Shen}}, \bibinfo {author} {\bibfnamefont {M.}~\bibnamefont {Qin}},\ and\
  \bibinfo {author} {\bibfnamefont {G.-M.}\ \bibnamefont {Zhang}},\ }\bibfield
  {title} {\bibinfo {title} {Effective bi-layer model hamiltonian and
  density-matrix renormalization\\ group study for the high-{$T_{\rm c}$}
  superconductivity\\ in {La$_{3}$Ni$_{2}$O$_{7}$} under high pressure},\
  }\href {https://doi.org/10.1088/0256-307X/40/12/127401} {\bibfield  {journal}
  {\bibinfo  {journal} {Chin. Phys. Lett.}\ }\textbf {\bibinfo {volume} {40}},\
  \bibinfo {pages} {127401} (\bibinfo {year} {2023})}\BibitemShut {NoStop}%
\bibitem [{\citenamefont {Yao}\ and\ \citenamefont
  {Carlson}(2007)}]{PhysRevB.75.012414}%
  \BibitemOpen
  \bibfield  {author} {\bibinfo {author} {\bibfnamefont {D.~X.}\ \bibnamefont
  {Yao}}\ and\ \bibinfo {author} {\bibfnamefont {E.~W.}\ \bibnamefont
  {Carlson}},\ }\bibfield  {title} {\bibinfo {title} {Spin-wave dispersion in
  half-doped {${\mathrm{La}}_{3/2}{\mathrm{Sr}}_{1/2}{\mathrm{NiO}}_{4}$}},\
  }\href {https://doi.org/10.1103/PhysRevB.75.012414} {\bibfield  {journal}
  {\bibinfo  {journal} {Phys. Rev. B}\ }\textbf {\bibinfo {volume} {75}},\
  \bibinfo {pages} {012414} (\bibinfo {year} {2007})}\BibitemShut {NoStop}%
\bibitem [{\citenamefont {Ryee}\ \emph {et~al.}(2020)\citenamefont {Ryee},
  \citenamefont {Yoon}, \citenamefont {Kim}, \citenamefont {Jeong},\ and\
  \citenamefont {Han}}]{PhysRevB.101.064513}%
  \BibitemOpen
  \bibfield  {author} {\bibinfo {author} {\bibfnamefont {S.}~\bibnamefont
  {Ryee}}, \bibinfo {author} {\bibfnamefont {H.}~\bibnamefont {Yoon}}, \bibinfo
  {author} {\bibfnamefont {T.~J.}\ \bibnamefont {Kim}}, \bibinfo {author}
  {\bibfnamefont {M.~Y.}\ \bibnamefont {Jeong}},\ and\ \bibinfo {author}
  {\bibfnamefont {M.~J.}\ \bibnamefont {Han}},\ }\bibfield  {title} {\bibinfo
  {title} {Induced magnetic two-dimensionality by hole doping in the
  superconducting infinite-layer nickelate
  {${\mathrm{Nd}}_{1-x}{\mathrm{Sr}}_{x}{\mathrm{NiO}}_{2}$}},\ }\href
  {https://doi.org/10.1103/PhysRevB.101.064513} {\bibfield  {journal} {\bibinfo
   {journal} {Phys. Rev. B}\ }\textbf {\bibinfo {volume} {101}},\ \bibinfo
  {pages} {064513} (\bibinfo {year} {2020})}\BibitemShut {NoStop}%
\bibitem [{\citenamefont {Qin}\ and\ \citenamefont
  {Yang}(2023)}]{PhysRevB.108.L140504}%
  \BibitemOpen
  \bibfield  {author} {\bibinfo {author} {\bibfnamefont {Q.}~\bibnamefont
  {Qin}}\ and\ \bibinfo {author} {\bibfnamefont {Y.-f.}\ \bibnamefont {Yang}},\
  }\bibfield  {title} {\bibinfo {title} {High-${T}_{c}$ superconductivity by
  mobilizing local spin singlets and possible route to higher ${T}_{c}$ in
  pressurized {${\mathrm{La}}_{3}{\mathrm{Ni}}_{2}{\mathrm{O}}_{7}$}},\ }\href
  {https://doi.org/10.1103/PhysRevB.108.L140504} {\bibfield  {journal}
  {\bibinfo  {journal} {Phys. Rev. B}\ }\textbf {\bibinfo {volume} {108}},\
  \bibinfo {pages} {L140504} (\bibinfo {year} {2023})}\BibitemShut {NoStop}%
\bibitem [{\citenamefont {W\'u}\ \emph {et~al.}(2024)\citenamefont {W\'u},
  \citenamefont {Luo}, \citenamefont {Yao},\ and\ \citenamefont
  {Wang}}]{wu2023charge}%
  \BibitemOpen
  \bibfield  {author} {\bibinfo {author} {\bibfnamefont {W.}~\bibnamefont
  {W\'u}}, \bibinfo {author} {\bibfnamefont {Z.}~\bibnamefont {Luo}}, \bibinfo
  {author} {\bibfnamefont {D.-X.}\ \bibnamefont {Yao}},\ and\ \bibinfo {author}
  {\bibfnamefont {M.}~\bibnamefont {Wang}},\ }\bibfield  {title} {\bibinfo
  {title} {Superexchange and charge transfer in the nickelate superconductor
  $\mathrm{La_3Ni_2O_7}$ under pressure},\ }\href
  {https://doi.org/10.1007/s11433-023-2300-4} {\bibfield  {journal} {\bibinfo
  {journal} {Science China Physics, Mechanics \& Astronomy}\ }\textbf {\bibinfo
  {volume} {67}},\ \bibinfo {pages} {117402} (\bibinfo {year}
  {2024})}\BibitemShut {NoStop}%
\bibitem [{\citenamefont {Botzel}\ \emph {et~al.}(2024)\citenamefont {Botzel},
  \citenamefont {Lechermann}, \citenamefont {Gondolf},\ and\ \citenamefont
  {Eremin}}]{botzel2024theory}%
  \BibitemOpen
  \bibfield  {author} {\bibinfo {author} {\bibfnamefont {S.}~\bibnamefont
  {Botzel}}, \bibinfo {author} {\bibfnamefont {F.}~\bibnamefont {Lechermann}},
  \bibinfo {author} {\bibfnamefont {J.}~\bibnamefont {Gondolf}},\ and\ \bibinfo
  {author} {\bibfnamefont {I.~M.}\ \bibnamefont {Eremin}},\ }\href@noop {}
  {\bibinfo {title} {Theory of magnetic excitations in multilayer nickelate
  superconductor {La$_{3}$Ni$_{2}$O$_{7}$}}} (\bibinfo {year} {2024}),\ \Eprint
  {https://arxiv.org/abs/2401.16151} {arXiv:2401.16151 [cond-mat.supr-con]}
  \BibitemShut {NoStop}%
\bibitem [{\citenamefont {Nomura}\ \emph {et~al.}(2019)\citenamefont {Nomura},
  \citenamefont {Hirayama}, \citenamefont {Tadano}, \citenamefont {Yoshimoto},
  \citenamefont {Nakamura},\ and\ \citenamefont {Arita}}]{PhysRevB.100.205138}%
  \BibitemOpen
  \bibfield  {author} {\bibinfo {author} {\bibfnamefont {Y.}~\bibnamefont
  {Nomura}}, \bibinfo {author} {\bibfnamefont {M.}~\bibnamefont {Hirayama}},
  \bibinfo {author} {\bibfnamefont {T.}~\bibnamefont {Tadano}}, \bibinfo
  {author} {\bibfnamefont {Y.}~\bibnamefont {Yoshimoto}}, \bibinfo {author}
  {\bibfnamefont {K.}~\bibnamefont {Nakamura}},\ and\ \bibinfo {author}
  {\bibfnamefont {R.}~\bibnamefont {Arita}},\ }\bibfield  {title} {\bibinfo
  {title} {Formation of a two-dimensional single-component correlated electron
  system and band engineering in the nickelate superconductor
  {${\mathrm{NdNiO}}_{2}$}},\ }\href
  {https://doi.org/10.1103/PhysRevB.100.205138} {\bibfield  {journal} {\bibinfo
   {journal} {Phys. Rev. B}\ }\textbf {\bibinfo {volume} {100}},\ \bibinfo
  {pages} {205138} (\bibinfo {year} {2019})}\BibitemShut {NoStop}%
\bibitem [{\citenamefont {Botana}\ and\ \citenamefont
  {Norman}(2020)}]{PhysRevX.10.011024}%
  \BibitemOpen
  \bibfield  {author} {\bibinfo {author} {\bibfnamefont {A.~S.}\ \bibnamefont
  {Botana}}\ and\ \bibinfo {author} {\bibfnamefont {M.~R.}\ \bibnamefont
  {Norman}},\ }\bibfield  {title} {\bibinfo {title} {Similarities and
  differences between {${\mathrm{LaNiO}}_{2}$} and {${\mathrm{CaCuO}}_{2}$} and
  implications for superconductivity},\ }\href
  {https://doi.org/10.1103/PhysRevX.10.011024} {\bibfield  {journal} {\bibinfo
  {journal} {Phys. Rev. X}\ }\textbf {\bibinfo {volume} {10}},\ \bibinfo
  {pages} {011024} (\bibinfo {year} {2020})}\BibitemShut {NoStop}%
\bibitem [{\citenamefont {Jiang}\ \emph {et~al.}(2020)\citenamefont {Jiang},
  \citenamefont {Berciu},\ and\ \citenamefont
  {Sawatzky}}]{PhysRevLett.124.207004}%
  \BibitemOpen
  \bibfield  {author} {\bibinfo {author} {\bibfnamefont {M.}~\bibnamefont
  {Jiang}}, \bibinfo {author} {\bibfnamefont {M.}~\bibnamefont {Berciu}},\ and\
  \bibinfo {author} {\bibfnamefont {G.~A.}\ \bibnamefont {Sawatzky}},\
  }\bibfield  {title} {\bibinfo {title} {Critical nature of the ni spin state
  in doped {${\mathrm{NdNiO}}_{2}$}},\ }\href
  {https://doi.org/10.1103/PhysRevLett.124.207004} {\bibfield  {journal}
  {\bibinfo  {journal} {Phys. Rev. Lett.}\ }\textbf {\bibinfo {volume} {124}},\
  \bibinfo {pages} {207004} (\bibinfo {year} {2020})}\BibitemShut {NoStop}%
\bibitem [{\citenamefont {Zhang}\ \emph
  {et~al.}(2020{\natexlab{a}})\citenamefont {Zhang}, \citenamefont {Yang},\
  and\ \citenamefont {Zhang}}]{PhysRevB.101.020501}%
  \BibitemOpen
  \bibfield  {author} {\bibinfo {author} {\bibfnamefont {G.-M.}\ \bibnamefont
  {Zhang}}, \bibinfo {author} {\bibfnamefont {Y.-f.}\ \bibnamefont {Yang}},\
  and\ \bibinfo {author} {\bibfnamefont {F.-C.}\ \bibnamefont {Zhang}},\
  }\bibfield  {title} {\bibinfo {title} {Self-doped mott insulator for parent
  compounds of nickelate superconductors},\ }\href
  {https://doi.org/10.1103/PhysRevB.101.020501} {\bibfield  {journal} {\bibinfo
   {journal} {Phys. Rev. B}\ }\textbf {\bibinfo {volume} {101}},\ \bibinfo
  {pages} {020501} (\bibinfo {year} {2020}{\natexlab{a}})}\BibitemShut
  {NoStop}%
\bibitem [{\citenamefont {Zhang}\ and\ \citenamefont
  {Vishwanath}(2020)}]{PhysRevResearch.2.023112}%
  \BibitemOpen
  \bibfield  {author} {\bibinfo {author} {\bibfnamefont {Y.-H.}\ \bibnamefont
  {Zhang}}\ and\ \bibinfo {author} {\bibfnamefont {A.}~\bibnamefont
  {Vishwanath}},\ }\bibfield  {title} {\bibinfo {title} {Type-{II} $t{\rm
  -}{J}$ model in superconducting nickelate
  {${\mathrm{Nd}}_{1\ensuremath{-}x}{\mathrm{Sr}}_{x}{\mathrm{NiO}}_{2}$}},\
  }\href {https://doi.org/10.1103/PhysRevResearch.2.023112} {\bibfield
  {journal} {\bibinfo  {journal} {Phys. Rev. Res.}\ }\textbf {\bibinfo {volume}
  {2}},\ \bibinfo {pages} {023112} (\bibinfo {year} {2020})}\BibitemShut
  {NoStop}%
\bibitem [{\citenamefont {Werner}\ and\ \citenamefont
  {Hoshino}(2020)}]{PhysRevB.101.041104}%
  \BibitemOpen
  \bibfield  {author} {\bibinfo {author} {\bibfnamefont {P.}~\bibnamefont
  {Werner}}\ and\ \bibinfo {author} {\bibfnamefont {S.}~\bibnamefont
  {Hoshino}},\ }\bibfield  {title} {\bibinfo {title} {Nickelate
  superconductors: Multiorbital nature and spin freezing},\ }\href
  {https://doi.org/10.1103/PhysRevB.101.041104} {\bibfield  {journal} {\bibinfo
   {journal} {Phys. Rev. B}\ }\textbf {\bibinfo {volume} {101}},\ \bibinfo
  {pages} {041104} (\bibinfo {year} {2020})}\BibitemShut {NoStop}%
\bibitem [{\citenamefont {Bernardini}\ \emph {et~al.}(2020)\citenamefont
  {Bernardini}, \citenamefont {Olevano},\ and\ \citenamefont
  {Cano}}]{PhysRevResearch.2.013219}%
  \BibitemOpen
  \bibfield  {author} {\bibinfo {author} {\bibfnamefont {F.}~\bibnamefont
  {Bernardini}}, \bibinfo {author} {\bibfnamefont {V.}~\bibnamefont
  {Olevano}},\ and\ \bibinfo {author} {\bibfnamefont {A.}~\bibnamefont
  {Cano}},\ }\bibfield  {title} {\bibinfo {title} {Magnetic penetration depth
  and ${T}_{c}$ in superconducting nickelates},\ }\href
  {https://doi.org/10.1103/PhysRevResearch.2.013219} {\bibfield  {journal}
  {\bibinfo  {journal} {Phys. Rev. Res.}\ }\textbf {\bibinfo {volume} {2}},\
  \bibinfo {pages} {013219} (\bibinfo {year} {2020})}\BibitemShut {NoStop}%
\bibitem [{\citenamefont {Hepting}\ \emph {et~al.}(2020)\citenamefont
  {Hepting}, \citenamefont {Li}, \citenamefont {Jia}, \citenamefont {Lu},
  \citenamefont {Paris}, \citenamefont {Tseng}, \citenamefont {Feng},
  \citenamefont {Osada}, \citenamefont {Been}, \citenamefont {Hikita},
  \citenamefont {Chuang}, \citenamefont {Hussain}, \citenamefont {Zhou},
  \citenamefont {Nag}, \citenamefont {Garcia-Fernandez}, \citenamefont {Rossi},
  \citenamefont {Huang}, \citenamefont {Huang}, \citenamefont {Shen},
  \citenamefont {Schmitt}, \citenamefont {Hwang}, \citenamefont {Moritz},
  \citenamefont {Zaanen}, \citenamefont {Devereaux},\ and\ \citenamefont
  {Lee}}]{RN16}%
  \BibitemOpen
  \bibfield  {author} {\bibinfo {author} {\bibfnamefont {M.}~\bibnamefont
  {Hepting}}, \bibinfo {author} {\bibfnamefont {D.}~\bibnamefont {Li}},
  \bibinfo {author} {\bibfnamefont {C.~J.}\ \bibnamefont {Jia}}, \bibinfo
  {author} {\bibfnamefont {H.}~\bibnamefont {Lu}}, \bibinfo {author}
  {\bibfnamefont {E.}~\bibnamefont {Paris}}, \bibinfo {author} {\bibfnamefont
  {Y.}~\bibnamefont {Tseng}}, \bibinfo {author} {\bibfnamefont
  {X.}~\bibnamefont {Feng}}, \bibinfo {author} {\bibfnamefont {M.}~\bibnamefont
  {Osada}}, \bibinfo {author} {\bibfnamefont {E.}~\bibnamefont {Been}},
  \bibinfo {author} {\bibfnamefont {Y.}~\bibnamefont {Hikita}}, \bibinfo
  {author} {\bibfnamefont {Y.~D.}\ \bibnamefont {Chuang}}, \bibinfo {author}
  {\bibfnamefont {Z.}~\bibnamefont {Hussain}}, \bibinfo {author} {\bibfnamefont
  {K.~J.}\ \bibnamefont {Zhou}}, \bibinfo {author} {\bibfnamefont
  {A.}~\bibnamefont {Nag}}, \bibinfo {author} {\bibfnamefont {M.}~\bibnamefont
  {Garcia-Fernandez}}, \bibinfo {author} {\bibfnamefont {M.}~\bibnamefont
  {Rossi}}, \bibinfo {author} {\bibfnamefont {H.~Y.}\ \bibnamefont {Huang}},
  \bibinfo {author} {\bibfnamefont {D.~J.}\ \bibnamefont {Huang}}, \bibinfo
  {author} {\bibfnamefont {Z.~X.}\ \bibnamefont {Shen}}, \bibinfo {author}
  {\bibfnamefont {T.}~\bibnamefont {Schmitt}}, \bibinfo {author} {\bibfnamefont
  {H.~Y.}\ \bibnamefont {Hwang}}, \bibinfo {author} {\bibfnamefont
  {B.}~\bibnamefont {Moritz}}, \bibinfo {author} {\bibfnamefont
  {J.}~\bibnamefont {Zaanen}}, \bibinfo {author} {\bibfnamefont {T.~P.}\
  \bibnamefont {Devereaux}},\ and\ \bibinfo {author} {\bibfnamefont {W.~S.}\
  \bibnamefont {Lee}},\ }\bibfield  {title} {\bibinfo {title} {Electronic
  structure of the parent compound of superconducting infinite-layer
  nickelates},\ }\href {https://doi.org/10.1038/s41563-019-0585-z} {\bibfield
  {journal} {\bibinfo  {journal} {Nature Materials}\ }\textbf {\bibinfo
  {volume} {19}},\ \bibinfo {pages} {381} (\bibinfo {year} {2020})}\BibitemShut
  {NoStop}%
\bibitem [{\citenamefont {Gu}\ \emph {et~al.}(2020)\citenamefont {Gu},
  \citenamefont {Zhu}, \citenamefont {Wang}, \citenamefont {Hu},\ and\
  \citenamefont {Chen}}]{RN17}%
  \BibitemOpen
  \bibfield  {author} {\bibinfo {author} {\bibfnamefont {Y.}~\bibnamefont
  {Gu}}, \bibinfo {author} {\bibfnamefont {S.}~\bibnamefont {Zhu}}, \bibinfo
  {author} {\bibfnamefont {X.}~\bibnamefont {Wang}}, \bibinfo {author}
  {\bibfnamefont {J.}~\bibnamefont {Hu}},\ and\ \bibinfo {author}
  {\bibfnamefont {H.}~\bibnamefont {Chen}},\ }\bibfield  {title} {\bibinfo
  {title} {A substantial hybridization between correlated {Ni-d} orbital and
  itinerant electrons in infinite-layer nickelates},\ }\href
  {https://doi.org/10.1038/s42005-020-0347-x} {\bibfield  {journal} {\bibinfo
  {journal} {Communications Physics}\ }\textbf {\bibinfo {volume} {3}},\
  \bibinfo {pages} {84} (\bibinfo {year} {2020})}\BibitemShut {NoStop}%
\bibitem [{\citenamefont {Chang}\ \emph {et~al.}(2020)\citenamefont {Chang},
  \citenamefont {Zhao},\ and\ \citenamefont {Ding}}]{RN18}%
  \BibitemOpen
  \bibfield  {author} {\bibinfo {author} {\bibfnamefont {J.}~\bibnamefont
  {Chang}}, \bibinfo {author} {\bibfnamefont {J.}~\bibnamefont {Zhao}},\ and\
  \bibinfo {author} {\bibfnamefont {Y.}~\bibnamefont {Ding}},\ }\bibfield
  {title} {\bibinfo {title} {Hund-{H}eisenberg model in superconducting
  infinite-layer nickelates},\ }\href
  {https://doi.org/10.1140/epjb/e2020-10343-7} {\bibfield  {journal} {\bibinfo
  {journal} {The European Physical Journal B}\ }\textbf {\bibinfo {volume}
  {93}},\ \bibinfo {pages} {220} (\bibinfo {year} {2020})}\BibitemShut
  {NoStop}%
\bibitem [{\citenamefont {Kitatani}\ \emph {et~al.}(2020)\citenamefont
  {Kitatani}, \citenamefont {Si}, \citenamefont {Janson}, \citenamefont
  {Arita}, \citenamefont {Zhong},\ and\ \citenamefont {Held}}]{RN19}%
  \BibitemOpen
  \bibfield  {author} {\bibinfo {author} {\bibfnamefont {M.}~\bibnamefont
  {Kitatani}}, \bibinfo {author} {\bibfnamefont {L.}~\bibnamefont {Si}},
  \bibinfo {author} {\bibfnamefont {O.}~\bibnamefont {Janson}}, \bibinfo
  {author} {\bibfnamefont {R.}~\bibnamefont {Arita}}, \bibinfo {author}
  {\bibfnamefont {Z.}~\bibnamefont {Zhong}},\ and\ \bibinfo {author}
  {\bibfnamefont {K.}~\bibnamefont {Held}},\ }\bibfield  {title} {\bibinfo
  {title} {Nickelate superconductors-a renaissance of the one-band hubbard
  model},\ }\href {https://doi.org/10.1038/s41535-020-00260-y} {\bibfield
  {journal} {\bibinfo  {journal} {npj Quantum Materials}\ }\textbf {\bibinfo
  {volume} {5}},\ \bibinfo {pages} {59} (\bibinfo {year} {2020})}\BibitemShut
  {NoStop}%
\bibitem [{\citenamefont {Krieger}\ \emph {et~al.}(2022)\citenamefont
  {Krieger}, \citenamefont {Martinelli}, \citenamefont {Zeng}, \citenamefont
  {Chow}, \citenamefont {Kummer}, \citenamefont {Arpaia}, \citenamefont
  {Moretti~Sala}, \citenamefont {Brookes}, \citenamefont {Ariando},
  \citenamefont {Viart}, \citenamefont {Salluzzo}, \citenamefont
  {Ghiringhelli},\ and\ \citenamefont {Preziosi}}]{PhysRevLett.129.027002}%
  \BibitemOpen
  \bibfield  {author} {\bibinfo {author} {\bibfnamefont {G.}~\bibnamefont
  {Krieger}}, \bibinfo {author} {\bibfnamefont {L.}~\bibnamefont {Martinelli}},
  \bibinfo {author} {\bibfnamefont {S.}~\bibnamefont {Zeng}}, \bibinfo {author}
  {\bibfnamefont {L.~E.}\ \bibnamefont {Chow}}, \bibinfo {author}
  {\bibfnamefont {K.}~\bibnamefont {Kummer}}, \bibinfo {author} {\bibfnamefont
  {R.}~\bibnamefont {Arpaia}}, \bibinfo {author} {\bibfnamefont
  {M.}~\bibnamefont {Moretti~Sala}}, \bibinfo {author} {\bibfnamefont {N.~B.}\
  \bibnamefont {Brookes}}, \bibinfo {author} {\bibfnamefont {A.}~\bibnamefont
  {Ariando}}, \bibinfo {author} {\bibfnamefont {N.}~\bibnamefont {Viart}},
  \bibinfo {author} {\bibfnamefont {M.}~\bibnamefont {Salluzzo}}, \bibinfo
  {author} {\bibfnamefont {G.}~\bibnamefont {Ghiringhelli}},\ and\ \bibinfo
  {author} {\bibfnamefont {D.}~\bibnamefont {Preziosi}},\ }\bibfield  {title}
  {\bibinfo {title} {Charge and spin order dichotomy in
  {${\mathrm{NdNiO}}_{2}$} driven by the capping layer},\ }\href
  {https://doi.org/10.1103/PhysRevLett.129.027002} {\bibfield  {journal}
  {\bibinfo  {journal} {Phys. Rev. Lett.}\ }\textbf {\bibinfo {volume} {129}},\
  \bibinfo {pages} {027002} (\bibinfo {year} {2022})}\BibitemShut {NoStop}%
\bibitem [{\citenamefont {Wu}\ \emph {et~al.}(2020)\citenamefont {Wu},
  \citenamefont {Di~Sante}, \citenamefont {Schwemmer}, \citenamefont {Hanke},
  \citenamefont {Hwang}, \citenamefont {Raghu},\ and\ \citenamefont
  {Thomale}}]{PhysRevB.101.060504}%
  \BibitemOpen
  \bibfield  {author} {\bibinfo {author} {\bibfnamefont {X.}~\bibnamefont
  {Wu}}, \bibinfo {author} {\bibfnamefont {D.}~\bibnamefont {Di~Sante}},
  \bibinfo {author} {\bibfnamefont {T.}~\bibnamefont {Schwemmer}}, \bibinfo
  {author} {\bibfnamefont {W.}~\bibnamefont {Hanke}}, \bibinfo {author}
  {\bibfnamefont {H.~Y.}\ \bibnamefont {Hwang}}, \bibinfo {author}
  {\bibfnamefont {S.}~\bibnamefont {Raghu}},\ and\ \bibinfo {author}
  {\bibfnamefont {R.}~\bibnamefont {Thomale}},\ }\bibfield  {title} {\bibinfo
  {title} {Robust ${d}_{{x}^{2}\ensuremath{-}{y}^{2}}$-wave superconductivity
  of infinite-layer nickelates},\ }\href
  {https://doi.org/10.1103/PhysRevB.101.060504} {\bibfield  {journal} {\bibinfo
   {journal} {Phys. Rev. B}\ }\textbf {\bibinfo {volume} {101}},\ \bibinfo
  {pages} {060504} (\bibinfo {year} {2020})}\BibitemShut {NoStop}%
\bibitem [{\citenamefont {Luo}\ \emph {et~al.}(2023{\natexlab{c}})\citenamefont
  {Luo}, \citenamefont {Yao},\ and\ \citenamefont
  {W\'u}}]{luo2023superconductivity}%
  \BibitemOpen
  \bibfield  {author} {\bibinfo {author} {\bibfnamefont {Z.}~\bibnamefont
  {Luo}}, \bibinfo {author} {\bibfnamefont {D.-X.}\ \bibnamefont {Yao}},\ and\
  \bibinfo {author} {\bibfnamefont {W.}~\bibnamefont {W\'u}},\ }\href@noop {}
  {\bibinfo {title} {Superconductivity in the two-orbital hubbard model of
  infinite-layer nickelates}} (\bibinfo {year} {2023}{\natexlab{c}}),\ \Eprint
  {https://arxiv.org/abs/2310.12250} {arXiv:2310.12250 [cond-mat.supr-con]}
  \BibitemShut {NoStop}%
\bibitem [{\citenamefont {Fan}\ \emph {et~al.}(2023)\citenamefont {Fan},
  \citenamefont {Zhang}, \citenamefont {Zhan}, \citenamefont {Lv},
  \citenamefont {Jiang}, \citenamefont {Normand},\ and\ \citenamefont
  {Xiang}}]{fan2023superconductivity}%
  \BibitemOpen
  \bibfield  {author} {\bibinfo {author} {\bibfnamefont {Z.}~\bibnamefont
  {Fan}}, \bibinfo {author} {\bibfnamefont {J.-F.}\ \bibnamefont {Zhang}},
  \bibinfo {author} {\bibfnamefont {B.}~\bibnamefont {Zhan}}, \bibinfo {author}
  {\bibfnamefont {D.}~\bibnamefont {Lv}}, \bibinfo {author} {\bibfnamefont
  {X.-Y.}\ \bibnamefont {Jiang}}, \bibinfo {author} {\bibfnamefont
  {B.}~\bibnamefont {Normand}},\ and\ \bibinfo {author} {\bibfnamefont
  {T.}~\bibnamefont {Xiang}},\ }\href@noop {} {\bibinfo {title}
  {Superconductivity in nickelate and cuprate superconductors with strong
  bilayer coupling}} (\bibinfo {year} {2023}),\ \Eprint
  {https://arxiv.org/abs/2312.17064} {arXiv:2312.17064 [cond-mat.supr-con]}
  \BibitemShut {NoStop}%
\bibitem [{\citenamefont {Shilenko}\ and\ \citenamefont
  {Leonov}(2023)}]{PhysRevB.108.125105}%
  \BibitemOpen
  \bibfield  {author} {\bibinfo {author} {\bibfnamefont {D.~A.}\ \bibnamefont
  {Shilenko}}\ and\ \bibinfo {author} {\bibfnamefont {I.~V.}\ \bibnamefont
  {Leonov}},\ }\bibfield  {title} {\bibinfo {title} {Correlated electronic
  structure, orbital-selective behavior, and magnetic correlations in
  double-layer {${\mathrm{La}}_{3}{\mathrm{Ni}}_{2}{\mathrm{O}}_{7}$} under
  pressure},\ }\href {https://doi.org/10.1103/PhysRevB.108.125105} {\bibfield
  {journal} {\bibinfo  {journal} {Phys. Rev. B}\ }\textbf {\bibinfo {volume}
  {108}},\ \bibinfo {pages} {125105} (\bibinfo {year} {2023})}\BibitemShut
  {NoStop}%
\bibitem [{\citenamefont {Yang}\ \emph
  {et~al.}(2023{\natexlab{a}})\citenamefont {Yang}, \citenamefont {Wang},\ and\
  \citenamefont {Wang}}]{PhysRevB.108.L140505}%
  \BibitemOpen
  \bibfield  {author} {\bibinfo {author} {\bibfnamefont {Q.-G.}\ \bibnamefont
  {Yang}}, \bibinfo {author} {\bibfnamefont {D.}~\bibnamefont {Wang}},\ and\
  \bibinfo {author} {\bibfnamefont {Q.-H.}\ \bibnamefont {Wang}},\ }\bibfield
  {title} {\bibinfo {title} {Possible ${s}_{\ifmmode\pm\else\textpm\fi{}}$-wave
  superconductivity in
  {${\mathrm{La}}_{3}{\mathrm{Ni}}_{2}{\mathrm{O}}_{7}$}},\ }\href
  {https://doi.org/10.1103/PhysRevB.108.L140505} {\bibfield  {journal}
  {\bibinfo  {journal} {Phys. Rev. B}\ }\textbf {\bibinfo {volume} {108}},\
  \bibinfo {pages} {L140505} (\bibinfo {year}
  {2023}{\natexlab{a}})}\BibitemShut {NoStop}%
\bibitem [{\citenamefont {Gu}\ \emph {et~al.}(2023)\citenamefont {Gu},
  \citenamefont {Le}, \citenamefont {Yang}, \citenamefont {Wu},\ and\
  \citenamefont {Hu}}]{gu2023effective}%
  \BibitemOpen
  \bibfield  {author} {\bibinfo {author} {\bibfnamefont {Y.}~\bibnamefont
  {Gu}}, \bibinfo {author} {\bibfnamefont {C.}~\bibnamefont {Le}}, \bibinfo
  {author} {\bibfnamefont {Z.}~\bibnamefont {Yang}}, \bibinfo {author}
  {\bibfnamefont {X.}~\bibnamefont {Wu}},\ and\ \bibinfo {author}
  {\bibfnamefont {J.}~\bibnamefont {Hu}},\ }\href@noop {} {\bibinfo {title}
  {Effective model and pairing tendency in bilayer ni-based superconductor
  {La$_3$Ni$_2$O$_7$}}} (\bibinfo {year} {2023}),\ \Eprint
  {https://arxiv.org/abs/2306.07275} {arXiv:2306.07275 [cond-mat.supr-con]}
  \BibitemShut {NoStop}%
\bibitem [{\citenamefont {Yang}\ \emph
  {et~al.}(2023{\natexlab{b}})\citenamefont {Yang}, \citenamefont {Sun},
  \citenamefont {Hu}, \citenamefont {Xie}, \citenamefont {Miao}, \citenamefont
  {Luo}, \citenamefont {Chen}, \citenamefont {Liang}, \citenamefont {Zhu},
  \citenamefont {Qu}, \citenamefont {Chen}, \citenamefont {Huo}, \citenamefont
  {Huang}, \citenamefont {Zhang}, \citenamefont {Zhang}, \citenamefont {Yang},
  \citenamefont {Wang}, \citenamefont {Peng}, \citenamefont {Mao},
  \citenamefont {Liu}, \citenamefont {Xu}, \citenamefont {Qian}, \citenamefont
  {Yao}, \citenamefont {Wang}, \citenamefont {Zhao},\ and\ \citenamefont
  {Zhou}}]{yang2023orbitaldependent}%
  \BibitemOpen
  \bibfield  {author} {\bibinfo {author} {\bibfnamefont {J.}~\bibnamefont
  {Yang}}, \bibinfo {author} {\bibfnamefont {H.}~\bibnamefont {Sun}}, \bibinfo
  {author} {\bibfnamefont {X.}~\bibnamefont {Hu}}, \bibinfo {author}
  {\bibfnamefont {Y.}~\bibnamefont {Xie}}, \bibinfo {author} {\bibfnamefont
  {T.}~\bibnamefont {Miao}}, \bibinfo {author} {\bibfnamefont {H.}~\bibnamefont
  {Luo}}, \bibinfo {author} {\bibfnamefont {H.}~\bibnamefont {Chen}}, \bibinfo
  {author} {\bibfnamefont {B.}~\bibnamefont {Liang}}, \bibinfo {author}
  {\bibfnamefont {W.}~\bibnamefont {Zhu}}, \bibinfo {author} {\bibfnamefont
  {G.}~\bibnamefont {Qu}}, \bibinfo {author} {\bibfnamefont {C.-Q.}\
  \bibnamefont {Chen}}, \bibinfo {author} {\bibfnamefont {M.}~\bibnamefont
  {Huo}}, \bibinfo {author} {\bibfnamefont {Y.}~\bibnamefont {Huang}}, \bibinfo
  {author} {\bibfnamefont {S.}~\bibnamefont {Zhang}}, \bibinfo {author}
  {\bibfnamefont {F.}~\bibnamefont {Zhang}}, \bibinfo {author} {\bibfnamefont
  {F.}~\bibnamefont {Yang}}, \bibinfo {author} {\bibfnamefont {Z.}~\bibnamefont
  {Wang}}, \bibinfo {author} {\bibfnamefont {Q.}~\bibnamefont {Peng}}, \bibinfo
  {author} {\bibfnamefont {H.}~\bibnamefont {Mao}}, \bibinfo {author}
  {\bibfnamefont {G.}~\bibnamefont {Liu}}, \bibinfo {author} {\bibfnamefont
  {Z.}~\bibnamefont {Xu}}, \bibinfo {author} {\bibfnamefont {T.}~\bibnamefont
  {Qian}}, \bibinfo {author} {\bibfnamefont {D.-X.}\ \bibnamefont {Yao}},
  \bibinfo {author} {\bibfnamefont {M.}~\bibnamefont {Wang}}, \bibinfo {author}
  {\bibfnamefont {L.}~\bibnamefont {Zhao}},\ and\ \bibinfo {author}
  {\bibfnamefont {X.~J.}\ \bibnamefont {Zhou}},\ }\href@noop {} {\bibinfo
  {title} {Orbital-dependent electron correlation in double-layer nickelate
  {La$_3$Ni$_2$O$_7$}}} (\bibinfo {year} {2023}{\natexlab{b}}),\ \Eprint
  {https://arxiv.org/abs/2309.01148} {arXiv:2309.01148 [cond-mat.supr-con]}
  \BibitemShut {NoStop}%
\bibitem [{\citenamefont {Zhang}\ \emph
  {et~al.}(2024{\natexlab{b}})\citenamefont {Zhang}, \citenamefont {Pei},
  \citenamefont {Wang}, \citenamefont {Zhao}, \citenamefont {Li}, \citenamefont
  {Cao}, \citenamefont {Zhu}, \citenamefont {Wu},\ and\ \citenamefont
  {Qi}}]{ZHANG2024147}%
  \BibitemOpen
  \bibfield  {author} {\bibinfo {author} {\bibfnamefont {M.}~\bibnamefont
  {Zhang}}, \bibinfo {author} {\bibfnamefont {C.}~\bibnamefont {Pei}}, \bibinfo
  {author} {\bibfnamefont {Q.}~\bibnamefont {Wang}}, \bibinfo {author}
  {\bibfnamefont {Y.}~\bibnamefont {Zhao}}, \bibinfo {author} {\bibfnamefont
  {C.}~\bibnamefont {Li}}, \bibinfo {author} {\bibfnamefont {W.}~\bibnamefont
  {Cao}}, \bibinfo {author} {\bibfnamefont {S.}~\bibnamefont {Zhu}}, \bibinfo
  {author} {\bibfnamefont {J.}~\bibnamefont {Wu}},\ and\ \bibinfo {author}
  {\bibfnamefont {Y.}~\bibnamefont {Qi}},\ }\bibfield  {title} {\bibinfo
  {title} {Effects of pressure and doping on ruddlesden-popper phases
  $\mathrm{La_{n+1}Ni_nO_{3n+1}}$},\ }\href
  {https://doi.org/https://doi.org/10.1016/j.jmst.2023.11.011} {\bibfield
  {journal} {\bibinfo  {journal} {Journal of Materials Science \& Technology}\
  }\textbf {\bibinfo {volume} {185}},\ \bibinfo {pages} {147} (\bibinfo {year}
  {2024}{\natexlab{b}})}\BibitemShut {NoStop}%
\bibitem [{\citenamefont {Yuan}\ \emph {et~al.}(2024)\citenamefont {Yuan},
  \citenamefont {Elghandour}, \citenamefont {Arneth}, \citenamefont {Dey},\
  and\ \citenamefont {Klingeler}}]{YUAN2024127511}%
  \BibitemOpen
  \bibfield  {author} {\bibinfo {author} {\bibfnamefont {N.}~\bibnamefont
  {Yuan}}, \bibinfo {author} {\bibfnamefont {A.}~\bibnamefont {Elghandour}},
  \bibinfo {author} {\bibfnamefont {J.}~\bibnamefont {Arneth}}, \bibinfo
  {author} {\bibfnamefont {K.}~\bibnamefont {Dey}},\ and\ \bibinfo {author}
  {\bibfnamefont {R.}~\bibnamefont {Klingeler}},\ }\bibfield  {title} {\bibinfo
  {title} {High-pressure crystal growth and investigation of the metal-to-metal
  transition of ruddlesden-popper trilayer nickelates
  {La}$_4${Ni}$_3${O}$_{10}$},\ }\href
  {https://doi.org/10.1016/j.jcrysgro.2023.127511} {\bibfield  {journal}
  {\bibinfo  {journal} {Journal of Crystal Growth}\ }\textbf {\bibinfo {volume}
  {627}},\ \bibinfo {pages} {127511} (\bibinfo {year} {2024})}\BibitemShut
  {NoStop}%
\bibitem [{\citenamefont {Li}\ \emph {et~al.}(2020)\citenamefont {Li},
  \citenamefont {He}, \citenamefont {Si}, \citenamefont {Zhu}, \citenamefont
  {Zhang},\ and\ \citenamefont {Wen}}]{RN23}%
  \BibitemOpen
  \bibfield  {author} {\bibinfo {author} {\bibfnamefont {Q.}~\bibnamefont
  {Li}}, \bibinfo {author} {\bibfnamefont {C.}~\bibnamefont {He}}, \bibinfo
  {author} {\bibfnamefont {J.}~\bibnamefont {Si}}, \bibinfo {author}
  {\bibfnamefont {X.}~\bibnamefont {Zhu}}, \bibinfo {author} {\bibfnamefont
  {Y.}~\bibnamefont {Zhang}},\ and\ \bibinfo {author} {\bibfnamefont {H.-H.}\
  \bibnamefont {Wen}},\ }\bibfield  {title} {\bibinfo {title} {Absence of
  superconductivity in bulk {Nd$_{1-x}$Sr$_x$NiO$_2$}},\ }\href
  {https://doi.org/10.1038/s43246-020-0018-1} {\bibfield  {journal} {\bibinfo
  {journal} {Communications Materials}\ }\textbf {\bibinfo {volume} {1}},\
  \bibinfo {pages} {16} (\bibinfo {year} {2020})}\BibitemShut {NoStop}%
\bibitem [{\citenamefont {Jiang}\ \emph {et~al.}(2024)\citenamefont {Jiang},
  \citenamefont {Hou}, \citenamefont {Fan}, \citenamefont {Lang},\ and\
  \citenamefont {Ku}}]{jiang2023pressure}%
  \BibitemOpen
  \bibfield  {author} {\bibinfo {author} {\bibfnamefont {R.}~\bibnamefont
  {Jiang}}, \bibinfo {author} {\bibfnamefont {J.}~\bibnamefont {Hou}}, \bibinfo
  {author} {\bibfnamefont {Z.}~\bibnamefont {Fan}}, \bibinfo {author}
  {\bibfnamefont {Z.-J.}\ \bibnamefont {Lang}},\ and\ \bibinfo {author}
  {\bibfnamefont {W.}~\bibnamefont {Ku}},\ }\bibfield  {title} {\bibinfo
  {title} {Pressure driven fractionalization of ionic spins results in
  cupratelike high-${T}_{c}$ superconductivity in
  ${\mathrm{la}}_{3}{\mathrm{ni}}_{2}{\mathrm{o}}_{7}$},\ }\href
  {https://doi.org/10.1103/PhysRevLett.132.126503} {\bibfield  {journal}
  {\bibinfo  {journal} {Phys. Rev. Lett.}\ }\textbf {\bibinfo {volume} {132}},\
  \bibinfo {pages} {126503} (\bibinfo {year} {2024})}\BibitemShut {NoStop}%
\bibitem [{\citenamefont {Li}\ \emph {et~al.}(2024{\natexlab{b}})\citenamefont
  {Li}, \citenamefont {Chen}, \citenamefont {Huang}, \citenamefont {Han},
  \citenamefont {Huo}, \citenamefont {Huang}, \citenamefont {Ma}, \citenamefont
  {Qiu}, \citenamefont {Chen}, \citenamefont {Hu}, \citenamefont {Chen},
  \citenamefont {Xie}, \citenamefont {Shen}, \citenamefont {Sun}, \citenamefont
  {Yao},\ and\ \citenamefont {Wang}}]{experimental}%
  \BibitemOpen
  \bibfield  {author} {\bibinfo {author} {\bibfnamefont {J.}~\bibnamefont
  {Li}}, \bibinfo {author} {\bibfnamefont {C.-Q.}\ \bibnamefont {Chen}},
  \bibinfo {author} {\bibfnamefont {C.}~\bibnamefont {Huang}}, \bibinfo
  {author} {\bibfnamefont {Y.}~\bibnamefont {Han}}, \bibinfo {author}
  {\bibfnamefont {M.}~\bibnamefont {Huo}}, \bibinfo {author} {\bibfnamefont
  {X.}~\bibnamefont {Huang}}, \bibinfo {author} {\bibfnamefont
  {P.}~\bibnamefont {Ma}}, \bibinfo {author} {\bibfnamefont {Z.}~\bibnamefont
  {Qiu}}, \bibinfo {author} {\bibfnamefont {J.}~\bibnamefont {Chen}}, \bibinfo
  {author} {\bibfnamefont {X.}~\bibnamefont {Hu}}, \bibinfo {author}
  {\bibfnamefont {L.}~\bibnamefont {Chen}}, \bibinfo {author} {\bibfnamefont
  {T.}~\bibnamefont {Xie}}, \bibinfo {author} {\bibfnamefont {B.}~\bibnamefont
  {Shen}}, \bibinfo {author} {\bibfnamefont {H.}~\bibnamefont {Sun}}, \bibinfo
  {author} {\bibfnamefont {D.-X.}\ \bibnamefont {Yao}},\ and\ \bibinfo {author}
  {\bibfnamefont {M.}~\bibnamefont {Wang}},\ }\bibfield  {title} {\bibinfo
  {title} {Structural transition, electric transport, and electronic structures
  in the compressed trilayer nickelate $\mathrm{La_4Ni_3O_{10}}$},\ }\href
  {https://doi.org/10.1007/s11433-023-2329-x} {\bibfield  {journal} {\bibinfo
  {journal} {Science China Physics, Mechanics \& Astronomy}\ }\textbf {\bibinfo
  {volume} {67}},\ \bibinfo {pages} {117403} (\bibinfo {year}
  {2024}{\natexlab{b}})}\BibitemShut {NoStop}%
\bibitem [{\citenamefont {Wang}\ \emph {et~al.}(2023)\citenamefont {Wang},
  \citenamefont {Li}, \citenamefont {Xie}, \citenamefont {Liu}, \citenamefont
  {Sun}, \citenamefont {Huang}, \citenamefont {Gao}, \citenamefont {Nakagawa},
  \citenamefont {Fu}, \citenamefont {Dong}, \citenamefont {Cao}, \citenamefont
  {Yu}, \citenamefont {Kawaguchi}, \citenamefont {Kadobayashi}, \citenamefont
  {Wang}, \citenamefont {Jin}, \citenamefont {kwang Mao},\ and\ \citenamefont
  {Liu}}]{wang2023structure}%
  \BibitemOpen
  \bibfield  {author} {\bibinfo {author} {\bibfnamefont {L.}~\bibnamefont
  {Wang}}, \bibinfo {author} {\bibfnamefont {Y.}~\bibnamefont {Li}}, \bibinfo
  {author} {\bibfnamefont {S.}~\bibnamefont {Xie}}, \bibinfo {author}
  {\bibfnamefont {F.}~\bibnamefont {Liu}}, \bibinfo {author} {\bibfnamefont
  {H.}~\bibnamefont {Sun}}, \bibinfo {author} {\bibfnamefont {C.}~\bibnamefont
  {Huang}}, \bibinfo {author} {\bibfnamefont {Y.}~\bibnamefont {Gao}}, \bibinfo
  {author} {\bibfnamefont {T.}~\bibnamefont {Nakagawa}}, \bibinfo {author}
  {\bibfnamefont {B.}~\bibnamefont {Fu}}, \bibinfo {author} {\bibfnamefont
  {B.}~\bibnamefont {Dong}}, \bibinfo {author} {\bibfnamefont {Z.}~\bibnamefont
  {Cao}}, \bibinfo {author} {\bibfnamefont {R.}~\bibnamefont {Yu}}, \bibinfo
  {author} {\bibfnamefont {S.~I.}\ \bibnamefont {Kawaguchi}}, \bibinfo {author}
  {\bibfnamefont {H.}~\bibnamefont {Kadobayashi}}, \bibinfo {author}
  {\bibfnamefont {M.}~\bibnamefont {Wang}}, \bibinfo {author} {\bibfnamefont
  {C.}~\bibnamefont {Jin}}, \bibinfo {author} {\bibfnamefont {H.}~\bibnamefont
  {kwang Mao}},\ and\ \bibinfo {author} {\bibfnamefont {H.}~\bibnamefont
  {Liu}},\ }\href@noop {} {\bibinfo {title} {Structure responsible for the
  superconducting state in {La$_3$Ni$_2$O$_7$} at high pressure and low
  temperature conditions}} (\bibinfo {year} {2023}),\ \Eprint
  {https://arxiv.org/abs/2311.09186} {arXiv:2311.09186 [cond-mat.supr-con]}
  \BibitemShut {NoStop}%
\bibitem [{\citenamefont {Geisler}\ \emph {et~al.}(2023)\citenamefont
  {Geisler}, \citenamefont {Hamlin}, \citenamefont {Stewart}, \citenamefont
  {Hennig},\ and\ \citenamefont {Hirschfeld}}]{geisler2023structural}%
  \BibitemOpen
  \bibfield  {author} {\bibinfo {author} {\bibfnamefont {B.}~\bibnamefont
  {Geisler}}, \bibinfo {author} {\bibfnamefont {J.~J.}\ \bibnamefont {Hamlin}},
  \bibinfo {author} {\bibfnamefont {G.~R.}\ \bibnamefont {Stewart}}, \bibinfo
  {author} {\bibfnamefont {R.~G.}\ \bibnamefont {Hennig}},\ and\ \bibinfo
  {author} {\bibfnamefont {P.~J.}\ \bibnamefont {Hirschfeld}},\ }\href@noop {}
  {\bibinfo {title} {Structural transitions, octahedral rotations, and
  electronic properties of $\mathrm{A_3Ni_2O_7}$ rare-earth nickelates under
  high pressure}} (\bibinfo {year} {2023}),\ \Eprint
  {https://arxiv.org/abs/2309.15078} {arXiv:2309.15078 [cond-mat.supr-con]}
  \BibitemShut {NoStop}%
\bibitem [{\citenamefont {Jung}\ \emph {et~al.}(2022)\citenamefont {Jung},
  \citenamefont {Kapeghian}, \citenamefont {Hanson}, \citenamefont {Pamuk},\
  and\ \citenamefont {Botana}}]{PhysRevB.105.085150}%
  \BibitemOpen
  \bibfield  {author} {\bibinfo {author} {\bibfnamefont {M.-C.}\ \bibnamefont
  {Jung}}, \bibinfo {author} {\bibfnamefont {J.}~\bibnamefont {Kapeghian}},
  \bibinfo {author} {\bibfnamefont {C.}~\bibnamefont {Hanson}}, \bibinfo
  {author} {\bibfnamefont {B.}~\bibnamefont {Pamuk}},\ and\ \bibinfo {author}
  {\bibfnamefont {A.~S.}\ \bibnamefont {Botana}},\ }\bibfield  {title}
  {\bibinfo {title} {Electronic structure of higher-order ruddlesden-popper
  nickelates},\ }\href {https://doi.org/10.1103/PhysRevB.105.085150} {\bibfield
   {journal} {\bibinfo  {journal} {Phys. Rev. B}\ }\textbf {\bibinfo {volume}
  {105}},\ \bibinfo {pages} {085150} (\bibinfo {year} {2022})}\BibitemShut
  {NoStop}%
\bibitem [{\citenamefont {Kresse}\ and\ \citenamefont
  {Furthmuller}(1996{\natexlab{a}})}]{VASP1}%
  \BibitemOpen
  \bibfield  {author} {\bibinfo {author} {\bibfnamefont {G.}~\bibnamefont
  {Kresse}}\ and\ \bibinfo {author} {\bibfnamefont {J.}~\bibnamefont
  {Furthmuller}},\ }\bibfield  {title} {\bibinfo {title} {Efficiency of
  ab-initio total energy calculations for metals and semiconductors using a
  plane-wave basis set},\ }\href
  {https://doi.org/https://doi.org/10.1016/0927-0256(96)00008-0} {\bibfield
  {journal} {\bibinfo  {journal} {Computational Materials Science}\ }\textbf
  {\bibinfo {volume} {6}},\ \bibinfo {pages} {15} (\bibinfo {year}
  {1996}{\natexlab{a}})}\BibitemShut {NoStop}%
\bibitem [{\citenamefont {Kresse}\ and\ \citenamefont
  {Furthmuller}(1996{\natexlab{b}})}]{VASP2}%
  \BibitemOpen
  \bibfield  {author} {\bibinfo {author} {\bibfnamefont {G.}~\bibnamefont
  {Kresse}}\ and\ \bibinfo {author} {\bibfnamefont {J.}~\bibnamefont
  {Furthmuller}},\ }\bibfield  {title} {\bibinfo {title} {Efficient iterative
  schemes for ab initio total-energy calculations using a plane-wave basis
  set},\ }\href {https://doi.org/10.1103/PhysRevB.54.11169} {\bibfield
  {journal} {\bibinfo  {journal} {Physical Review B}\ }\textbf {\bibinfo
  {volume} {54}},\ \bibinfo {pages} {11169} (\bibinfo {year}
  {1996}{\natexlab{b}})}\BibitemShut {NoStop}%
\bibitem [{\citenamefont {Blochl}(1994)}]{PAW1}%
  \BibitemOpen
  \bibfield  {author} {\bibinfo {author} {\bibfnamefont {P.~E.}\ \bibnamefont
  {Blochl}},\ }\bibfield  {title} {\bibinfo {title} {Projector augmented-wave
  method},\ }\href {https://doi.org/10.1103/PhysRevB.50.17953} {\bibfield
  {journal} {\bibinfo  {journal} {Physical Review B}\ }\textbf {\bibinfo
  {volume} {50}},\ \bibinfo {pages} {17953} (\bibinfo {year}
  {1994})}\BibitemShut {NoStop}%
\bibitem [{\citenamefont {Kresse}\ and\ \citenamefont {Joubert}(1999)}]{PAW2}%
  \BibitemOpen
  \bibfield  {author} {\bibinfo {author} {\bibfnamefont {G.}~\bibnamefont
  {Kresse}}\ and\ \bibinfo {author} {\bibfnamefont {D.}~\bibnamefont
  {Joubert}},\ }\bibfield  {title} {\bibinfo {title} {From ultrasoft
  pseudopotentials to the projector augmented-wave method},\ }\href
  {https://doi.org/10.1103/PhysRevB.59.1758} {\bibfield  {journal} {\bibinfo
  {journal} {Physical Review B}\ }\textbf {\bibinfo {volume} {59}},\ \bibinfo
  {pages} {1758} (\bibinfo {year} {1999})}\BibitemShut {NoStop}%
\bibitem [{\citenamefont {Kohn}\ and\ \citenamefont {Sham}(1965)}]{LDA}%
  \BibitemOpen
  \bibfield  {author} {\bibinfo {author} {\bibfnamefont {W.}~\bibnamefont
  {Kohn}}\ and\ \bibinfo {author} {\bibfnamefont {L.~J.}\ \bibnamefont
  {Sham}},\ }\bibfield  {title} {\bibinfo {title} {Self-consistent equations
  including exchange and correlation effects},\ }\href
  {https://doi.org/10.1103/PhysRev.140.A1133} {\bibfield  {journal} {\bibinfo
  {journal} {Physical Review}\ }\textbf {\bibinfo {volume} {140}},\ \bibinfo
  {pages} {A1133} (\bibinfo {year} {1965})}\BibitemShut {NoStop}%
\bibitem [{\citenamefont {Mostofi}\ \emph {et~al.}(2008)\citenamefont
  {Mostofi}, \citenamefont {Yates}, \citenamefont {Lee}, \citenamefont {Souza},
  \citenamefont {Vanderbilt},\ and\ \citenamefont {Marzari}}]{w90}%
  \BibitemOpen
  \bibfield  {author} {\bibinfo {author} {\bibfnamefont {A.~A.}\ \bibnamefont
  {Mostofi}}, \bibinfo {author} {\bibfnamefont {J.~R.}\ \bibnamefont {Yates}},
  \bibinfo {author} {\bibfnamefont {Y.-S.}\ \bibnamefont {Lee}}, \bibinfo
  {author} {\bibfnamefont {I.}~\bibnamefont {Souza}}, \bibinfo {author}
  {\bibfnamefont {D.}~\bibnamefont {Vanderbilt}},\ and\ \bibinfo {author}
  {\bibfnamefont {N.}~\bibnamefont {Marzari}},\ }\bibfield  {title} {\bibinfo
  {title} {wannier90: A tool for obtaining maximally-localised wannier
  functions},\ }\href
  {https://doi.org/https://doi.org/10.1016/j.cpc.2007.11.016} {\bibfield
  {journal} {\bibinfo  {journal} {Computer Physics Communications}\ }\textbf
  {\bibinfo {volume} {178}},\ \bibinfo {pages} {685} (\bibinfo {year}
  {2008})}\BibitemShut {NoStop}%
\bibitem [{\citenamefont {Li}\ \emph {et~al.}(2017)\citenamefont {Li},
  \citenamefont {Zhou}, \citenamefont {Nummy}, \citenamefont {Zhang},
  \citenamefont {Pardo}, \citenamefont {Pickett}, \citenamefont {Mitchell},\
  and\ \citenamefont {Dessau}}]{NCARPES}%
  \BibitemOpen
  \bibfield  {author} {\bibinfo {author} {\bibfnamefont {H.}~\bibnamefont
  {Li}}, \bibinfo {author} {\bibfnamefont {X.}~\bibnamefont {Zhou}}, \bibinfo
  {author} {\bibfnamefont {T.}~\bibnamefont {Nummy}}, \bibinfo {author}
  {\bibfnamefont {J.}~\bibnamefont {Zhang}}, \bibinfo {author} {\bibfnamefont
  {V.}~\bibnamefont {Pardo}}, \bibinfo {author} {\bibfnamefont {W.~E.}\
  \bibnamefont {Pickett}}, \bibinfo {author} {\bibfnamefont {J.~F.}\
  \bibnamefont {Mitchell}},\ and\ \bibinfo {author} {\bibfnamefont {D.~S.}\
  \bibnamefont {Dessau}},\ }\bibfield  {title} {\bibinfo {title} {Fermiology
  and electron dynamics of trilayer nickelate {La$_4$Ni$_3$O$_{10}$}},\ }\href
  {https://doi.org/10.1038/s41467-017-00777-0} {\bibfield  {journal} {\bibinfo
  {journal} {Nature Communications}\ }\textbf {\bibinfo {volume} {8}},\
  \bibinfo {pages} {704} (\bibinfo {year} {2017})}\BibitemShut {NoStop}%
\bibitem [{\citenamefont {Christiansson}\ \emph {et~al.}(2023)\citenamefont
  {Christiansson}, \citenamefont {Petocchi},\ and\ \citenamefont
  {Werner}}]{PhysRevLett.131.206501}%
  \BibitemOpen
  \bibfield  {author} {\bibinfo {author} {\bibfnamefont {V.}~\bibnamefont
  {Christiansson}}, \bibinfo {author} {\bibfnamefont {F.}~\bibnamefont
  {Petocchi}},\ and\ \bibinfo {author} {\bibfnamefont {P.}~\bibnamefont
  {Werner}},\ }\bibfield  {title} {\bibinfo {title} {Correlated electronic
  structure of {${\mathrm{La}}_{3}{\text{Ni}}_{2}{\mathrm{O}}_{7}$} under
  pressure},\ }\href {https://doi.org/10.1103/PhysRevLett.131.206501}
  {\bibfield  {journal} {\bibinfo  {journal} {Phys. Rev. Lett.}\ }\textbf
  {\bibinfo {volume} {131}},\ \bibinfo {pages} {206501} (\bibinfo {year}
  {2023})}\BibitemShut {NoStop}%
\bibitem [{\citenamefont {Sakakibara}\ \emph {et~al.}(2024)\citenamefont
  {Sakakibara}, \citenamefont {Ochi}, \citenamefont {Nagata}, \citenamefont
  {Ueki}, \citenamefont {Sakurai}, \citenamefont {Matsumoto}, \citenamefont
  {Terashima}, \citenamefont {Hirose}, \citenamefont {Ohta}, \citenamefont
  {Kato}, \citenamefont {Takano},\ and\ \citenamefont
  {Kuroki}}]{sakakibara2023theoretical}%
  \BibitemOpen
  \bibfield  {author} {\bibinfo {author} {\bibfnamefont {H.}~\bibnamefont
  {Sakakibara}}, \bibinfo {author} {\bibfnamefont {M.}~\bibnamefont {Ochi}},
  \bibinfo {author} {\bibfnamefont {H.}~\bibnamefont {Nagata}}, \bibinfo
  {author} {\bibfnamefont {Y.}~\bibnamefont {Ueki}}, \bibinfo {author}
  {\bibfnamefont {H.}~\bibnamefont {Sakurai}}, \bibinfo {author} {\bibfnamefont
  {R.}~\bibnamefont {Matsumoto}}, \bibinfo {author} {\bibfnamefont
  {K.}~\bibnamefont {Terashima}}, \bibinfo {author} {\bibfnamefont
  {K.}~\bibnamefont {Hirose}}, \bibinfo {author} {\bibfnamefont
  {H.}~\bibnamefont {Ohta}}, \bibinfo {author} {\bibfnamefont {M.}~\bibnamefont
  {Kato}}, \bibinfo {author} {\bibfnamefont {Y.}~\bibnamefont {Takano}},\ and\
  \bibinfo {author} {\bibfnamefont {K.}~\bibnamefont {Kuroki}},\ }\bibfield
  {title} {\bibinfo {title} {Theoretical analysis on the possibility of
  superconductivity in the trilayer ruddlesden-popper
  nickelate{La$_4$Ni$_3$O$_{10}$} under pressure and its experimental
  examination: Comparison with {La$_3$Ni$_2$O$_7$}},\ }\href
  {https://doi.org/10.1103/PhysRevB.109.144511} {\bibfield  {journal} {\bibinfo
   {journal} {Phys. Rev. B}\ }\textbf {\bibinfo {volume} {109}},\ \bibinfo
  {pages} {144511} (\bibinfo {year} {2024})}\BibitemShut {NoStop}%
\bibitem [{\citenamefont {Kamihara}\ \emph {et~al.}(2006)\citenamefont
  {Kamihara}, \citenamefont {Hiramatsu}, \citenamefont {Hirano}, \citenamefont
  {Kawamura}, \citenamefont {Yanagi}, \citenamefont {Kamiya},\ and\
  \citenamefont {Hosono}}]{RN26}%
  \BibitemOpen
  \bibfield  {author} {\bibinfo {author} {\bibfnamefont {Y.}~\bibnamefont
  {Kamihara}}, \bibinfo {author} {\bibfnamefont {H.}~\bibnamefont {Hiramatsu}},
  \bibinfo {author} {\bibfnamefont {M.}~\bibnamefont {Hirano}}, \bibinfo
  {author} {\bibfnamefont {R.}~\bibnamefont {Kawamura}}, \bibinfo {author}
  {\bibfnamefont {H.}~\bibnamefont {Yanagi}}, \bibinfo {author} {\bibfnamefont
  {T.}~\bibnamefont {Kamiya}},\ and\ \bibinfo {author} {\bibfnamefont
  {H.}~\bibnamefont {Hosono}},\ }\bibfield  {title} {\bibinfo {title}
  {Iron-based layered superconductor: {LaOFeP}},\ }\href
  {https://doi.org/10.1021/ja063355c} {\bibfield  {journal} {\bibinfo
  {journal} {Journal of the American Chemical Society}\ }\textbf {\bibinfo
  {volume} {128}},\ \bibinfo {pages} {10012} (\bibinfo {year}
  {2006})}\BibitemShut {NoStop}%
\bibitem [{\citenamefont {Leb\`egue}(2007)}]{PhysRevB.75.035110}%
  \BibitemOpen
  \bibfield  {author} {\bibinfo {author} {\bibfnamefont {S.}~\bibnamefont
  {Leb\`egue}},\ }\bibfield  {title} {\bibinfo {title} {Electronic structure
  and properties of the {F}ermi surface of the superconductor {LaOFeP}},\
  }\href {https://doi.org/10.1103/PhysRevB.75.035110} {\bibfield  {journal}
  {\bibinfo  {journal} {Phys. Rev. B}\ }\textbf {\bibinfo {volume} {75}},\
  \bibinfo {pages} {035110} (\bibinfo {year} {2007})}\BibitemShut {NoStop}%
\bibitem [{\citenamefont {Leonov}(2024)}]{leonov2024electronic}%
  \BibitemOpen
  \bibfield  {author} {\bibinfo {author} {\bibfnamefont {I.~V.}\ \bibnamefont
  {Leonov}},\ }\href@noop {} {\bibinfo {title} {Electronic structure and
  magnetic correlations in trilayer nickelate superconductor
  {La$_4$Ni$_3$O$_{10}$} under pressure}} (\bibinfo {year} {2024}),\ \Eprint
  {https://arxiv.org/abs/2401.07350} {arXiv:2401.07350 [cond-mat.str-el]}
  \BibitemShut {NoStop}%
\bibitem [{\citenamefont {Zhang}\ \emph
  {et~al.}(2020{\natexlab{b}})\citenamefont {Zhang}, \citenamefont {Phelan},
  \citenamefont {Botana}, \citenamefont {Chen}, \citenamefont {Zheng},
  \citenamefont {Krogstad}, \citenamefont {Wang}, \citenamefont {Qiu},
  \citenamefont {Rodriguez-Rivera}, \citenamefont {Osborn}, \citenamefont
  {Rosenkranz}, \citenamefont {Norman},\ and\ \citenamefont
  {Mitchell}}]{2020intertwined}%
  \BibitemOpen
  \bibfield  {author} {\bibinfo {author} {\bibfnamefont {J.}~\bibnamefont
  {Zhang}}, \bibinfo {author} {\bibfnamefont {D.}~\bibnamefont {Phelan}},
  \bibinfo {author} {\bibfnamefont {A.~S.}\ \bibnamefont {Botana}}, \bibinfo
  {author} {\bibfnamefont {Y.~S.}\ \bibnamefont {Chen}}, \bibinfo {author}
  {\bibfnamefont {H.}~\bibnamefont {Zheng}}, \bibinfo {author} {\bibfnamefont
  {M.}~\bibnamefont {Krogstad}}, \bibinfo {author} {\bibfnamefont {S.~G.}\
  \bibnamefont {Wang}}, \bibinfo {author} {\bibfnamefont {Y.}~\bibnamefont
  {Qiu}}, \bibinfo {author} {\bibfnamefont {J.~A.}\ \bibnamefont
  {Rodriguez-Rivera}}, \bibinfo {author} {\bibfnamefont {R.}~\bibnamefont
  {Osborn}}, \bibinfo {author} {\bibfnamefont {S.}~\bibnamefont {Rosenkranz}},
  \bibinfo {author} {\bibfnamefont {M.~R.}\ \bibnamefont {Norman}},\ and\
  \bibinfo {author} {\bibfnamefont {J.~F.}\ \bibnamefont {Mitchell}},\
  }\bibfield  {title} {\bibinfo {title} {Intertwined density waves in a
  metallic nickelate},\ }\href {https://doi.org/10.1038/s41467-020-19836-0}
  {\bibfield  {journal} {\bibinfo  {journal} {Nat Commun}\ }\textbf {\bibinfo
  {volume} {11}},\ \bibinfo {pages} {6003} (\bibinfo {year}
  {2020}{\natexlab{b}})}\BibitemShut {NoStop}%
\bibitem [{\citenamefont {Norman}\ \emph {et~al.}(2023)\citenamefont {Norman},
  \citenamefont {Botana}, \citenamefont {Karp}, \citenamefont {Hampel},
  \citenamefont {LaBollita}, \citenamefont {Millis}, \citenamefont {Fabbris},
  \citenamefont {Shen},\ and\ \citenamefont {Dean}}]{PhysRevB.107.165124}%
  \BibitemOpen
  \bibfield  {author} {\bibinfo {author} {\bibfnamefont {M.~R.}\ \bibnamefont
  {Norman}}, \bibinfo {author} {\bibfnamefont {A.~S.}\ \bibnamefont {Botana}},
  \bibinfo {author} {\bibfnamefont {J.}~\bibnamefont {Karp}}, \bibinfo {author}
  {\bibfnamefont {A.}~\bibnamefont {Hampel}}, \bibinfo {author} {\bibfnamefont
  {H.}~\bibnamefont {LaBollita}}, \bibinfo {author} {\bibfnamefont {A.~J.}\
  \bibnamefont {Millis}}, \bibinfo {author} {\bibfnamefont {G.}~\bibnamefont
  {Fabbris}}, \bibinfo {author} {\bibfnamefont {Y.}~\bibnamefont {Shen}},\ and\
  \bibinfo {author} {\bibfnamefont {M.~P.~M.}\ \bibnamefont {Dean}},\
  }\bibfield  {title} {\bibinfo {title} {Orbital polarization, charge transfer,
  and fluorescence in reduced-valence nickelates},\ }\href
  {https://doi.org/10.1103/PhysRevB.107.165124} {\bibfield  {journal} {\bibinfo
   {journal} {Phys. Rev. B}\ }\textbf {\bibinfo {volume} {107}},\ \bibinfo
  {pages} {165124} (\bibinfo {year} {2023})}\BibitemShut {NoStop}%
\bibitem [{\citenamefont {Georges}\ \emph {et~al.}(2013)\citenamefont
  {Georges}, \citenamefont {Medici},\ and\ \citenamefont
  {Mravlje}}]{annurev-conmatphys-020911-125045}%
  \BibitemOpen
  \bibfield  {author} {\bibinfo {author} {\bibfnamefont {A.}~\bibnamefont
  {Georges}}, \bibinfo {author} {\bibfnamefont {L.~d.}\ \bibnamefont
  {Medici}},\ and\ \bibinfo {author} {\bibfnamefont {J.}~\bibnamefont
  {Mravlje}},\ }\bibfield  {title} {\bibinfo {title} {Strong correlations from
  {H}und's coupling},\ }\href
  {https://doi.org/10.1146/annurev-conmatphys-020911-125045} {\bibfield
  {journal} {\bibinfo  {journal} {Annual Review of Condensed Matter Physics}\
  }\textbf {\bibinfo {volume} {4}},\ \bibinfo {pages} {137} (\bibinfo {year}
  {2013})}\BibitemShut {NoStop}%
\bibitem [{\citenamefont {Sun}\ \emph {et~al.}(2021)\citenamefont {Sun},
  \citenamefont {Li}, \citenamefont {Cai}, \citenamefont {Yang}, \citenamefont
  {Guo}, \citenamefont {Gu}, \citenamefont {Zhu},\ and\ \citenamefont
  {Nie}}]{PhysRevB.104.184518}%
  \BibitemOpen
  \bibfield  {author} {\bibinfo {author} {\bibfnamefont {W.}~\bibnamefont
  {Sun}}, \bibinfo {author} {\bibfnamefont {Y.}~\bibnamefont {Li}}, \bibinfo
  {author} {\bibfnamefont {X.}~\bibnamefont {Cai}}, \bibinfo {author}
  {\bibfnamefont {J.}~\bibnamefont {Yang}}, \bibinfo {author} {\bibfnamefont
  {W.}~\bibnamefont {Guo}}, \bibinfo {author} {\bibfnamefont {Z.}~\bibnamefont
  {Gu}}, \bibinfo {author} {\bibfnamefont {Y.}~\bibnamefont {Zhu}},\ and\
  \bibinfo {author} {\bibfnamefont {Y.}~\bibnamefont {Nie}},\ }\bibfield
  {title} {\bibinfo {title} {Electronic and transport properties in
  ruddlesden-popper neodymium nickelates
  {${\mathrm{Nd}}_{n+1}{\mathrm{Ni}}_{n}{\mathrm{O}}_{3n+1}$ ($n=1-5$)}},\
  }\href {https://doi.org/10.1103/PhysRevB.104.184518} {\bibfield  {journal}
  {\bibinfo  {journal} {Phys. Rev. B}\ }\textbf {\bibinfo {volume} {104}},\
  \bibinfo {pages} {184518} (\bibinfo {year} {2021})}\BibitemShut {NoStop}%
\bibitem [{\citenamefont {Chen}\ \emph {et~al.}(2024)\citenamefont {Chen},
  \citenamefont {Choi}, \citenamefont {Jiang}, \citenamefont {Mei},
  \citenamefont {Jiang}, \citenamefont {Li}, \citenamefont {Agrestini},
  \citenamefont {Garcia-Fernandez}, \citenamefont {Huang}, \citenamefont {Sun},
  \citenamefont {Shen}, \citenamefont {Wang}, \citenamefont {Hu}, \citenamefont
  {Lu}, \citenamefont {Zhou},\ and\ \citenamefont {Feng}}]{chen2024electronic}%
  \BibitemOpen
  \bibfield  {author} {\bibinfo {author} {\bibfnamefont {X.}~\bibnamefont
  {Chen}}, \bibinfo {author} {\bibfnamefont {J.}~\bibnamefont {Choi}}, \bibinfo
  {author} {\bibfnamefont {Z.}~\bibnamefont {Jiang}}, \bibinfo {author}
  {\bibfnamefont {J.}~\bibnamefont {Mei}}, \bibinfo {author} {\bibfnamefont
  {K.}~\bibnamefont {Jiang}}, \bibinfo {author} {\bibfnamefont
  {J.}~\bibnamefont {Li}}, \bibinfo {author} {\bibfnamefont {S.}~\bibnamefont
  {Agrestini}}, \bibinfo {author} {\bibfnamefont {M.}~\bibnamefont
  {Garcia-Fernandez}}, \bibinfo {author} {\bibfnamefont {X.}~\bibnamefont
  {Huang}}, \bibinfo {author} {\bibfnamefont {H.}~\bibnamefont {Sun}}, \bibinfo
  {author} {\bibfnamefont {D.}~\bibnamefont {Shen}}, \bibinfo {author}
  {\bibfnamefont {M.}~\bibnamefont {Wang}}, \bibinfo {author} {\bibfnamefont
  {J.}~\bibnamefont {Hu}}, \bibinfo {author} {\bibfnamefont {Y.}~\bibnamefont
  {Lu}}, \bibinfo {author} {\bibfnamefont {K.-J.}\ \bibnamefont {Zhou}},\ and\
  \bibinfo {author} {\bibfnamefont {D.}~\bibnamefont {Feng}},\ }\href@noop {}
  {\bibinfo {title} {Electronic and magnetic excitations in
  {La$_3$Ni$_2$O$_7$}}} (\bibinfo {year} {2024}),\ \Eprint
  {https://arxiv.org/abs/2401.12657} {arXiv:2401.12657 [cond-mat.supr-con]}
  \BibitemShut {NoStop}%
\bibitem [{\citenamefont {Tam}\ \emph {et~al.}(2015)\citenamefont {Tam},
  \citenamefont {Yao},\ and\ \citenamefont {Ku}}]{tamyuting2015}%
  \BibitemOpen
  \bibfield  {author} {\bibinfo {author} {\bibfnamefont {Y.~T.}\ \bibnamefont
  {Tam}}, \bibinfo {author} {\bibfnamefont {D.~X.}\ \bibnamefont {Yao}},\ and\
  \bibinfo {author} {\bibfnamefont {W.}~\bibnamefont {Ku}},\ }\bibfield
  {title} {\bibinfo {title} {Itinerancy-enhanced quantum fluctuation of
  magnetic moments in iron-based superconductors},\ }\href
  {https://doi.org/10.1103/PhysRevLett.115.117001} {\bibfield  {journal}
  {\bibinfo  {journal} {Phys Rev Lett}\ }\textbf {\bibinfo {volume} {115}},\
  \bibinfo {pages} {117001} (\bibinfo {year} {2015})}\BibitemShut {NoStop}%
\bibitem [{\citenamefont {Zaanen}\ \emph {et~al.}(1985)\citenamefont {Zaanen},
  \citenamefont {Sawatzky},\ and\ \citenamefont
  {Allen}}]{ZSA-PhysRevLett.55.418}%
  \BibitemOpen
  \bibfield  {author} {\bibinfo {author} {\bibfnamefont {J.}~\bibnamefont
  {Zaanen}}, \bibinfo {author} {\bibfnamefont {G.~A.}\ \bibnamefont
  {Sawatzky}},\ and\ \bibinfo {author} {\bibfnamefont {J.~W.}\ \bibnamefont
  {Allen}},\ }\bibfield  {title} {\bibinfo {title} {Band gaps and electronic
  structure of transition-metal compounds},\ }\href
  {https://doi.org/10.1103/PhysRevLett.55.418} {\bibfield  {journal} {\bibinfo
  {journal} {Phys. Rev. Lett.}\ }\textbf {\bibinfo {volume} {55}},\ \bibinfo
  {pages} {418} (\bibinfo {year} {1985})}\BibitemShut {NoStop}%
\bibitem [{\citenamefont {Zhang}\ and\ \citenamefont
  {Rice}(1988)}]{zhang-rice1988}%
  \BibitemOpen
  \bibfield  {author} {\bibinfo {author} {\bibfnamefont {F.~C.}\ \bibnamefont
  {Zhang}}\ and\ \bibinfo {author} {\bibfnamefont {T.~M.}\ \bibnamefont
  {Rice}},\ }\bibfield  {title} {\bibinfo {title} {Effective {H}amiltonian for
  the superconducting {C}u oxides},\ }\href
  {https://doi.org/10.1103/PhysRevB.37.3759} {\bibfield  {journal} {\bibinfo
  {journal} {Phys. Rev. B}\ }\textbf {\bibinfo {volume} {37}},\ \bibinfo
  {pages} {3759} (\bibinfo {year} {1988})}\BibitemShut {NoStop}%
\bibitem [{\citenamefont {Kowalski}\ \emph {et~al.}(2021)\citenamefont
  {Kowalski}, \citenamefont {Dash}, \citenamefont {Semon}, \citenamefont
  {Senechal},\ and\ \citenamefont {Tremblay}}]{pnas_chargetransfer}%
  \BibitemOpen
  \bibfield  {author} {\bibinfo {author} {\bibfnamefont {N.}~\bibnamefont
  {Kowalski}}, \bibinfo {author} {\bibfnamefont {S.~S.}\ \bibnamefont {Dash}},
  \bibinfo {author} {\bibfnamefont {P.}~\bibnamefont {Semon}}, \bibinfo
  {author} {\bibfnamefont {D.}~\bibnamefont {Senechal}},\ and\ \bibinfo
  {author} {\bibfnamefont {A.~M.}\ \bibnamefont {Tremblay}},\ }\bibfield
  {title} {\bibinfo {title} {Oxygen hole content, charge-transfer gap,
  covalency, and cuprate superconductivity},\ }\bibfield  {journal} {\bibinfo
  {journal} {Proc Natl Acad Sci U S A}\ }\textbf {\bibinfo {volume} {118}},\
  \href {https://doi.org/10.1073/pnas.2106476118} {10.1073/pnas.2106476118}
  (\bibinfo {year} {2021})\BibitemShut {NoStop}%
\bibitem [{\citenamefont {Been}\ \emph {et~al.}(2021)\citenamefont {Been},
  \citenamefont {Lee}, \citenamefont {Hwang}, \citenamefont {Cui},
  \citenamefont {Zaanen}, \citenamefont {Devereaux}, \citenamefont {Moritz},\
  and\ \citenamefont {Jia}}]{PhysRevX.11.011050}%
  \BibitemOpen
  \bibfield  {author} {\bibinfo {author} {\bibfnamefont {E.}~\bibnamefont
  {Been}}, \bibinfo {author} {\bibfnamefont {W.-S.}\ \bibnamefont {Lee}},
  \bibinfo {author} {\bibfnamefont {H.~Y.}\ \bibnamefont {Hwang}}, \bibinfo
  {author} {\bibfnamefont {Y.}~\bibnamefont {Cui}}, \bibinfo {author}
  {\bibfnamefont {J.}~\bibnamefont {Zaanen}}, \bibinfo {author} {\bibfnamefont
  {T.}~\bibnamefont {Devereaux}}, \bibinfo {author} {\bibfnamefont
  {B.}~\bibnamefont {Moritz}},\ and\ \bibinfo {author} {\bibfnamefont
  {C.}~\bibnamefont {Jia}},\ }\bibfield  {title} {\bibinfo {title} {Electronic
  structure trends across the rare-earth series in superconducting
  infinite-layer nickelates},\ }\href
  {https://doi.org/10.1103/PhysRevX.11.011050} {\bibfield  {journal} {\bibinfo
  {journal} {Phys. Rev. X}\ }\textbf {\bibinfo {volume} {11}},\ \bibinfo
  {pages} {011050} (\bibinfo {year} {2021})}\BibitemShut {NoStop}%
\bibitem [{\citenamefont {Lu}\ \emph {et~al.}(2023)\citenamefont {Lu},
  \citenamefont {Pan}, \citenamefont {Yang},\ and\ \citenamefont
  {Wu}}]{lu2023interplay}%
  \BibitemOpen
  \bibfield  {author} {\bibinfo {author} {\bibfnamefont {C.}~\bibnamefont
  {Lu}}, \bibinfo {author} {\bibfnamefont {Z.}~\bibnamefont {Pan}}, \bibinfo
  {author} {\bibfnamefont {F.}~\bibnamefont {Yang}},\ and\ \bibinfo {author}
  {\bibfnamefont {C.}~\bibnamefont {Wu}},\ }\href@noop {} {\bibinfo {title}
  {Interplay of two $e_g$ orbitals in superconducting {La$_3$Ni$_2$O$_7$} under
  pressure}} (\bibinfo {year} {2023}),\ \Eprint
  {https://arxiv.org/abs/2310.02915} {arXiv:2310.02915 [cond-mat.supr-con]}
  \BibitemShut {NoStop}%
\bibitem [{\citenamefont {Lawrence}\ \emph {et~al.}(1981)\citenamefont
  {Lawrence}, \citenamefont {Riseborough},\ and\ \citenamefont
  {Parks}}]{JMLawrence_1981}%
  \BibitemOpen
  \bibfield  {author} {\bibinfo {author} {\bibfnamefont {J.~M.}\ \bibnamefont
  {Lawrence}}, \bibinfo {author} {\bibfnamefont {P.~S.}\ \bibnamefont
  {Riseborough}},\ and\ \bibinfo {author} {\bibfnamefont {R.~D.}\ \bibnamefont
  {Parks}},\ }\bibfield  {title} {\bibinfo {title} {Valence fluctuation
  phenomena},\ }\href {https://doi.org/10.1088/0034-4885/44/1/001} {\bibfield
  {journal} {\bibinfo  {journal} {Reports on Progress in Physics}\ }\textbf
  {\bibinfo {volume} {44}},\ \bibinfo {pages} {1} (\bibinfo {year}
  {1981})}\BibitemShut {NoStop}%
\bibitem [{\citenamefont {Tian}\ \emph {et~al.}(2023)\citenamefont {Tian},
  \citenamefont {Chen}, \citenamefont {Wang}, \citenamefont {He},\ and\
  \citenamefont {Lu}}]{tian2023correlation}%
  \BibitemOpen
  \bibfield  {author} {\bibinfo {author} {\bibfnamefont {Y.-H.}\ \bibnamefont
  {Tian}}, \bibinfo {author} {\bibfnamefont {Y.}~\bibnamefont {Chen}}, \bibinfo
  {author} {\bibfnamefont {J.-M.}\ \bibnamefont {Wang}}, \bibinfo {author}
  {\bibfnamefont {R.-Q.}\ \bibnamefont {He}},\ and\ \bibinfo {author}
  {\bibfnamefont {Z.-Y.}\ \bibnamefont {Lu}},\ }\href@noop {} {\bibinfo {title}
  {Correlation effects and concomitant two-orbital $s_\pm$-wave
  superconductivity in {La$_3$Ni$_2$O$_7$} under high pressure}} (\bibinfo
  {year} {2023}),\ \Eprint {https://arxiv.org/abs/2308.09698} {arXiv:2308.09698
  [cond-mat.supr-con]} \BibitemShut {NoStop}%
\bibitem [{\citenamefont {Zhang}\ \emph
  {et~al.}(2023{\natexlab{b}})\citenamefont {Zhang}, \citenamefont {Lin},
  \citenamefont {Moreo},\ and\ \citenamefont {Dagotto}}]{PhysRevB.108.L180510}%
  \BibitemOpen
  \bibfield  {author} {\bibinfo {author} {\bibfnamefont {Y.}~\bibnamefont
  {Zhang}}, \bibinfo {author} {\bibfnamefont {L.-F.}\ \bibnamefont {Lin}},
  \bibinfo {author} {\bibfnamefont {A.}~\bibnamefont {Moreo}},\ and\ \bibinfo
  {author} {\bibfnamefont {E.}~\bibnamefont {Dagotto}},\ }\bibfield  {title}
  {\bibinfo {title} {Electronic structure, dimer physics, orbital-selective
  behavior, and magnetic tendencies in the bilayer nickelate superconductor
  {${\mathrm{La}}_{3}{\mathrm{Ni}}_{2}{\mathrm{O}}_{7}$} under pressure},\
  }\href {https://doi.org/10.1103/PhysRevB.108.L180510} {\bibfield  {journal}
  {\bibinfo  {journal} {Phys. Rev. B}\ }\textbf {\bibinfo {volume} {108}},\
  \bibinfo {pages} {L180510} (\bibinfo {year}
  {2023}{\natexlab{b}})}\BibitemShut {NoStop}%
\bibitem [{\citenamefont {Yang}\ \emph
  {et~al.}(2023{\natexlab{c}})\citenamefont {Yang}, \citenamefont {Zhang},\
  and\ \citenamefont {Zhang}}]{PhysRevB.108.L201108}%
  \BibitemOpen
  \bibfield  {author} {\bibinfo {author} {\bibfnamefont {Y.-f.}\ \bibnamefont
  {Yang}}, \bibinfo {author} {\bibfnamefont {G.-M.}\ \bibnamefont {Zhang}},\
  and\ \bibinfo {author} {\bibfnamefont {F.-C.}\ \bibnamefont {Zhang}},\
  }\bibfield  {title} {\bibinfo {title} {Interlayer valence bonds and
  two-component theory for high-${T}_{c}$ superconductivity of
  {${\mathrm{La}}_{3}{\mathrm{Ni}}_{2}{\mathrm{O}}_{7}$} under pressure},\
  }\href {https://doi.org/10.1103/PhysRevB.108.L201108} {\bibfield  {journal}
  {\bibinfo  {journal} {Phys. Rev. B}\ }\textbf {\bibinfo {volume} {108}},\
  \bibinfo {pages} {L201108} (\bibinfo {year}
  {2023}{\natexlab{c}})}\BibitemShut {NoStop}%
\bibitem [{str(2023)}]{strange327}%
  \BibitemOpen
  \href@noop {} {\bibinfo {title} {High-temperature superconductivity with
  zero-resistance and strange metal behaviour in {La$_3$Ni$_2$O$_7$}}}
  (\bibinfo {year} {2023}),\ \Eprint {https://arxiv.org/abs/2307.14819}
  {arXiv:2307.14819 [cond-mat.str-el]} \BibitemShut {NoStop}%
\bibitem [{\citenamefont {Dagotto}\ and\ \citenamefont {Rice}(1996)}]{rnxx}%
  \BibitemOpen
  \bibfield  {author} {\bibinfo {author} {\bibfnamefont {E.}~\bibnamefont
  {Dagotto}}\ and\ \bibinfo {author} {\bibfnamefont {T.~M.}\ \bibnamefont
  {Rice}},\ }\bibfield  {title} {\bibinfo {title} {Surprises on the way from
  one- to two-dimensional quantum magnets: The ladder materials},\ }\href
  {https://doi.org/10.1126/science.271.5249.618} {\bibfield  {journal}
  {\bibinfo  {journal} {Science}\ }\textbf {\bibinfo {volume} {271}},\ \bibinfo
  {pages} {618} (\bibinfo {year} {1996})}\BibitemShut {NoStop}%
\bibitem [{\citenamefont {Cheng}\ \emph {et~al.}(2022)\citenamefont {Cheng},
  \citenamefont {Li}, \citenamefont {Xiong}, \citenamefont {Wu}, \citenamefont
  {Sandvik},\ and\ \citenamefont {Yao}}]{RN27}%
  \BibitemOpen
  \bibfield  {author} {\bibinfo {author} {\bibfnamefont {J.-Q.}\ \bibnamefont
  {Cheng}}, \bibinfo {author} {\bibfnamefont {J.}~\bibnamefont {Li}}, \bibinfo
  {author} {\bibfnamefont {Z.}~\bibnamefont {Xiong}}, \bibinfo {author}
  {\bibfnamefont {H.-Q.}\ \bibnamefont {Wu}}, \bibinfo {author} {\bibfnamefont
  {A.~W.}\ \bibnamefont {Sandvik}},\ and\ \bibinfo {author} {\bibfnamefont
  {D.-X.}\ \bibnamefont {Yao}},\ }\bibfield  {title} {\bibinfo {title}
  {Fractional and composite excitations of antiferromagnetic quantum spin
  trimer chains},\ }\href {https://doi.org/10.1038/s41535-021-00416-4}
  {\bibfield  {journal} {\bibinfo  {journal} {npj Quantum Materials}\ }\textbf
  {\bibinfo {volume} {7}},\ \bibinfo {pages} {3} (\bibinfo {year}
  {2022})}\BibitemShut {NoStop}%
\bibitem [{\citenamefont {Wang}\ \emph
  {et~al.}(2024{\natexlab{b}})\citenamefont {Wang}, \citenamefont {Ouyang},
  \citenamefont {He},\ and\ \citenamefont {Lu}}]{wang2024nonfermi}%
  \BibitemOpen
  \bibfield  {author} {\bibinfo {author} {\bibfnamefont {J.-X.}\ \bibnamefont
  {Wang}}, \bibinfo {author} {\bibfnamefont {Z.}~\bibnamefont {Ouyang}},
  \bibinfo {author} {\bibfnamefont {R.-Q.}\ \bibnamefont {He}},\ and\ \bibinfo
  {author} {\bibfnamefont {Z.-Y.}\ \bibnamefont {Lu}},\ }\href@noop {}
  {\bibinfo {title} {Non-{F}ermi liquid and {H}und correlation in
  $\mathrm{La_4Ni_3O_{10}}$ under high pressure}} (\bibinfo {year}
  {2024}{\natexlab{b}}),\ \Eprint {https://arxiv.org/abs/2402.02581}
  {arXiv:2402.02581 [cond-mat.str-el]} \BibitemShut {NoStop}%
\bibitem [{\citenamefont {Zhang}\ \emph
  {et~al.}(2024{\natexlab{c}})\citenamefont {Zhang}, \citenamefont {Lin},
  \citenamefont {Moreo}, \citenamefont {Maier},\ and\ \citenamefont
  {Dagotto}}]{zhang2024prediction}%
  \BibitemOpen
  \bibfield  {author} {\bibinfo {author} {\bibfnamefont {Y.}~\bibnamefont
  {Zhang}}, \bibinfo {author} {\bibfnamefont {L.-F.}\ \bibnamefont {Lin}},
  \bibinfo {author} {\bibfnamefont {A.}~\bibnamefont {Moreo}}, \bibinfo
  {author} {\bibfnamefont {T.~A.}\ \bibnamefont {Maier}},\ and\ \bibinfo
  {author} {\bibfnamefont {E.}~\bibnamefont {Dagotto}},\ }\href@noop {}
  {\bibinfo {title} {Prediction of $s^\pm$-wave superconductivity enhanced by
  electronic doping in trilayer nickelates $\mathrm{La_4Ni_3O_{10}}$ under
  pressure}} (\bibinfo {year} {2024}{\natexlab{c}}),\ \Eprint
  {https://arxiv.org/abs/2402.05285} {arXiv:2402.05285 [cond-mat.supr-con]}
  \BibitemShut {NoStop}%
\bibitem [{\citenamefont {Yang}\ \emph {et~al.}(2024)\citenamefont {Yang},
  \citenamefont {Jiang}, \citenamefont {Wang}, \citenamefont {Lu},\ and\
  \citenamefont {Wang}}]{yang2024effective}%
  \BibitemOpen
  \bibfield  {author} {\bibinfo {author} {\bibfnamefont {Q.-G.}\ \bibnamefont
  {Yang}}, \bibinfo {author} {\bibfnamefont {K.-Y.}\ \bibnamefont {Jiang}},
  \bibinfo {author} {\bibfnamefont {D.}~\bibnamefont {Wang}}, \bibinfo {author}
  {\bibfnamefont {H.-Y.}\ \bibnamefont {Lu}},\ and\ \bibinfo {author}
  {\bibfnamefont {Q.-H.}\ \bibnamefont {Wang}},\ }\href@noop {} {\bibinfo
  {title} {Effective model and $s_\pm$-wave superconductivity in trilayer
  nickelate $\mathrm{La_4Ni_3O_{10}}$}} (\bibinfo {year} {2024}),\ \Eprint
  {https://arxiv.org/abs/2402.05447} {arXiv:2402.05447 [cond-mat.supr-con]}
  \BibitemShut {NoStop}%
\bibitem [{\citenamefont {Lu}\ \emph {et~al.}(2024)\citenamefont {Lu},
  \citenamefont {Pan}, \citenamefont {Yang},\ and\ \citenamefont
  {Wu}}]{lu2024superconductivity}%
  \BibitemOpen
  \bibfield  {author} {\bibinfo {author} {\bibfnamefont {C.}~\bibnamefont
  {Lu}}, \bibinfo {author} {\bibfnamefont {Z.}~\bibnamefont {Pan}}, \bibinfo
  {author} {\bibfnamefont {F.}~\bibnamefont {Yang}},\ and\ \bibinfo {author}
  {\bibfnamefont {C.}~\bibnamefont {Wu}},\ }\href@noop {} {\bibinfo {title}
  {Superconductivity in $\mathrm{La_4Ni_3O_{10}}$ under pressure}} (\bibinfo
  {year} {2024}),\ \Eprint {https://arxiv.org/abs/2402.06450} {arXiv:2402.06450
  [cond-mat.supr-con]} \BibitemShut {NoStop}%
\bibitem [{\citenamefont {LaBollita}\ \emph {et~al.}(2024)\citenamefont
  {LaBollita}, \citenamefont {Kapeghian}, \citenamefont {Norman},\ and\
  \citenamefont {Botana}}]{labollita2024electronic}%
  \BibitemOpen
  \bibfield  {author} {\bibinfo {author} {\bibfnamefont {H.}~\bibnamefont
  {LaBollita}}, \bibinfo {author} {\bibfnamefont {J.}~\bibnamefont
  {Kapeghian}}, \bibinfo {author} {\bibfnamefont {M.~R.}\ \bibnamefont
  {Norman}},\ and\ \bibinfo {author} {\bibfnamefont {A.~S.}\ \bibnamefont
  {Botana}},\ }\href@noop {} {\bibinfo {title} {Electronic structure and
  magnetic tendencies of trilayer $\mathrm{La_4Ni_3O_{10}}$ under pressure:
  structural transition, molecular orbitals, and layer differentiation}}
  (\bibinfo {year} {2024}),\ \Eprint {https://arxiv.org/abs/2402.05085}
  {arXiv:2402.05085 [cond-mat.supr-con]} \BibitemShut {NoStop}%
\bibitem [{\citenamefont {Tian}\ \emph {et~al.}(2024)\citenamefont {Tian},
  \citenamefont {Ma}, \citenamefont {Ming}, \citenamefont {Zheng},\ and\
  \citenamefont {Li}}]{tian2024effective}%
  \BibitemOpen
  \bibfield  {author} {\bibinfo {author} {\bibfnamefont {P.-F.}\ \bibnamefont
  {Tian}}, \bibinfo {author} {\bibfnamefont {H.-T.}\ \bibnamefont {Ma}},
  \bibinfo {author} {\bibfnamefont {X.}~\bibnamefont {Ming}}, \bibinfo {author}
  {\bibfnamefont {X.-J.}\ \bibnamefont {Zheng}},\ and\ \bibinfo {author}
  {\bibfnamefont {H.}~\bibnamefont {Li}},\ }\href@noop {} {\bibinfo {title}
  {Effective model and electron correlations in trilayer nickelate
  superconductor $\mathrm{La_4Ni_3O_{10}}$}} (\bibinfo {year} {2024}),\ \Eprint
  {https://arxiv.org/abs/2402.02351} {arXiv:2402.02351 [cond-mat.str-el]}
  \BibitemShut {NoStop}%
\bibitem [{\citenamefont {Zhang}\ \emph
  {et~al.}(2024{\natexlab{d}})\citenamefont {Zhang}, \citenamefont {Sun},
  \citenamefont {Liu}, \citenamefont {Liu}, \citenamefont {Chen},\ and\
  \citenamefont {Yang}}]{zhang2024spmwave}%
  \BibitemOpen
  \bibfield  {author} {\bibinfo {author} {\bibfnamefont {M.}~\bibnamefont
  {Zhang}}, \bibinfo {author} {\bibfnamefont {H.}~\bibnamefont {Sun}}, \bibinfo
  {author} {\bibfnamefont {Y.-B.}\ \bibnamefont {Liu}}, \bibinfo {author}
  {\bibfnamefont {Q.}~\bibnamefont {Liu}}, \bibinfo {author} {\bibfnamefont
  {W.-Q.}\ \bibnamefont {Chen}},\ and\ \bibinfo {author} {\bibfnamefont
  {F.}~\bibnamefont {Yang}},\ }\href@noop {} {\bibinfo {title} {The
  $s^\pm$-wave superconductivity in the pressurized $\mathrm{La_4Ni_3O_{10}}$}}
  (\bibinfo {year} {2024}{\natexlab{d}}),\ \Eprint
  {https://arxiv.org/abs/2402.07902} {arXiv:2402.07902 [cond-mat.supr-con]}
  \BibitemShut {NoStop}%
\end{thebibliography}%

\end{document}